\documentclass[prb,twocolumn,superscriptaddress,showpacs,notitlepage,floatfix,bibnotes,nofootinbib]{revtex4-2}
\usepackage{amsmath,amsthm,amsfonts,amssymb,mathtools,bbm}
\usepackage{graphicx}
\usepackage{floatrow}
\usepackage[caption=false]{subfig}
\usepackage{tikz}
\usepackage{tikz-cd}
\usepackage{bm}
\usepackage{float}
\usepackage{xcolor}
\usepackage{braket}

\usepackage[pdfencoding=auto, psdextra]{hyperref}
\hypersetup{linktocpage}
\hypersetup{colorlinks=true,citecolor=blue,linkcolor=blue, urlcolor=blue}

\makeatletter
\def\l@subsubsection#1#2{}
\makeatother

\newcommand{\ZZ}{\mathbb{Z}}
\newcommand{\CC}{\mathbb{C}}
\newcommand{\ee}{\mathsf{e}}
\newcommand{\mm}{\mathsf{m}}
\newcommand{\DD}{\mathsf{D}}
\newcommand{\fol}{\mathrm{fol}}

\newcommand{\TC}{\mathrm{TC}}
\newcommand{\XC}{\mathrm{XC}}
\newcommand{\SPT}{\mathrm{SPT}}
\newcommand{\PIM}{\mathrm{PIM}}
\newcommand{\dd}{\partial}

\newcommand{\ie}{\begin{equation}\begin{aligned}}
\newcommand{\fe}{\end{aligned}\end{equation}}

\newcommand{\ii}{\mathrm{i}}

\definecolor{NT}{rgb}{0.8,0,0.8}

\newcommand{\bb}[1]{\boldsymbol{#1}}

\graphicspath{{./Figures/}}

\newcommand{\nocontentsline}[3]{}
\let\origcontentsline\addcontentsline
\newcommand\stoptoc{\let\addcontentsline\nocontentsline}
\newcommand\resumetoc{\let\addcontentsline\origcontentsline}

\begin{document}

\date{\today}

\title{There and Back Again: A Gauging Nexus between Topological and Fracton Phases}

\author{Pranay Gorantla}
\affiliation{Leinweber Institute for Theoretical Physics \& Enrico Fermi Institute, University of Chicago, Chicago, IL 60637, USA}
\affiliation{Department of Physics, Princeton University, Princeton, NJ 08544, USA}
\affiliation{Theoretical Sciences Visiting Program, Okinawa Institute of Science and Technology Graduate University, Onna, 904-0495, Japan}
\author{Abhinav Prem}
\affiliation{Physics Program, Bard College, 30 Campus Road, Annandale-on-Hudson, NY 12504, USA}
\affiliation{School of Natural Sciences, Institute for Advanced Study, Princeton, NJ 08540, USA}
\affiliation{Theoretical Sciences Visiting Program, Okinawa Institute of Science and Technology Graduate University, Onna, 904-0495, Japan}
\author{Nathanan Tantivasadakarn}
\affiliation{C. N. Yang Institute for Theoretical Physics, Stony Brook University, Stony Brook, NY 11794, USA}
\affiliation{Walter Burke Institute for Theoretical Physics and Department of Physics, California Institute of Technology, Pasadena, CA, 91125, USA}
\author{Dominic J. Williamson}
\affiliation{School of Physics, The University of Sydney, NSW 2006, Australia}
\date{\today}

\begin{abstract}
Coupled layer constructions are a valuable tool for capturing the universal properties of certain interacting quantum phases of matter in terms of the simpler data that characterizes the underlying layers. In the study of fracton phases, the X-Cube model in 3+1D can be realized via such a construction by starting with a stack of 2+1D Toric Codes and turning on a coupling which condenses a composite ``particle-string" object. In a recent work [\href{https://doi.org/10.1103/qq9n-16hk}{Phys. Rev. B \textbf{112}, 125124 (2025)}], we have demonstrated that in fact, the particle-string can be viewed as a symmetry defect of a topological 1-form symmetry. In this paper, we study the result of gauging this symmetry in depth. We unveil a rich gauging web relating the X-Cube model to symmetry protected topological (SPT) phases protected by a mix of subsystem and higher-form symmetries, subsystem symmetry fractionalization in the 3+1D Toric Code, and non-trivial extensions of topological symmetries by subsystem symmetries. Our work emphasizes the importance of topological symmetries in non-topological, geometric phases of matter. 
\end{abstract}

\maketitle

\tableofcontents


\section{Introduction}
\label{sec:intro}

\begin{table*}[t]
    \centering
    \begin{tabular}{|c|l|c|l|}
    \hline
    Symmetry     & Description & Dual symmetry & Description\\
    \hline
      $G^{(k)}$ & $k$-form symmetry & $\hat G^{(2-k)}$ & $(2-k)$-form symmetry  \\
       $\hat{G}^{(1),\text{fol}} $ & 3-foliated $1$-form symmetry &  $\hat{G}^{(0),\text{fol}} $ & 3-foliated $0$-form symmetry\\
       $G^{f}$ & Fracton Wilson symmetry & $\hat G^{\hat f}$ & Fracton planar subsystem symmetry\\
    
       $G^{\ell}$ &  Lineon Wilson symmetry & $\hat G^{\hat \ell}$ & Lineon planar subsystem symmetry\\
      \hline
    \end{tabular}
    \caption{Descriptions of various 3+1D symmetries that appear in the gauging web Fig.~\ref{fig:gaugingweb}. In each row, gauging a symmetry yields the corresponding dual symmetry, and gauging the dual symmetry yields the original symmetry. The fracton(lineon) Wilson symmetry refers to the symmetry whose corresponding defect is a fracton(lineon), as in the X-Cube model. The corresponding dual symmetry is a 3-foliated planar symmetry whose corresponding charges are fractons(lineons).  }
    \label{tab:symmetries}
\end{table*}

\begin{table*}[t]
    \centering
    \begin{tabular}{|c|l|c|c|c|c|c|}
    \hline
    Theory     & Name & $\ee$-Symmetry & $\mm$-Symmetry & $\log$ GSD on $T^3$ & Hamiltonian & Lagrangian\\
    \hline
$\mathcal T_\text{fol}$ & 3-foliated stacks of $2+1$D Toric Codes& $G_\ee^{(1),\text{fol}}$ &  $G_\mm^{(1),\text{fol}}$ & $6L$ & Eq.~\eqref{eq:Hfol}&Eq.~\eqref{3dfollag}\\
$\mathcal T_{\text{TC},\mm}$ & $3+1$D $\hat{G}^{\hat{\ell}}_\mm$-enriched Toric Code& $G_\ee^{(2)}$ & $\hat{G}^{\hat{\ell}}_\mm \times G_m^{(1)}$ & $3$&Eq.~\eqref{eq:HTCm}&Eq.~\eqref{3dfollag-gaugelsym}\\
$\mathcal T_{\text{XC},\ee}$ & $3+1$D $\hat G_\ee^{(1)}$-enriched X-Cube& $G_\ee^{\ell} \times \hat G_\ee^{(1)}$   & $G^f_\mm$  & $6L-3$&Eq.~\eqref{eq:HXCe}&Eq.~\eqref{3dfollag-gauge1-form}\\
$\mathcal T_{0,\mm}$ & Trivial 3+1D SPT (product state)&  ---  &$\hat{G}_\mm^{(0),\text{fol}}$ & $0$&Eq.~\eqref{eq:H0m}&Eq.~\eqref{3dfollag-gaugefol1-form}\\
$\mathcal T_\text{SPT}$ & Nontrivial $3+1$D SPT&  $\hat{G}^{(1)}_\ee$  &   $\hat{G}^{\hat{\ell}}_\mm$ & $0$&Eq.~\eqref{eq:HSPT}&Eq.~\eqref{3dlagspt-p}\\
      \hline
    \end{tabular}
    \caption{Properties of the theories that participate in the gauging web Fig.~\ref{fig:gaugingweb}. For the ground state degeneracy, we place the theory on an $L\times L\times L$ periodic cubic lattice.}
    \label{tab:theories}
\end{table*}

Gapped fracton phases of matter arise as locally stable quantum phases in three (and higher) spatial dimensions (3+1D)~\cite{fractonreview,fractonreview2}. These phases exhibit gapped excitations which, in isolation, have restricted spatial mobility i.e., they can move only along sub-dimensional manifolds (planes or lines) or are fully immobile. Remarkably, while these phases have locally indistinguishable ground states, their ground-state degeneracy depends not only on the spatial topology but also depends sensitively on certain geometric data~\cite{haah,haah2,fracton1,fracton2,williamson,shirleygeneral,slagle3,symmetric}. This geometric sensitivity renders the low-energy description of gapped fracton phases beyond a conventional topological quantum field theory description. Rather, the universal properties of these phases are captured by foliated TQFTs~\cite{SlagleXcubeQFT,Slagle21}, so-called exotic field theories~\cite{SeibergSymmetry,Seiberg:2020bhn,seiberg2021zn,gorantla2021villain}, or more generally, by topological defect networks~\cite{SlagleSMN,Wen2020,defectnetworks,Juven2020,Song2023}.

By now, several families of exactly solvable models have been constructed~\cite{yoshida,twisted,cagenet,premgauging,bulmashgauging,Tantivasadakarnsearching20,FractonCSBF20,JWfracton20,Shirley20,TJV1,TJV2,HsinSlagle21,defectnetworks,gorantla2023graphs,boesl2025}, providing a systematic approach for studying fracton phases~\cite{DuaSorting} and leading to several illuminating perspectives on fracton order. One perspective is that fracton order can be thought of as arising from a gauge theory where the gauged symmetry is a subsystem symmetry~\cite{fracton2,williamson,yizhi1,strongsspt,spurious,williamsonSET,shirleygauging,Stephen2020,Shirley23}, i.e.,  a symmetry which acts on rigid submanifolds. A second perspective views fracton order as resulting from a coupled layer construction~\cite{sagar,han,cagenet,balents,designer,defectnetworks,Williamson2020a,SullivanPlanarpstring,Ebisu:2023idd,Ebisu:2024cke}. In many of these constructions, fracton phases can be obtained by starting from decoupled stacks of lower dimensional topological phases, and then coupling them together by condensing certain composite ``$p$-strings," which are extended one-dimensional strings comprised of point-like anyon excitations.

In a recent work~\cite{GPTW25}, we introduced a reformulation of the coupled layer construction in which the $p$-string is viewed as a 1-form symmetry defect in the 3+1D stack of 2+1D topological orders. Although the stack of 2+1D topological orders is not topological---since we must specify where in space these stacks live---the $p$-string is nonetheless a defect for a purely topological symmetry, which is a 1-form symmetry in 3+1D. Phrased thusly, $p$-string condensation serves as a transition from the phase of stacks, where the 1-form symmetry is spontaneously broken, into the X-Cube phase, where this 1-form symmetry is restored. 

A concrete result of our newly introduced point of view is an interpretation of the following ``numerological" observation: the ground state degeneracy (GSD) of a 3-foliated stack of 2+1D Toric Codes on a three-torus is $2^{6L}$. Part of this GSD can be attributed to the fact that the aforementioned topological 1-form symmetry is spontaneously broken; condensing the $p$-string restores this 1-form symmetry and hence, the GSD should be reduced. Now, the GSD of the X-Cube model~\cite{fracton1} is $2^{6L-3}$, which differs from that of the stack by $2^3$. This difference can be explained precisely by the GSD of the 3+1D Toric Code! Indeed, the 3+1D Toric Code is a phase that is characterized by the spontaneous breaking of exactly the 1-form symmetry which is restored when obtaining the X-Cube model from the 2+1D Toric Code stacks via $p$-string condensation.

One might then naively ask whether there is a sense in which a decoupled theory consisting of the X-Cube model times the 3+1D Toric Code yields the stack of 2+1D Toric Codes. The obvious answer is no, since the stack contains no fully mobile excitations while the 3+1D Toric Code does. Nevertheless, in this work we demonstrate that the stack of 2+1D Toric Codes can in fact be realized as a \textit{twisted gauge theory} of the X-Cube and 3+1D Toric Code. In particular, it is obtained upon gauging a non-trivial symmetry protected topological (SPT) state protected by both a planar subsystem symmetry with lineon charges, and a topological 1-form symmetry.

More generally, we consider 3+1D stacks composed of 2+1D layers of $G = \mathbb{Z}_N$ gauge theories and, using lattice models and continuum field theory, show that these theories sit at the nexus between 3+1D SPT, topologically ordered, and fracton ordered phases. In particular, these distinct phases can be accessed by gauging particular subgroups of the global (foliated) symmetry present in the decoupled stacks. As with gauging conventional topological symmetries~\cite{Vafa:1989ih,Tachikawa20,GaiottoKulp21}, the gauged theories host dual symmetries, which we further gauge to reveal a rich gauging web relating 3+1D topological (including SPTs and topological orders) and fracton phases. Given the length of the paper, we summarize our results below.


\subsection{Summary of main results}

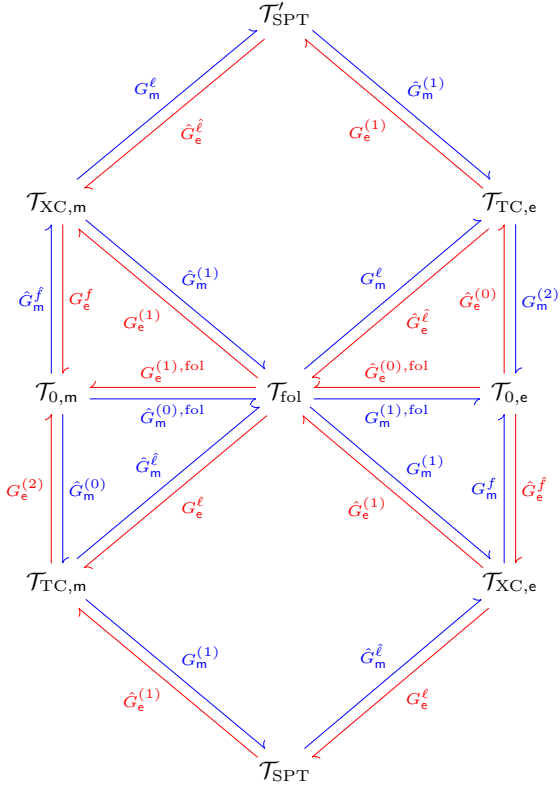
\begin{figure}[t]
    \begin{tikzcd}[sep=2cm]
        &\mathcal T'_\text{SPT}
        \arrow[dl,harpoon,red,shift left =1,"\hat G^{\hat{\ell}}_\ee"]
        \arrow[dr,harpoon,blue,shift left =1,"\hat G^{(1)}_\mm"]&
        \\
        \mathcal T_{\text{XC},\mm}
        \arrow[ur,harpoon,blue,shift left=1,"G^{\ell}_\mm"]
        \arrow[d,harpoon,red,shift left =1," G^{f}_\ee"]
        \arrow[dr,harpoon,blue,shift left =1,"\hat G^{(1)}_\mm"]
        &&\mathcal T_{\text{TC},\ee}
        \arrow[ul,harpoon,red,shift left =1,"G^{(1)}_\ee"]
        \arrow[dl,harpoon,red,shift left =1,"\hat G^{\hat{\ell}}_\ee"]
        \arrow[d,harpoon,blue,shift left =1,"G^{(2)}_\mm"]
        \\
        \mathcal T_{0,\mm}
        \arrow[u,harpoon,blue,shift left =1,"\hat G^{\hat{f}}_\mm"]
        \arrow[r,harpoon',swap, blue,shift right =1,"\hat G^{(0),\text{fol}}_\mm"]
        \arrow[d,harpoon,blue,shift left =1," \hat G^{(0)}_\mm"]&
        \mathcal T_\text{fol}
        \arrow[ul,harpoon,red,shift left =1,"G^{(1)}_\ee"]
        \arrow[l,harpoon',swap, red,shift right =1,"G^{(1),\text{fol}}_\ee"]
        \arrow[dl,harpoon,red,shift left =1,"G^{\ell}_\ee"]
        \arrow[r,harpoon',swap, blue,shift right =1,"G^{(1),\text{fol}}_\mm"]
        \arrow[dr,harpoon,blue,shift left =1,"G^{(1)}_\mm"]
        \arrow[ur,harpoon,blue,shift left=1,"G^{\ell}_\mm"]
        &\mathcal T_{0,\ee}
        \arrow[u,harpoon,red,shift left =1," \hat G^{(0)}_\ee"]
        \arrow[d,harpoon,red,shift left =1," \hat G^{\hat{f}}_\ee"]
        \arrow[l,harpoon',swap, red,shift right =1,"\hat G^{(0),\text{fol}}_\ee"]
        \\
        \mathcal T_{\text{TC},\mm}
        \arrow[u,harpoon,red,shift left =1," G^{(2)}_\ee"]
        \arrow[ur,harpoon,blue,shift left =1,"\hat G^{\hat{\ell}}_\mm"]
        \arrow[dr,harpoon,blue,shift left =1,"G^{(1)}_\mm"]&&
        \mathcal T_{\text{XC},\ee}
        \arrow[ul,harpoon,red,shift left =1,"\hat G^{(1)}_\ee"]
        \arrow[dl,harpoon,red,shift left =1,"G^{\ell}_\ee"]
        \arrow[u,harpoon,blue,shift left =1,"  G^{f}_\mm"]
        \\
        &\mathcal T_\text{SPT}
        \arrow[ul,harpoon,red,shift left =1,"\hat G^{(1)}_\ee"]
        \arrow[ur,harpoon,blue,shift left =1,"\hat G^{\hat{\ell}}_\mm"]&
    \end{tikzcd}
    \caption{The gauging web of topological and subsystem symmetries. Starting from a foliated stack of 2+1D Toric Codes ($\mathcal T_\text{fol}$), gauging the 1-form symmetry whose defect is the p-string results in the X-Cube model (XC). Alternatively, gauging the lineon Wilson symmetry results in the 3+1D Toric Code (TC). These two models are enriched by the corresponding dual symmetries (denoted by hats). Gauging both these symmetries lands us in a non-trivial SPT phase protected by the dual 1-form symmetry and the lineon planar subsystem symmetry. Finally, rotating this diagram by $180^\circ$ exchanges $\ee$ and $\mm$, which is ultimately related to the $\ee$-$\mm$ duality of the 2+1D Toric Code.}
    \label{fig:gaugingweb}
\end{figure}

We motivate our results from the perspective of lattice models, focusing first on the $G=\ZZ_2$ case. In Sec.~\ref{sec:symreview} we review the symmetries present in the 3+1D Toric Code and the X-Cube model, along with the corresponding dual symmetries that are obtained after gauging the original symmetries. These symmetries are summarized in Table~\ref{tab:symmetries}. In Sec.~\ref{sec:gaugingweblattice}, we start from a 3-foliated stack of 2+1D Toric Codes ($\mathcal T_\text{fol}$) and show that its total symmetry group contains certain symmetries present in the 3+1D Toric Code and the X-Cube model as subgroups. Respectively, these are a topological 1-form symmetry and the symmetry generated by the lineon Wilson operators. We then proceed to gauge the symmetries present in $\mathcal T_\text{fol}$, and analyze the resulting gauged theories. This results in an intricate web between distinct theories, which are related by gauging different subgroups of the total symmetry group of $\mathcal{T}_\text{fol}$. This gauging web is shown in Fig.~\ref{fig:gaugingweb}, and we summarize the various theories and their corresponding symmetries in Table~\ref{tab:theories}. Our conventions for denoting the various symmetry groups are summarized in Table~\ref{tab:symmetries}. As mentioned above, one of these theories is a novel SPT phase that is protected by the combination of a planar subsystem symmetry and a topological 1-form symmetry. Along the way, we also uncover new symmetry extensions between subsystem and topological symmetries, which are not simply captured by conventional group extensions. This is summarized below, along with a qualitative description of the symmetry extension.

\begin{widetext}
\begin{subequations}
\label{eq:extall}
    \begin{align}
            &1\to \hat G^{(0)}\to \hat G^{(0),\text{fol}} \to \hat{G}^{\hat{\ell}}\to 1\qquad&&(x,y,z \text{ lineons fuse into a 0-form charge}) \label{eq:ext1}\\
         &1\to G^{\ell}\to G^{(1),\text{fol}} \to G^{(2)} \to 1 \qquad && (\text{2-form lines meeting at a corner gives lineon}) \label{eq:ext2}\\
             &1\to \hat{G}^{\hat{f}}\to \hat G^{(0),\text{fol}} \to \hat G^{(1)} \to 1 \qquad &&(\text{2 planar 1-forms differ by the planar subsystem layers in between}) \label{eq:ext3}\\
        &1\to G^{(1)}\to G^{(1),\text{fol}}\to G^{f}\to 1\qquad &&(\text{2 fractons are connected by a $p$-string}) \label{eq:ext4}
\end{align}    
\end{subequations}
\end{widetext}

In Sec.~\ref{sec:cont-gaugingweb}, we provide a complementary field theory perspective on the above results, while generalizing from $\ZZ_2$ to $\ZZ_N$ at the same time. Finally, in Sec.~\ref{sec:outlook}, we summarize our results and present some future directions. 

\subsection{Related work}

Here, we highlight related works which have also discussed relations between topological and fracton models. The possible phase diagram relating the stacks of 2+1D Toric Codes to the X-Cube model was pointed out early on in the fracton literature in Refs.~\onlinecite{sagar,han}. An exactly solvable phase diagram was later obtained in Ref.~\onlinecite{zhu23} by tuning the tensor network realization of these phases. It was also appreciated in this work that the term that induces the p-string condensation from the Toric Code stacks to the X-Cube model is the same kind of term that induces the condensation of the flux-loop ($m$-loop) in the Toric Code to trivial transition. More generally, transitions out of the $\mathbb{Z}_N$ X-Cube model were discussed in Ref.~\onlinecite{lake2021sub}, which also highlighted the effect of subsystem symmetries on the phase transitions. Work by one of the present authors~\cite{domhigherform} also studied transitions out of fracton phases by using gauging dualities to relate them to spontaneous (higher-form) symmetry breaking phases of decoupled stacks of lower dimensional phases, some of which are exactly solvable.

Lastly, let us mention that several works have previously studied non-trivial symmetry extensions between topological and non-topological symmetries as well as related foliated field theories~\cite{Stephen2020,Stephen22,TJV1, TJV2,HsinSlagle21,designer,hsin2025fft}. In each of these works, the symmetry extension is non-trivial in the conventional, mathematical sense. For example, in the hybrid Toric Code studied in Ref.~\onlinecite{TJV1}, the symmetry that was gauged can be considered as the following group extension
\begin{align}
    1 \rightarrow \ZZ_2^L \rightarrow \ZZ_4 \times \ZZ_2^{L-1} \rightarrow \ZZ_2 \rightarrow 1
\end{align}
where $\ZZ_2^L$ denotes a subsystem symmetry of $L$ layers, and $\ZZ_2$ denotes a global $0$-form symmetry. In this sense, these previously considered symmetry extensions are more akin to extensions between topological symmetries in different dimensions, as encountered when studying higher-group gauge theories. 
 
In contrast, we emphasize that the symmetry extensions we discuss in this paper correspond to trivial extensions as groups, and hence lie beyond those discussed previously. Consider for example the extension in Eq.~\eqref{eq:ext1}. As groups,
$\hat G^{(0),\text{fol}} = \ZZ_2^{3L}$, $\hat G^{(0)} = \ZZ_2$, and $\hat{G}^{\hat{\ell}} = \ZZ_2^{3L-1}$, which is clearly a trivial extension. However, physically, we are not able to make this association, since the two groups act on completely different geometrical spaces. It is in this sense that the extensions we list in Eq.~\eqref{eq:extall} are non-trivial, highlighting the importance of keeping track of the non-trivial geometric structure associated with these symmetries.


\section{Review: Symmetries and Gauging in the Toric Code and X-Cube models}
\label{sec:symreview}

In this Section, we review the symmetries, their gauging, and the corresponding dual symmetries in both the 3+1D Toric Code and X-Cube models. On the lattice, all Hamiltonians considered in this paper are Pauli stabilizer Hamiltonians, consisting of a sum of commuting Pauli terms. As such, we use the terms Hamiltonians and stabilizers interchangeably. Throughout, we consider a cubic lattice of size $L \times L \times L$. For convenience, we focus on the $G=\ZZ_2$ case in this Section, where we use Pauli operators $X$,$Z$ satisfying the Pauli algebra $Z^2 = 1, X^2 = 1, ZX=-XZ$. This naturally generalizes to the $G = \mathbb{Z}_N$ case by using clock and shift operators satisfying $Z^N = 1, X^N = 1, ZX = e^{2\pi i/N}XZ$ and by appropriately placing daggers in all the expressions to follow.

{We follow a convention where the notation $e \in \gamma$ is used to denote containment of objects $e,\gamma,$ that are of the same spatial dimension. The expression $e \in \gamma$ then refers to any edge $e$ contained along a path $\gamma$, both of which are one-dimensional. When referring to containment or adjacency of objects that have different spatial dimensions, we use the symbol $\subset$ or $\supset$ to refer to two objects that touch each other. For example, $e\subset c$ refers to all edges $e$ that touch a cube $c$. The direction of containment in $\subset$ is towards the object with a larger spatial dimension.}

For each model, we complement the discussion on the lattice with a corresponding discussion in the continuum using field theory. We work in the Euclidean signature with $\tau$ denoting the Euclidean time and $x,y,z,\ldots$ denoting the spatial directions. We assume periodic boundary conditions in both space and time directions. Throughout, $\mu,\nu,\ldots$ run over spacetime indices, whereas $i,j,\ldots$ run over spatial indices. The stabilizers in the lattice Hamiltonian can be interpreted as the equations of motion for the $0^\text{th}$ component of fields in the continuum field theory. See Ref.~\onlinecite{SlagleXcubeQFT} for a review of this correspondence for the Toric Code and X-Cube models.

\subsection{3+1D Toric Code}

The Toric Code Hamiltonian is defined on the Hilbert space $\mathcal H_e = \bigotimes_e \CC^2$, corresponding to a qubit (or spin-$1/2$) living on each edge of the cubic lattice. The Hamiltonian reads
\begin{align}
\label{eq:3DTC}
    H^0_{\text{TC},\mm} = -\sum_v A_v -\sum_p B_p
\end{align}
where
  \begin{equation}
\begin{aligned}
    A_v &= \prod_{e \supset v}
    X_e =  \raisebox{-0.5\height}{\includegraphics{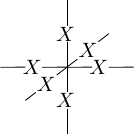}}\\
    B_p &= \prod_{e \subset p} Z_e = \raisebox{-0.5\height}{\includegraphics{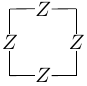}}, \raisebox{-0.5\height}{\includegraphics{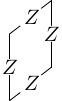}}, \raisebox{-0.5\height}{\includegraphics{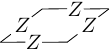}}.
\end{aligned}
\end{equation}
The Hamiltonian enjoys two types of (non-topological) higher-form symmetries\footnote{These symmetries can be made topological by restricting the Hilbert space to a non-tensor product Hilbert space, spanned by $+1$ eigenstates of the locally-generated symmetry operators.} Here, we label symmetries as $\ee$ or $\mm$ if they are created by applying $Z$ or $X$ Pauli operators, respectively. First, for any closed surface on the dual lattice $\hat{\Sigma}$ we may define the operator
\begin{align}
\label{eq:G1m}
    \eta^{(1)}_\mm(\hat{\Sigma}) = \prod_{e \in \hat \Sigma} X_e,
\end{align}
which we call a 1-form symmetry $G_\mm^{(1)}$. On a torus, we may choose a convenient basis for the 1-form symmetry as follows
  \begin{equation}
\begin{aligned}
    \eta^{(1)}_\mm(\delta v) &= A_v,\\
    \eta^{(1)}_\mm(\hat{\Sigma}_{yz}^{(i)}) &= \prod_{e_x \in \hat{\Sigma}_{yz}^{(i)}} X_{e_x},\\
    \eta^{(1)}_\mm(\hat{\Sigma}_{xz}^{(j)}) &= \prod_{e_y \in \hat{\Sigma}_{xz}^{(j)}} X_{e_y},\\
    \eta^{(1)}_\mm(\hat{\Sigma}_{xy}^{(k)}) &= \prod_{e_z \in \hat{\Sigma}_{xy}^{(k)}} X_{e_z},
\end{aligned}
\label{eq:G1mbasis}
\end{equation}
Here, $\delta v$ denotes the edges surrounding the vertex $v$ (i.e. the coboundary of $v$), and $\hat{\Sigma}_{yz}^{(i)}$ denotes all the $x$ edges with $x$ coordinate fixed to be $i+1/2$ for $i \in \ZZ$.
Here, the $\eta^{(1)}_\mm(\delta v)$ operators for all vertices form a basis surfaces that are boundaries, while $ \eta^{(1)}_\mm(\hat{\Sigma}_{yz}^{(i)})$, $\eta^{(1)}_\mm(\hat{\Sigma}_{xz}^{(j)})$, and $\eta^{(1)}_\mm(\hat{\Sigma}_{xy}^{(k)})$ form a basis for the non-trivial 2-cycles of the torus. The value of $i,j,k$ is arbitrary, since $\eta^{(1)}_\mm(\hat{\Sigma}_{xz}^{(i)})$ and $\eta^{(1)}_\mm(\hat{\Sigma}_{xz}^{(i')})$ differ by a product of $\eta^{(1)}_\mm(\delta v)$.

Furthermore, for any closed loop $\gamma$ on the direct lattice we also have
\begin{align}
\label{eq:G2e}
    \eta^{(2)}_\ee(\gamma) = \prod_{e \in \gamma} Z_e
\end{align}
corresponding to a 2-form symmetry $G_\ee^{(2)}$. The ground states of the Toric Code spontaneously break these higher form symmetries. Similarly to the 1-form symmetry above, we may chose a basis given by all the contractible 2-cycles corresponding to $B_p$, along with the three non-contractible ones.

The low energy physics of these ground states and the 1-form symmetries is captured by the 3+1D $\mathbb Z_N$ gauge theory with $N=2$:
\ie\label{3dznlag}
\mathcal L_{\TC,\mm}^0 = \frac{\ii N}{2\pi} b_\mm da_\ee~,
\fe
where $a_\ee$ and $b_\mm$ are $U(1)$ 1-form and 2-form gauge fields respectively, with corresponding gauge symmetries $a_\ee \sim a_\ee + d\alpha_\ee$ and $b_\mm \sim b_\mm + d \beta_\mm$ (here, $\alpha_\ee$ is a compact scalar, while $\beta_\mm$ is a 1-form field with its own gauge redundancy $\beta_\mm \sim \beta_\mm + d\lambda_\mm$). This theory has a $\mathbb Z_N$ 1-form global symmetry $G^{(1)}_\mm$ generated by the Wilson surface operators $\eta_\mm^{(1)}(\Sigma) = \exp(\ii \oint_{\Sigma} b_\mm)$, where $\Sigma$ is a closed surface, and a $\mathbb Z_N$ 2-form global symmetry $G^{(2)}_\ee$ generated by the Wilson line operators $\eta_\ee^{(2)}(\gamma) = \exp(\ii \oint_\gamma a_\ee)$, where $\gamma$ is a closed curve. These operators satisfy
\ie
&\eta_\ee^{(2)}(\gamma)^N = \eta_\mm^{(1)}(\Sigma)^N= 1~,
\\
&\eta_\ee^{(2)}(\gamma_i) \eta_\mm^{(1)}(\Sigma_{jk}) = e^{-2\pi \ii \epsilon_{ijk}/N} \eta_\mm^{(1)}(\Sigma_{jk}) \eta_\ee^{(2)}(\gamma_i)~,
\fe
where $\gamma_i$ is the non-trivial 1-cycle along the $i$-th direction, $\Sigma_{jk}$ is the non-trivial 2-cycle along the $jk$-plane, and $\epsilon_{ijk}$ is the Levi-Civita symbol.

We may trivialize the Toric Code Hamiltonian by gauging the symmetries present. On the lattice, this can be realized by a generalized Kramers-Wannier map~\cite{Cobanera11,levin2012braiding,HaegemanGaugingpaper,Yoshida15,fracton2,williamson,kubica,shirleygauging,Radicevic19,TantivasadakarnSearching,JWfracton20,TJV1,TJV2,Tantivasadakarn2021,Tantivasadakarn2022,Dolev21,domhigherform,Rakovszky23}, which maps a theory with symmetry $G$ to a theory with a dual symmetry, denoted $\hat G$. We denote the map that gauges a symmetry $G$ as $\mathsf D^G$, 

which has the property that for any symmetry operator $\eta \in G$, and for any dual symmetry operator $\hat \eta \in \hat G$
\begin{align}
    \mathsf D^G \eta &= \mathsf D^G , &   \hat \eta \mathsf D^G= \mathsf D^G.
\end{align}
Conversely, we may gauge $\hat G$ to return back to the original theory via $\mathsf D^{\hat G} \equiv (\mathsf D^G)^\dagger$.

To gauge the 1-form symmetry $G_\mm^{(1)}$, we define $\mathsf D^{G_\mm^{(1)}}: \mathcal H_e \rightarrow \mathcal H_p = \bigotimes_p \CC^2$, which satisfies
\begin{align}
    \mathsf D^{G_\mm^{(1)}} X_e &= \left( \prod_{p \supset e} X_p\right) \mathsf D^{G_\mm^{(1)}}, &\mathsf D^{G_\mm^{(1)}} \left(\prod_{e \subset p} Z_e\right)&=  Z_p \mathsf D^{G_\mm^{(1)}} \, .
\end{align}
As a shorthand, we may denote the action of such a map as
  \begin{equation}
 \begin{aligned}
\label{eq:GaugeG1m}
   X_e &\xmapsto{\DD^{G_\mm^{(1)}}} \prod_{p \supset e} X_p,\\
   \prod_{e \subset p} Z_e &\xmapsto{\DD^{G_\mm^{(1)}}} Z_p.
\end{aligned}
    \end{equation}

We denote the corresponding dual symmetry $\hat G_\ee^{(1)} \equiv \widehat{G_\mm^{(1)}}$, which is defined for any closed surface on the direct lattice $\Sigma$ as
\begin{align}
    \hat \eta_\ee^{(1)}(\Sigma) = \prod_{p \in \Sigma} Z_p.
\end{align}
Again, on a torus, we may choose a basis for the 1-form symmetry as
  \begin{equation}
\begin{aligned}
    \hat \eta^{(1)}_\ee(\partial c) &= \prod_{p \subset c} Z_p,\\
   \hat \eta^{(1)}_\ee({\Sigma}_{yz}^{(i)}) &= \prod_{p_{yz} \in \hat{\Sigma}_{yz}^{(i)}} Z_{p_{yz}},\\
    \hat \eta^{(1)}_\ee({\Sigma}_{xz}^{(j)}) &= \prod_{p_{xz} \in \hat{\Sigma}_{xz}^{(j)}} Z_{p_{xz}},\\
    \hat \eta^{(1)}_\ee({\Sigma}_{xy}^{(k)}) &= \prod_{p_{xy} \in \hat{\Sigma}_{xy}^{(k)}} Z_{p_{xy}},
\end{aligned}
\label{eq:G1ebasis}
\end{equation}
where $\Sigma_{yz}^{(i)}$ denotes all the $yz$ plaquettes with $x$ coordinate $i$. Up to locally generated terms $\hat \eta^{(1)}_\ee(\partial c)$, which form a basis of all surfaces that are boundaries, $\Sigma_{yz}^{(i)} \sim \Sigma_{yz}^{(i')}$. The corresponding Hamiltonian after gauging is the parent Hamiltonian for the trivial product state with $\hat G_e^{(1)}$ symmetry:
\begin{align}
  H^0_{\text{TC},\mm}  \xmapsto{\DD^{G_\mm^{(1)}}}   H^{(1)}_{\ee,0} = -\sum_p Z_p.
  \label{eq:trivial1form}
\end{align}

In the continuum, the 1-form symmetry $G^{(1)}_\mm$ can be gauged by coupling the theory in Eq.~\eqref{3dznlag} to a $\mathbb Z_N$ 2-form gauge field:
\ie\label{3dznlag-gauge1-form}
\mathcal L_{\ee,0}^{(1)} = \frac{\ii N}{2\pi} [b_\mm (da_\ee - b_\ee) + a_\mm db_\ee]~,
\fe
where $b_\ee$ is a $U(1)$ 2-form gauge field and $a_\mm$ is a Lagrange multiplier (1-form gauge field) that constrains $b_\ee$ to be a $\mathbb Z_N$ 2-form gauge field. The gauge symmetry is given by
\ie
&a_\ee \sim a_\ee + \beta_\ee~,\quad b_\ee \sim b_\ee + d\beta_\ee~,
\\
&b_\mm \sim b_\mm + d\beta_\mm~,\quad a_\mm \sim a_\mm + \beta_\mm~.
\fe
The gauged theory Eq.~\eqref{3dznlag-gauge1-form} has a dual 1-form symmetry generated by the Wilson surface operator $\hat \eta_\ee^{(1)}(\Sigma) = \exp(\ii\oint_\Sigma b_\ee)$, where $\Sigma$ is a closed surface. However, all gauge invariant operators in this theory are trivial, so this theory describes a trivially gapped phase. Indeed, integrating out $b_\ee$ sets $b_\mm = da_\mm$, which trivializes the Lagrangian. Moreover, gauging a $\mathbb Z_N$ 1-form symmetry in a trivially gapped phase gives the $\mathbb Z_N$ 2-form gauge theory. This is consistent with the result of gauging the dual 1-form symmetry in Eq.~\eqref{3dznlag-gauge1-form}:
\ie
\frac{\ii N}{2\pi} [b_\mm (da_\ee - b_\ee) + b_\ee (da_\mm - b'_\mm) + a'_\ee db'_\mm],
\fe
which reduces to Eq.~\eqref{3dznlag} after integrating out $b'_\mm$, which sets $b_\ee = da'_\ee$, and redefining $a_\ee \rightarrow a_\ee + a'_\ee$.

To instead gauge the 2-form symmetry $G_\ee^{(2)}$, we define $\DD^{G_\ee^{(2)}}: \mathcal H_e \rightarrow \mathcal H_v = \bigotimes_v \CC^2$ which acts as
  \begin{equation}
 \begin{aligned}
 \label{eq:GaugeG2e}
     Z_e  &\xmapsto{\mathsf D^{G_\ee^{(2)}}} \prod_{v \subset e} Z_v,\\
     \prod_{e \supset v} X_e&\xmapsto{\mathsf D^{G_\ee^{(2)}}}  X_v.
\end{aligned}
    \end{equation}
The corresponding dual symmetry is a 0-form symmetry $\hat G_\mm^{(0)} \equiv \widehat{G_\ee^{(2)}}$ defined as
\begin{align}
    \hat\eta_\mm^{(0)} = \prod_v X_v,
\end{align}
and the dual Hamiltonian is the trivial product state Hamiltonian with $\hat G_\mm^{(0)}$ symmetry.
\begin{align}
  H^0_{\text{TC},\mm}  \xmapsto{D^{G_\ee^{(2)}}}  H^{(0)}_{\mm,0} = -\sum_v X_v.
  \label{eq:trivial0form}
\end{align}

In the continuum, the 2-form symmetry $G^{(2)}_\ee$ of Eq.~\eqref{3dznlag} can be gauged by coupling the theory to a $\mathbb Z_N$ 3-form gauge field:
\ie\label{3dznlag-gauge2-form}
\mathcal L_{\mm,0}^{(0)} = \frac{\ii N}{2\pi} [a_\ee (db_\mm - c_\mm) + \phi_\ee dc_\mm]~,
\fe
where $c_\mm$ is a $U(1)$ 3-form gauge field and $\phi_\ee$ is a Lagrange multiplier (compact scalar) that constrains $c_\mm$ to be a $\mathbb Z_N$ 3-form gauge field. The gauge symmetry is given by
\ie
&a_\ee \sim a_\ee + d\alpha_\ee ~,\quad \phi_\ee \sim \phi_\ee + \alpha_\ee~,
\\
&b_\mm \sim b_\mm + \gamma_\mm~,\quad c_\mm \sim c_\mm + d\gamma_\mm~.
\fe
The gauged theory Eq.~\eqref{3dznlag-gauge2-form} has a dual 0-form symmetry generated by the Wilson operator $\hat\eta_\mm^{(0)}(M) = \exp(\ii\oint_M \hat c_\mm)$, where $M$ is a 3-cycle. However, once again, all the gauge invariant operators in this theory are trivial, so it describes a trivially gapped phase. Relatedly, gauging the dual 0-form symmetry in Eq.~\eqref{3dznlag-gauge2-form} takes us back to the $\mathbb Z_N$ gauge theory in Eq.~\eqref{3dznlag}.

\subsection{X-Cube}
The X-Cube Hamiltonian is also defined on the same Hilbert space $\mathcal H_e$ as the Toric Code model. The Hamiltonian reads
\begin{align}
    H^0_{\text{XC},\ee} = -\sum_v (A_{v,yz} + A_{v,xz} +A_{v,xy})  -\sum_c W_c,
    \label{eq:HXC0}
\end{align}
where
  \begin{equation}
 \begin{aligned}
    A_{v,yz} &= \prod_{e_y,e_z \supset v } X_e &&= \raisebox{-0.5\height}{\includegraphics{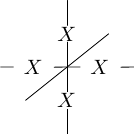}},\\
    A_{v,xz} &= \prod_{e_x,e_z \supset v } X_e &&= \raisebox{-0.5\height}{\includegraphics{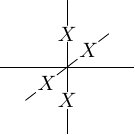}},\\
    A_{v,xy}&=  \prod_{e_x,e_y \supset v } X_e &&= \raisebox{-0.5\height}{\includegraphics{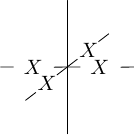}},\\
    W_c &= \prod_{e \subset c} Z_e &&=\raisebox{-0.5\height}{\includegraphics{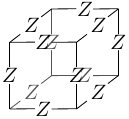}}.
    \label{eq:XC0terms}
\end{aligned}
    \end{equation}
Here, $e_x,e_y,e_z$ denote the edges oriented in the $x,y,z$ directions, respectively.

The Hamiltonian enjoys two types of symmetries. The first is $G^f_\mm$, where the symmetries are generated from Wilson operators of the fractons, which correspond to stabilizer defects for the cube term $W_c$ in the X-Cube Hamiltonian. The locally generated operators are of the form $A_{v,yz}$, $A_{v,xz}$, and $A_{v,xy}$, for all vertices $v$, while the non-local ones are of the form
\begin{align}
\label{eq:Gfm}
    \eta^f_\mm(\hat \Gamma_{y,z,i})= \prod_{e \in \hat \Gamma_{y,z,i}} X_e,
\end{align}
where $\hat \Gamma_{y,z,i}$ is a collection of $e_y$ edges along the $z$ direction with $x$-coordinate $i$. This is defined analogously for $\hat \Gamma_{z,y,i}$, $\hat \Gamma_{x,z,j}$, $\hat \Gamma_{z,x,j}$, $\hat \Gamma_{x,y,k}$, and $\hat \Gamma_{y,x,k}$. Unlike the Toric Code case, the different choices of $i,j,k$ are not necessarily equivalent. Up to the locally-generated operators, there are $6L-3$ independent non-local symmetry operators.

 The second type of symmetry is $G^{\ell}_\ee$ generated from the Wilson operators of the lineons, which correspond to stabilizer defects of the $A_{v,ij}$ terms in the X-Cube Hamiltonian. The locally generated symmetry operators are products of $\eta^{\ell}_\ee(c) = W_c$, while the non-local symmetry operators are generated by
  \begin{equation}
 \begin{aligned}
    \eta^{\ell}_\ee(\Gamma_{x}^{(j_0,k_0)}) &=\prod_{e_x \in \Gamma_{x}^{(j_0,k_0)}} Z_{e_x},\\
    \eta^{\ell}_\ee(\Gamma_{y}^{(i_0,k_0)}) &=\prod_{e_x \in \Gamma_{y}^{(i_0,k_0)}} Z_{e_y},\\
    \eta^{\ell}_\ee(\Gamma_{z}^{(i_0,j_0)}) &=\prod_{e_x \in \Gamma_{z}^{(i_0,j_0)}} Z_{e_z},
\end{aligned}
\label{eq:Gle}
    \end{equation}
where $\Gamma_{x}^{(j_0,k_0)}$ is a rigid line along the $x$-axis with $y$-coordinates $j_0$ and $z$ coordinates $k_0$. This is defined analogously for $\Gamma_{y}^{(i_0,k_0)}$, and $\Gamma_{z}^{(i_0,j_0)}$. There are also $6L-3$ independent non-local symmetry operators, up to $W_c$. Together, these symmetry operators lead to a nontrivial ground state degeneracy of $2^{6L-3}$, where this degeneracy is locally stable~\cite{fracton1}.

The low energy physics of these ground states is captured by an exotic field theory, specifically by a $\mathbb Z_N$ rank-2 hollow tensor gauge theory~\cite{seiberg2021zn}\footnote{see also Ref.~\onlinecite{gorantla2021villain} for a discussion of the lattice counterparts of these gauge fields.} with $N=2$:
\ie
\label{3dxclag}
\mathcal L_{\XC,\ee}^0 &= \frac{\ii N}{2\pi} \bigg[\sum_{\text{cyclic} \atop i,j,k} a_{ij}^\mm (\partial_\tau a_\ee^{ij} - \partial_k a^{k(ij)}_{\ee,\tau}) + a^\mm_\tau \sum_{i<j} \partial_i \partial_j a_\ee^{ij}\bigg]~.
\fe
where the gauge symmetry acts as
\ie
&a^{i(jk)}_{\ee,\tau} \sim a^{i(jk)}_{\ee,\tau} + \partial_\tau \alpha^{i(jk)}_\ee~,\quad a_\ee^{ij} \sim a_\ee^{ij} + \partial_k \alpha_\ee^{k(ij)}~,
\\
&a^\mm_\tau \sim a^\mm_\tau + \partial_\tau \alpha^\mm~,\quad a^\mm_{ij} \sim a^\mm_{ij} + \partial_i \partial_j \alpha^\mm~.
\fe
The two types of subsystem symmetries, $G^f_\mm$ and $G^\ell_\ee$, are generated by the following Wilson strip and line operators:
\ie
\eta^f_{\mm}(\Gamma_z^{x_0;y_1,y_2}) &= \exp\left(\ii\int_{y_1}^{y_2}dy \oint dz~ a^\mm_{yz}(x_0,y,z)\right)~,
\\
\eta^\ell_{\ee}(\Gamma^{x_0,y_0}_z) &= \exp\left(\ii\oint dz~ a^{xy}_\ee(x_0,y_0,z)\right)~,
\fe
and their variants in the other directions. Here, $\Gamma^{x_0;y_1,y_2}_z$ is a strip in the $yz$ plane $x=x_0$ with width $[y_1,y_2]$ and extended in the $z$ direction\footnote{After regularizing on the lattice, the continuum version of $\eta^f_\mm$ is related to the lattice version as follows
\ie
\eta^{f,\mathrm{cont}}_\mm(\Gamma^{x_0;y_1,y_2}_z) = \prod_{y_1\le y < y_2} \eta^{f,\mathrm{lat}}_\mm(\hat\Gamma_{y,z,x_0})
\fe
}, whereas $\Gamma^{x_0,y_0}_z$ is the line in the $z$ direction at $(x,y)=(x_0,y_0)$. If we regularize this model on a spatial lattice with $L$ sites in the three spatial directions, then the number of independent operators of both types is $6L-3$, consistent with the preceding discussion of the lattice model.

To gauge $G^f_\mm$, we define the map $\mathsf D^{G^f_\mm}: \mathcal H_e \rightarrow\mathcal H_c = \bigotimes_c \CC^2$, which maps
  \begin{equation}
 \begin{aligned}
        X_e &\xmapsto{\mathsf D^{G^f_\mm}} \prod_{c \supset e} X_c,\\
    \prod_{e\subset c} Z_e &\xmapsto{\mathsf D^{G^f_\mm}} Z_c.
\end{aligned}
\label{eq:GaugeGfm}
    \end{equation}
The dual symmetry, denoted $\hat{G}^{\hat{f}}_\ee \equiv \widehat{G^f_\mm}$ is a subsystem planar symmetry, whose charges are fractons. The symmetry operators are given by
\begin{equation}
\begin{aligned}
   \hat{\eta}^{\hat{f}}_\ee(\hat \Sigma_{yz}^{(i_0)}) = \prod_{c \subset \hat \Sigma_{yz}^{(i_0)}} Z_c,\\
    \hat{\eta}^{\hat{f}}_\ee(\hat \Sigma_{xz}^{(j_0)}) = \prod_{c \subset \hat \Sigma_{xz}^{(j_0)}} Z_c,\\
   \hat{\eta}^{\hat{f}}_\ee(\hat \Sigma_{xy}^{(k_0)}) = \prod_{c \subset \hat \Sigma_{xy}^{(k_0)}} Z_c,
   \label{eq:hatGhatfe}
\end{aligned}
\end{equation}
where $\hat \Sigma_{yz}^{(i_0)}$ consist of all cubes in the $yz$ plane with $x$-coordinate $i_0+1/2$, and similarly defined for $\hat \Sigma_{xz}^{(j_0)}$ and $\hat \Sigma_{xy}^{(k_0)}$. There are $3L-2$ independent symmetry generators on the torus coming from the constraint that
\begin{align}
    \prod_{i_0}  \hat{\eta}^{\hat{f}}_\ee(\hat \Sigma_{yz}^{(i_0)}) =  \prod_{j_0} \hat{\eta}^{\hat{f}}_\ee(\hat \Sigma_{xz}^{(j_0)}) = \prod_{k_0} \hat{\eta}^{\hat{f}}_\ee(\hat \Sigma_{xy}^{(k_0)}).
\end{align}
The dual Hamiltonian is the trivial product state Hamiltonian with $\hat{G}^{\hat{f}}_\ee$ symmetry
\begin{align}
    H^0_{\text{XC},\ee} \xmapsto{\mathsf D^{G^f_\mm}} H^{\hat{f}}_{\ee,0}= -\sum_c Z_c.
\end{align}

In the continuum, we can gauge the fracton subsystem symmetry $G^f_\mm$ by coupling the theory Eq.~\eqref{3dxclag} to $\mathbb Z_N$ tensor gauge fields $(b_{\ee,\tau}^{ij}, b_\ee)$:
\ie\label{3dxclag-gaugefsym}
&\mathcal L_{\ee,0}^{\hat f} = \frac{\ii N}{2\pi} \bigg[\sum_{\text{cyclic} \atop i,j,k} a_{ij}^\mm (\partial_\tau a_\ee^{ij} - \partial_k a^{k(ij)}_{\ee,\tau} - b_{\ee,\tau}^{ij})
\\
&+ a^\mm_\tau \big(\sum_{i<j} \partial_i \partial_j a_\ee^{ij} - b_\ee\big) - \phi^\mm\big(\partial_\tau b_\ee - \sum_{i<j} \partial_i\partial_j b_{\ee,\tau}^{ij}\big)\bigg]~,
\fe
where $\phi^\mm$ is a Lagrange multiplier (compact scalar) that constrains $(b_{\ee,\tau}^{ij}, b_\ee)$ to be a $\mathbb Z_N$ tensor gauge field. The gauge symmetry acts as
\ie
&a_{\ee,\tau}^{k(ij)} \sim a_{\ee,\tau}^{k(ij)} + \hat\beta_{\ee,\tau}^{k(ij)}~,\quad b_{\ee,\tau}^{ij} \sim b_{\ee,\tau}^{ij}+\partial_\tau \beta_\ee^{ij}-\partial_k \hat\beta_{\ee,\tau}^{k(ij)}~,
\\
&a_\ee^{ij} \sim a_\ee^{ij} + \beta_\ee^{ij}~,\quad b_\ee \sim b_\ee+\sum_{i<j} \partial_i \partial_j \beta_\ee^{ij}~,
\\
&a^\mm_\tau \sim a^\mm_\tau + \partial_\tau \alpha^\mm~,\quad \phi^\mm \sim \phi^\mm + \alpha^\mm~,
\\
&a^\mm_{ij} \sim a^\mm_{ij} + \partial_i \partial_j \alpha^\mm~.
\fe

The gauged theory Eq.~\eqref{3dxclag-gaugefsym} has a dual subsystem symmetry, denoted as $\hat G^{\hat f}_\ee$, generated by the Wilson slab operators
\ie
\hat \eta_\ee^{\hat f}(\Sigma_{yz}^{x_1,x_2}) = \exp\bigg(\ii\int_{x_1}^{x_2} dx \oint dy \oint dz~ b_\ee\bigg)~,
\fe
and its variants in the other directions. Here, $\Sigma_{yz}^{x_1,x_2}$ is the slab stretched along the $yz$ plane with width $[x_1,x_2]$. However, all gauge invariant operators in this theory, including the dual subsystem symmetry operators, act trivially, so this describes a trivially gapped phase. Moreover, gauging the dual subsystem symmetry in the gauged theory Eq.~\eqref{3dxclag-gaugefsym} takes us back to the original theory in Eq.~\eqref{3dxclag}. This can be done by coupling the gauged theory Eq.~\eqref{3dxclag-gaugefsym} to ``fracton gauge fields'' $(a'^\mm_\tau,a'^\mm_{ij})$.

To instead gauge $G^{\ell}_e$, we define the map $\mathsf D^{G^{\ell}_\ee}: \mathcal H_e \rightarrow\mathcal H_v = \bigotimes_v (\CC^2)^{\otimes 2}$, which maps
  \begin{equation}
 \begin{aligned}
Z_{e_x} &\xmapsto{\mathsf D^{G^{\ell}_\ee}}\prod_{v \subset e_x} IZ_v =\raisebox{-0.5\height}{\includegraphics{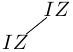}},\\
Z_{e_y} &\xmapsto{\mathsf D^{G^{\ell}_\ee}}\prod_{v \subset e_y} ZI_v = \raisebox{-0.5\height}{\includegraphics{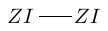}},\\
Z_{e_z} &\xmapsto{\mathsf D^{G^{\ell}_\ee}}\prod_{v \subset e_z} ZZ_v = \raisebox{-0.5\height}{\includegraphics{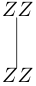}},\\
A^{yz}_v &\xmapsto{\mathsf D^{G^{\ell}_\ee}} XI_v, \\
A^{xz}_v &\xmapsto{\mathsf D^{G^{\ell}_\ee}} IX_v, \\
A^{xy}_v  &\xmapsto{\mathsf D^{G^{\ell}_\ee}} XX_v.
\end{aligned}
\label{eq:GaugeGle}
    \end{equation}

The dual symmetry, denoted $\hat{G}^{\hat{\ell}}_\mm \equiv \widehat{G^{\ell}_\ee}$ is a subsystem planar symmetry, whose charges are lineons. The symmetry operators are given by
  \begin{equation}
\begin{aligned}
\hat{\eta}^{\hat{\ell}}_\mm(\Sigma_{yz}^{(i_0)}) = \prod_{v \subset \Sigma_{yz}^{(i_0)}} XI_{v},\\
\hat{\eta}^{\hat{\ell}}_\mm(\Sigma_{xz}^{(j_0)}) = \prod_{v \subset \Sigma_{xz}^{(j_0)}} IX_{v},\\
\hat{\eta}^{\hat{\ell}}_\mm(\Sigma_{xy}^{(k_0)}) = \prod_{v \subset \Sigma_{xy}^{(k_0)}} XX_{v},
\end{aligned}
\end{equation}
where $\Sigma_{yz}^{(i_0)}$, consists of all vertices with $x$ coordinate $i_0$ and defined similarly for  $\Sigma_{xz}^{(j_0)}$, and $\Sigma_{yz}^{(k_0)}$.
 There are $3L-1$ independent generators on the torus coming from the fact that
 \begin{align}
    \prod_{i_0} \hat{\eta}^{\hat{\ell}}_\mm(\Sigma_{yz}^{(i_0)}) \prod_{j_0} \hat{\eta}^{\hat{\ell}}_\mm(\Sigma_{xz}^{(j_0)}) \prod_{k_0} \hat{\eta}^{\hat{\ell}}_\mm(\Sigma_{xy}^{(k_0)}) =1.
    \label{eq:lineonsymmetryconstraint}
 \end{align}
 The dual Hamiltonian is the trivial product state Hamiltonian with $\hat{G}^{\hat{\ell}}_\mm$ symmetry
\begin{align}
\label{eq:triviallineon}
    H^0_{\text{XC},\ee} \xmapsto{\mathsf D^{G^{\ell}_\ee}} H^{\hat{\ell}}_{\mm,0} =  -\sum_v (XI_v + IX_v +XX_v).
\end{align}

In the continuum, we can gauge the lineon subsystem symmetry $G^\ell_\ee$ by coupling the theory Eq.~\eqref{3dxclag} to $\mathbb Z_N$ tensor gauge fields $(b^\mm_{\tau ij}, b^\mm_{[ij]k})$:
\ie\label{3dxclag-gaugelsym}
\mathcal L_{\mm,0}^{\hat \ell} &= \frac{\ii N}{2\pi} \bigg[\sum_{i<j} a^{ij}_\ee (\partial_\tau a^\mm_{ij} - \partial_i \partial_j  a^\mm_\tau - b_{\tau ij}^\mm)
\\
&+ \sum_{\text{cyclic} \atop i,j,k} \Big( a_{\ee,\tau}^{[ij]k} (\partial_i a^\mm_{jk} - \partial_j a^\mm_{ik} - b^\mm_{[ij]k})
\\
&\qquad - \phi_\ee^{[ij]k} \big( \partial_\tau b^\mm_{[ij]k} -  \partial_i b^\mm_{\tau jk} + \partial_j b^\mm_{\tau ik} \big) \Big) \bigg]~,
\fe
where $\phi_\ee^{[ij]k}$ is a Lagrange multiplier (compact scalar) that constrains $(b^\mm_{\tau ij}, b^\mm_{[ij]k})$ to be a $\mathbb Z_N$ tensor gauge field. The gauge symmetry acts as
\ie
&a_{\ee,\tau}^{k(ij)} \sim a_{\ee,\tau}^{k(ij)} + \partial_\tau \alpha_\ee^{k(ij)}~,\quad \phi_\ee^{k(ij)} \sim \phi_\ee^{k(ij)} + \alpha_\ee^{k(ij)}~,
\\
&a_\ee^{ij} \sim a_\ee^{ij} + \partial_k \alpha_\ee^{k(ij)}~,
\\
&a^\mm_\tau \sim a^\mm_\tau + \beta^\mm_\tau~,\quad b_{\tau ij}^\mm \sim b_{\tau ij}^\mm + \partial_\tau \beta^\mm_{ij} - \partial_i \partial_j \beta^\mm_\tau~,
\\
&a^\mm_{ij} \sim a^\mm_{ij} + \beta^\mm_{ij}~,\quad b^\mm_{[ij]k} \sim b^\mm_{[ij]k} + \partial_i \beta^\mm_{jk} - \partial_j \beta^\mm_{ik}~.
\fe

The gauged theory Eq.~\eqref{3dxclag-gaugelsym} has a dual subsystem symmetry, denoted as $\hat G^{\hat \ell}_\mm$, generated by the Wilson slab operators
\ie
\hat \eta_\mm^{\hat \ell}(\Sigma_{yz}^{x_1,x_2}) = \exp\bigg(\ii\int_{x_1}^{x_2} dx \oint dy \oint dz~ b^\mm_{[yz]x} \bigg)~,
\fe
and its variants in the other directions. However, all gauge invariant operators in this theory, including the dual subsystem symmetry operators, act trivially, so this describes a trivially gapped phase. Moreover, gauging the dual subsystem symmetry in the gauged theory Eq.~\eqref{3dxclag-gaugelsym} takes us back to the original theory in Eq.~\eqref{3dxclag}. This is can be done by coupling the gauged theory Eq.~\eqref{3dxclag-gaugelsym} to ``lineon gauge fields'' $(a'^{k(ij)}_{\ee,\tau},a'^{ij}_\ee)$.


\section{Gauging web on the lattice}
\label{sec:gaugingweblattice}

With the review of symmetries in the Toric Code and the X-Cube models concluded, we now describe our gauging web from the lattice perspective. In particular, we introduce lattice models with topological and fracton order, and then describe the gauging maps between them.

\subsection{Decoupled 2+1D Toric Code stacks}
\label{sec:decouple}

Our starting point is a stack of 2+1D $\ZZ_2$ Toric Codes (the generalization to $\ZZ_N$ is considered in Sec.~\ref{sec:cont-gaugingweb}). These stacks come in three foliations along the $yz$, $xz$ and $xy$ planes, allowing us to write down the lattice model on a cubic lattice. These decoupled layers together form a lattice model with two qubits per edge. The Hamiltonian is
\begin{align}
\label{eq:Hfol}
    H_\text{fol} = &-\sum_v (A^\fol_{v,yz}  + A^\fol_{v,xz} +A^\fol_{v,xy} )- \sum_{p} B^\text{fol}_{p},
\end{align}
where
  \begin{equation}
\begin{aligned}
A^\fol_{v,yz} &=  \raisebox{-0.5\height}{\includegraphics{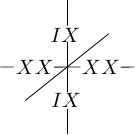}},& B^\text{fol}_{p_{yz}} &=  \raisebox{-0.5\height}{\includegraphics{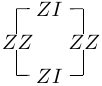}},\\
 A^\fol_{v,xz} &=  \raisebox{-0.5\height}{\includegraphics{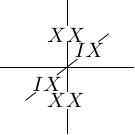}},& B^\text{fol}_{p_{xz}} &=  \raisebox{-0.5\height}{\includegraphics{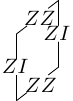}},\\  
A^\fol_{v,xy} &=  \raisebox{-0.5\height}{\includegraphics{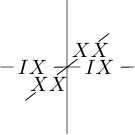}},& B^\text{fol}_{p_{xy}} &=  \raisebox{-0.5\height}{\includegraphics{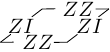}}.
\end{aligned}
\label{eq:TCstacks}
\end{equation}
Here, we remark that we are presenting the Hamiltonian in an unusual coupled basis to simplify the calculation that follows. The basis transformation that takes us back to the decoupled basis is given by
\begin{align}
\label{eq:decoupledbasistransform}
    U = \prod_e  CX_e,
\end{align}
where $CX$ is the CNOT gate which acts as
  \begin{equation}
\begin{aligned}
    XI &\xmapsto{CX} XX, & IX &\xmapsto{CX} IX,\\
    ZI &\xmapsto{CX} ZI, & IZ &\xmapsto{CX} ZZ.
\end{aligned}
\end{equation}

Under this map,
  \begin{equation}
\begin{aligned}
\label{eq:stackbasistransformed}
A^\fol_{v,yz} &\xmapsto{U} \raisebox{-0.5\height}{\includegraphics{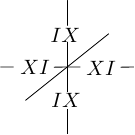}},&B^\text{fol}_{p_{yz}} &\xmapsto{U}  \raisebox{-0.5\height}{\includegraphics{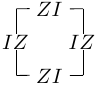}},\\
A^\fol_{v,xz} &\xmapsto{U}  \raisebox{-0.5\height}{\includegraphics{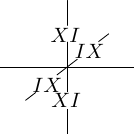}},&B^\text{fol}_{p_{xz}} &\xmapsto{U}  \raisebox{-0.5\height}{\includegraphics{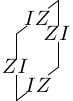}},\\  
A^\fol_{v,xy} &\xmapsto{U}  \raisebox{-0.5\height}{\includegraphics{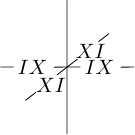}},&B^\text{fol}_{p_{xy}} &\xmapsto{U}  \raisebox{-0.5\height}{\includegraphics{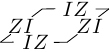}},
\end{aligned}
\end{equation}
where we see that we recover the decoupled Hamiltonians in each foliation.

Each foliation containing a 2+1D Toric Code comes with its own 1-form symmetry, generated by the Wilson lines of the $\ee$ and $\mm$ anyons. We call the total symmetry generated by all the $\ee$ anyons of the entire system a foliated 1-form symmetry $G_\ee^{(1),\text{fol}}$. Specifically, the contractible Wilson lines are generated by $ B_{p_{yz}}$, $B_{p_{xz}}$, and $B_{p_{xy}}$, while the non-contractible Wilson lines are generated by
  \begin{equation}
\begin{aligned}
    \eta^{(1),\text{fol}}_{\ee,yz}(\Gamma_{y}^{(i_0,k)}) &= \prod_{e_y \subset \Gamma_{y}^{(i_0,k)}} ZI_{e_y},\\
    \eta^{(1),\text{fol}}_{\ee,yz}(\Gamma_{z}^{(i_0,j)}) &= \prod_{e_z \subset \Gamma_{z}^{(i_0,j)}} ZZ_{e_z},\\
    \eta^{(1),\text{fol}}_{\ee,xz}(\Gamma_{z}^{(j_0,k)}), &= \prod_{e_z \subset \Gamma_{z}^{(j_0,k)}} ZI_{e_z}, \\
    \eta^{(1),\text{fol}}_{\ee,xz}(\Gamma_{x}^{(j_0,i)}) &= \prod_{e_x \subset \Gamma_{x}^{(j_0,i)}} ZZ_{e_x},\\
    \eta^{(1),\text{fol}}_{\ee,xy}(\Gamma_{x}^{(k_0,j)}) &= \prod_{e_x \subset \Gamma_{x}^{(k_0,j)}} ZI_{e_x},\\
    \eta^{(1),\text{fol}}_{\ee,xy}(\Gamma_{y}^{(k_0,i)}) &= \prod_{e_y \subset \Gamma_{y}^{(k_0,i)}} ZZ_{e_y},
\end{aligned}
\end{equation}
where $\Gamma_{y,i_0,k}$ is a rigid line along the direct lattice in the $y$ direction with $x$-coordinate $i_0$ and $z$ coordinate $k$. Here, different choices of $i_0$ give rise to different operators, while the different choices of $k$ are equivalent up to a product of $B_{p_{yz}}$ operators). There are $6L$ such independent non-local operators, up to those that are locally generated. 

Likewise, the symmetry generated by all the $\mm$ anyons is denoted $G_\mm^{(1),\text{fol}}$. The contractible Wilson lines are generated by $ A^\fol_{v,yz}$, $A^\fol_{v,xz}$, and $A^\fol_{v,xy}$, while the non-contractible Wilson lines are generated by
\begin{equation}
\begin{aligned}
    \eta^{(1),\text{fol}}_{\mm,yz}(\hat \Gamma_y^{(i_0,k)}) &= \prod_{e_z \subset\hat \Gamma_y^{(i_0,k)}} IX_e,\\
    \eta^{(1),\text{fol}}_{\mm,yz}(\hat \Gamma_z^{(i_0,j)}) &= \prod_{e_y \subset\hat \Gamma_z^{(i_0,j)} }XX_e,\\
    \eta^{(1),\text{fol}}_{\mm,xz}(\hat \Gamma_z^{(j_0,i)}) &= \prod_{e_x \subset\hat \Gamma_z^{(j_0,i)}} IX_e,\\
    \eta^{(1),\text{fol}}_{\mm,xz}(\hat \Gamma_x^{(j_0,k)}) &= \prod_{e_z \subset\hat \Gamma_x^{(j_0,k)}} XX_e,\\
    \eta^{(1),\text{fol}}_{\mm,xy}(\hat \Gamma_x^{(k_0,j)}) &= \prod_{e_y \subset\hat \Gamma_x^{(k_0,j)}} IX_e,\\
    \eta^{(1),\text{fol}}_{\mm,xy}(\hat \Gamma_y^{(k_0,i)}) &= \prod_{e_x \subset\hat \Gamma_y^{(k_0,i)}} XX_e.
\end{aligned}
\end{equation}
where $\Gamma_x^{(j_0,k)}$ consists of $e_z$ edges along the $x$ direction with $y$ coordinates $j_0$ and $z$ coordinates $k+1/2$, and are defined similarly for the other sets.

Refs.~\onlinecite{han,sagar} studied two deformations of the above model:
\begin{align}
    H = H_\text{fol} - h_X \sum_e XI_e - h_Z \sum_e IZ_e \, .
\end{align}
Note that in the decoupled basis, these perturbations are $XX_e$ and $ZZ_e$, respectively. First, keeping $h_Z = 0$ and turning on a large $h_X$ forces the condensation of a composite object called a $p$-string, with the condensed phase lying in the same phase as the X-Cube model. On the other hand, keeping $h_X = 0$ and turning on a large $h_Z$ results in the 3+1D Toric Code model. In addition, turning both $h_X$ and $h_Z$ to be large simultaneously results in a product state given by $\ket{+} \otimes \ket{0}$ on each edge.

Let us briefly provide some intuition behind how these condensations work. First, consider the action of the perturbation $ZZ_e$ on the ground state of the stack Hamiltonian e.g., apply $ZZ_e$ on an edge in in the $x$ direction. This creates a pair of excitations, each of which is a composite of two $\ee$ anyons: one from the $xy$ plane and one from the $xz$ plane $\ee_x \equiv  \ee_{xy}\ee_{xz}$. This composite particle behaves exactly like a lineon in the X-Cube model: it is clear that $\ee_x$ can only move in the $x$ direction and, moreover, the fusion of three such particles coming from three different directions is
\begin{align}
    \ee_x \times \ee_y \times \ee_z = \ee_{xy}\ee_{xz} \times \ee_{xy}\ee_{yz} \times \ee_{xz}\ee_{yz} =1 \, .
\end{align}
Thus, the $ZZ_e$ term creates lineons akin to those in the X-Cube model. It is therefore clear that by turning on a large $ZZ_e$ perturbation, we are condensing lineons in $\mathcal T_\text{stack}$. What happens to the remaining particles? Initially in the stacked theory, $\ee_{xy}$ is only mobile in the $xy$ plane. It can only turn into $\ee_{xz}$ at the cost of leaving behind a lineon $\ee_x$. Thus, we see that if all the lineons are condensed,  $\ee_{xy}$ is now free to move in all directions and becomes fully mobile in three dimensions, corresponding precisely to the $\ee$ anyon in the 3+1D Toric Code.

Next, we can consider the action of the perturbation $XX_e$ on the ground state of the stack Hamiltonian e.g., apply $XX_e$ on an edge in the $x$ direction. This creates a composite $p$-string excitation, composed of $\mm$ anyons supported on the two layers intersecting the edge: a pair of $\mm$ anyons in the $xy$ plane and a pair in the $xz$ plane. The end points of open $p$-strings behave precisely as fractons in the X-Cube model as they are clearly created at the end-points of membrane operators and cannot move without creating additional excitations. Thus, turning on a large $XX_e$ perturbation condenses the $p$-strings in $\mathcal T_\text{stack}$, with the condensed theory in the same phase as the X-Cube model. Since $\ee$ anyons supported on a single layer braid non-trivially with the $p$-string, they are confined in the condensed phase; however, pairs of $\ee$ anyons from intersecting layers braid trivially with the $p$-string and remain as deconfined lineon excitations, matching the lineons in the X-Cube model. See Refs.~\onlinecite{han,sagar,cagenet} for details on the $p$-string condensation procedure.

In the following, we instead analyze what happens when we modify the procedure from condensation (as induced by varying a term in the Hamiltonian) to a gauging process (which induces the algebraic procedure of anyon condensation); the latter can be thought of as the condensation of a bound state of the original condensation with a symmetry charge of the dual symmetry. Consequently, the resulting model also possess a dual symmetry. The dual symmetry necessarily enriches both the X-Cube and Toric Code models, because we must be able to gauge these dual symmetries to recover the foliated stacks of 2+1D Toric Codes. We uncover the nature of this interesting symmetry enrichment below, involving a mix of topological and non-topological symmetries.

First, we identify two important subgroups of the symmetries present. The first is the subgroup of $G_\ee^{(1),\text{fol}}$ generated by
\begin{equation}
\begin{aligned}
  \prod_{p \subset c} B^\text{fol}_{p} &= \prod_{e \subset c} IZ_{e} \\
  &= \mathbbm 1 \otimes  \eta^{\ell}_\ee(c),\\
\eta^{(1),\text{fol}}_{e,xz}(\Gamma_{x}^{(j_0,k_0)}) \eta^{(1),\text{fol}}_{e,xy}(\Gamma_{x}^{(j_0,k_0)}) &= \prod_{e_x \in \Gamma_{x}^{(j_0,k_0)}} IZ_{e_x} \\ 
&= \mathbbm 1 \otimes \eta^{\ell}_\ee(\Gamma_{x}^{(j_0,k_0)}),\\
\eta^{(1),\text{fol}}_{e,yz}(\Gamma_{y}^{(i_0,k_0)}) \eta^{(1),\text{fol}}_{e,xy}(\Gamma_{y}^{(i_0,k_0)}) &= \prod_{e_y \in \Gamma_{y}^{(i_0,k_0)}} IZ_{e_y} \\
&= \mathbbm 1 \otimes \eta^{\ell}_\ee(\Gamma_{y}^{(i_0,k_0)}) ,\\
 \eta^{(1),\text{fol}}_{e,xz}(\Gamma_{z}^{(i_0,j_0)}) \eta^{(1),\text{fol}}_{e,yz}(\Gamma_{z}^{(i_0,j_0)}) &= \prod_{e_z \in \Gamma_{z}^{(i_0,j_0)}} I Z_{e_z} \\
&= \mathbbm 1 \otimes \eta^{\ell}_\ee(\Gamma_{z}^{(i_0,j_0)}),
\end{aligned}
\end{equation}
which we identify as the symmetry generated by lineons $G^{\ell}_\ee$ represented in Eq.~\eqref{eq:Gle} acting on the second qubit of each edge.

The second is the subgroup of $G_\mm^{(1),\text{fol}}$ generated by
  \begin{equation}
\begin{aligned}
     A^\fol_{v,yz}A^\fol_{v,xz}A^\fol_{v,xy} &= \prod_{e \supset v} XI_v \\
     & =\eta^{(1)}_\mm(\delta v) \otimes \mathbbm 1,\\
  \prod_{j_0}  \eta^{(1),\text{fol}}_{\mm,xz}(\hat \Gamma_z^{(j_0,i)}) \prod_{k_0} \eta^{(1),\text{fol}}_{\mm,xy}(\hat \Gamma_y^{(k_0,i)})  &= \prod_{e_x \in \hat{\Sigma}_{yz}^{(i)}} XI_{e_x} \\
  &=  \eta^{(1)}_\mm(\hat{\Sigma}_{yz}^{(i)}) \otimes \mathbbm 1, \\
  \prod_{i_0}   \eta^{(1),\text{fol}}_{\mm,yz}(\hat \Gamma_z^{(i_0,j)}) \prod_{k_0}\eta^{(1),\text{fol}}_{\mm,xy}(\hat \Gamma_x^{(k_0,j)}) &=   \prod_{e_y \in \hat{\Sigma}_{xz}^{(j)}} XI_{e_y} \\
  &= \eta^{(1)}_\mm(\hat{\Sigma}_{xz}^{(j)})  \otimes \mathbbm 1, \\
   \prod_{i_0}\eta^{(1),\text{fol}}_{\mm,yz}(\hat \Gamma_y^{(i_0,k)}) \prod_{j_0}     \eta^{(1),\text{fol}}_{\mm,xz}(\hat \Gamma_x^{(j_0,k)}) &=  \prod_{e_z \in \hat{\Sigma}_{xy}^{(k)}} XI_{e_z} \\
   &=  \eta^{(1)}_\mm(\hat{\Sigma}_{xy}^{(k)}) \otimes \mathbbm 1,
\end{aligned}
\end{equation}
which we identify as the symmetry generated by the Wilson membranes of all the $p$-strings. In fact, as the notation suggests, this is exactly the 1-form symmetry $G^{(1)}_\mm$ represented in Eq.~\eqref{eq:G1mbasis} acting on the first qubit of each edge.

Let us finally comment that in this basis, $G^{\ell}_\ee$ only acts on the first qubit, while $G^{(1)}_\mm$ only act on the second qubit. Thus, from the lattice perspective, it is clear that these symmetries do not have a mutual anomaly and can be gauged simultaneously. However, we choose to gauge them sequentially to analyze the resulting intermediate models.

\subsection{Gauging $G^\ell_\ee$: Toric Code}

Let us gauge $G^\ell_\ee$ by applying the map $\DD^{G^\ell_\ee}$ in Eq.~\eqref{eq:GaugeGle} to the second qubit. We rename the operators under this mapping as follows $\mathcal O^\fol \xmapsto{\mathbbm 1 \otimes \DD^{G^\ell_\ee}} \mathcal O^{\TC,\mm}$. The resulting Hamiltonian is
\begin{align}
 H_{\TC,\mm} = -\sum_v (A^{\TC,\mm}_{v,yz}+A^{\TC,\mm}_{v,xz}+A^{\TC,\mm}_{v,xy}) - \sum_p B^{\TC,\mm}_{p},
    \label{eq:HTCm}
\end{align}
where
  \begin{equation}
\begin{aligned}
\label{eq:SSET}
    A^{\TC,\mm}_{v,yz} &= XX_v \prod_{e_y\supset v} X_e &&=\raisebox{-0.5\height}{\includegraphics{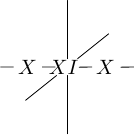}}, \\
    A^{\TC,\mm}_{v,xz} &= XI_v \prod_{e_z\supset v} X_e&&=\raisebox{-0.5\height}{\includegraphics{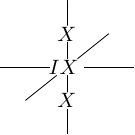}},\\
    A^{\TC,\mm}_{v,xy} &=  IX_v \prod_{e_x\supset v} X_e&&=\raisebox{-0.5\height}{\includegraphics{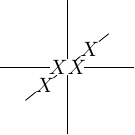}},\\
    B^{\TC,\mm}_{p_{yz}} &=\prod_{e\supset p} Z_e \prod_{v\supset p} ZZ_v &&=\raisebox{-0.5\height}{\includegraphics{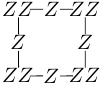}},  \\
    B^{\TC,\mm}_{p_{xz}} &=\prod_{e\supset p} Z_e \prod_{v\supset p} IZ_v&&=\raisebox{-0.5\height}{\includegraphics{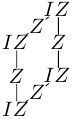}},   \\
    B^{\TC,\mm}_{p_{xy}} &=\prod_{e\supset p} Z_e \prod_{v\supset p} ZI_v 
    &&=\raisebox{-0.5\height}{\includegraphics{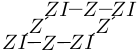}}.
\end{aligned}
\end{equation}
First, let us show that this model is equivalent to the Toric Code up to a basis transformation. Consider the unitary transformation
\begin{align}
    U_\text{SSET} = \prod_v\left [ \prod_{e_y \supset v}CX_{v_1,e_y} \prod_{e_z \supset v} CX_{v_2,e_z}\right]
    \label{eq:U_SSET}.
\end{align}
Conjugating the above unitary we see that
  \begin{equation}
\begin{aligned}
\label{eq:TC1}
    A^{\TC,\mm}_{v,yz} &\xmapsto{U_\text{SSET} }  XI_v,\\
    A^{\TC,\mm}_{v,xz} &\xmapsto{U_\text{SSET}} IX_v,\\
    A^{\TC,\mm}_{v,xy} & \xmapsto{U_\text{SSET} }  XX_v \prod_{e \supset v} X_e = XX_v \times A_v,\\
   B^{\TC,\mm}_{p}  &\xmapsto{U_\text{SSET}} \prod_{e\supset p} Z_e = B_p.
\end{aligned}
\end{equation}
Since all the terms in the Hamiltonian commute, we may restrict to the subspace where $A^{\TC}_{v,yz}= A^{\TC}_{v,xy}=1$. That is, where $XI_v = IX_v =1$ on every vertex. We thus find that the terms of the Hamiltonian is exactly the 3D Toric Code in Eq.~\eqref{eq:3DTC}.

However, consider the dual subsystem symmetry after gauging, $G^{\hat{\ell}}_\mm$. Let us show that the Hamiltonian $H_{\text{TC},\mm}$ defined by terms given in Eq.~\eqref{eq:SSET} exhibits a symmetry enrichment by $G^{\hat{\ell}}_\mm$. More specifically, we show that the model exhibits subsystem symmetry fractionalization, a natural generalization of the subsystem symmetry enriched Toric Code in 2+1D considered in Ref.~\onlinecite{Stephen22}. Hence, the Hamiltonian realizes a Subsystem Symmetry-Enriched Topological (SSET) order.

Recall that the planar subsystem symmetries $G^{\hat{\ell}}_\mm$ satisfy a constraint given by Eq.~\eqref{eq:lineonsymmetryconstraint}. However, let us restrict the symmetry action to a region. In particular, consider the symmetry in the region $M = ([i_1,i_2] \times  [j_1,j_2] \times  \in [k_1,k_2]) \cap \ZZ^3$. The planar symmetry operators restricted to that region is now

\begin{align}
    \hat{\eta}^{\hat{\ell}}_\mm(\Sigma_{yz}^{(i_0)})_\text{restricted} = \prod_{v \in \{i_0\} \times [j_1,j_2] \times[k_1,k_2] } XI_{v},\\
\hat{\eta}^{\hat{\ell}}_\mm(\Sigma_{xz}^{(j_0)})_\text{restricted} = \prod_{v \in [i_1,i_2] \times\{j_0\} \times[k_1,k_2] } IX_{v},\\
\hat{\eta}^{\hat{\ell}}_\mm(\Sigma_{xy}^{(k_0)})_\text{restricted} = \prod_{v \in [i_1,i_2]\times [j_1,j_2]  \times\{k_0\}} XX_{v}.
\end{align}

In the ground state of the Hamiltonian, since we have $A^{\TC}_{v,yz} = A^{\TC}_{v,xz} = A^{\TC}_{v,xy}=1$, this is equivalent to
\begin{align}
    \hat{\eta}^{\hat{\ell}}_\mm(\Sigma_{yz}^{(i_0)})_\text{restricted} \sim \prod_{v \in \{i_0\} \times [j_1,j_2] \times[k_1,k_2] } \prod_{e_z \supset v} X_e,\\
\hat{\eta}^{\hat{\ell}}_\mm(\Sigma_{xz}^{(j_0)})_\text{restricted} \sim \prod_{v \in [i_1,i_2] \times\{j_0\} \times[k_1,k_2] } \prod_{e_x \supset v} X_e,\\
\hat{\eta}^{\hat{\ell}}_\mm(\Sigma_{xy}^{(k_0)})_\text{restricted} \sim \prod_{v \in [i_1,i_2]\times [j_1,j_2]  \times\{k_0\}} \prod_{e_y \supset v} X_e.
\end{align}

Finally, taking the product of all the planar subsystem symmetries within the region gives
\begin{widetext}
\begin{align}
        \prod_{i_0=i_1}^{i_2} \hat{\eta}^{\hat{\ell}}_\mm(\Sigma_{yz}^{(i_0)})_\text{restricted} \prod_{j_0=j_1}^{j_2} \hat{\eta}^{\hat{\ell}}_\mm(\Sigma_{xz}^{(j_0)})_\text{restricted} \prod_{k_0=k_1}^{k_2} \hat{\eta}^{\hat{\ell}}_\mm(\Sigma_{xy}^{(k_0)})_\text{restricted} \sim 
       \prod_{v \in M } \prod_{e \supset v} X_e =\prod_{e \supset \delta M} X_e .
\end{align}
\end{widetext}
which is a closed membrane operator of $\mm$ in the 3+1D Toric Code around the region $M$, which detects the charge $\ee$.

The $\ee$ excitations are charged under the subsystem symmetry and therefore cannot turn unless they leave behind a planar subsystem charge. For example, consider applying an open string of $Z_e$ in the $x$ direction to create an $\ee$ excitation at each end point. In order for $e$ to turn the corner at a vertex $v$ into the $y$ direction, it must leave behind an operator $ZI_v$, which is charged under the subsystem symmetry $G^{\hat \ell}_\mm$. Likewise, to turn into the $z$ direction it must leave behind a $IZ_v$ charge. In fact, these charges are exactly lineons, and are the analogues of the gauge charges of $G^{\ell}_\ee$.

\subsection{Gauging $G^\ell_\ee$ and $G^{(1)}_\mm$: SPT}
Starting from $ H_{\TC,\mm}$ in Eq.~\eqref{eq:HTCm}, we may further gauge $G^{(1)}_\mm$ using the mapping $D^{G_\mm^{(1)}}$ in Eq.~\eqref{eq:GaugeG1m} on the first qubit on every edge. We rename the operators under this mapping as follows $\mathcal O^{\TC,\mm} \xmapsto{ \DD^{G^{(1)}_\mm} \otimes \mathbbm 1} \mathcal O^{\SPT}$

Note that gauging $G^{(1)}_\mm$ and $G^{\ell}_\ee$ commute since they act on different subspaces. Thus, we may also obtain the result from gauging $G^{\ell}_\ee \times G^{(1)}_\mm$ in $H_\text{fol}$ via $D^{G_\mm^{(1)}} \otimes D^{G^{\ell}_\ee}$. The resulting Hamiltonian is
\begin{align}
 H_{\SPT} = -\sum_v (A^{\SPT}_{v,yz}+A^{\SPT}_{v,xz}+A^{\SPT}_{v,xy}) - \sum_p B^{\SPT}_{p},
    \label{eq:HSPT}
\end{align}
where
  \begin{equation}
\begin{aligned}
\label{eq:SPTlineon1form}
    A^{\SPT}_{v,yz} &= XI_v \prod_{e_y\supset v} \prod_{p \supset e_y} X_p&&=\raisebox{-0.5\height}{\includegraphics{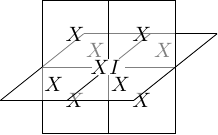}},\\
     A^{\SPT}_{v,xz} &= IX_v \prod_{e_z\supset v} \prod_{p \supset e_z} X_p&&=\raisebox{-0.5\height}{\includegraphics{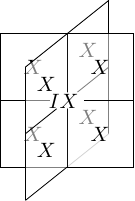}},\\
     A^{\SPT}_{v,xy} &=  XX_v \prod_{e_x\supset v} \prod_{p \supset e_x} X_p&&=\raisebox{-0.5\height}{\includegraphics{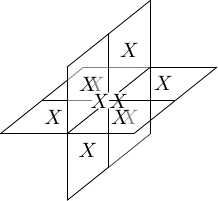}},\\
    B^{\SPT}_{p_{yz}} & =Z_p\prod_{v\supset p} ZZ_v &&=\raisebox{-0.5\height}{\includegraphics{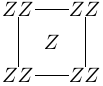}}, \\
    B^{\SPT}_{p_{xz}} &= Z_p\prod_{v\supset p} IZ_v &&=\raisebox{-0.5\height}{\includegraphics{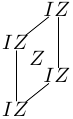}}, \\
    B^{\SPT}_{p_{xy}} &= Z_p\prod_{v\supset p} ZI_v &&=\raisebox{-0.5\height}{\includegraphics{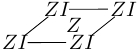}} . 
\end{aligned}
\end{equation}
The Hamiltonian above respects the symmetry $\hat G^{\hat{\ell}}_\mm$ acting on the vertices, and $\hat G^{(1)}_\ee$ acting on the plaquettes. Moreover, the Hamiltonian above has a unique ground state: there are 5 qubits per unit cell, with two independent vertex terms (all three product to the identity) and three plaquette terms. In fact, we may disentangle the Hamiltonian to obtain a product state Hamiltonian via the unitary
\begin{align}
        U_\text{Cluster} = \prod_v \left [\prod_{p \supset e_x \supset v} CX_{v_1,p} \prod_{p \supset e_y \supset v} CX_{v_2,p}\right],
\end{align}
which commutes with the symmetry as a whole, but cannot be decomposed into local gates that commute with the symmetry. The result of applying this unitary gives the Hamiltonian $H^{\hat{\ell}}_{\ee,0} \otimes H^{(1)}_{\mm,0}$ defined in Eqs.~\eqref{eq:triviallineon} and \eqref{eq:trivial1form}. In fact, the form of the unitary above explicitly shows that the ground state is a cluster state after performing a Hadamard on all the plaquettes.

Let us show that the resulting state is in fact a non-trivial SPT state protected by $\hat G^{(1)}_\ee \times \hat G^{\hat{\ell}}_\mm$. We may see this from a decorated domain wall interpretation, where the symmetry defect of one symmetry traps a charge of the other. For example, the defect of the 1-form symmetry hosts lineons on its corners such that a closed membrane of the 1-form symmetry gives a lineon cage configuration.\footnote{This configuration is able to detect the fracton if we gauged the lineon symmetry.} Conversely, the defect of the $\hat G^{\hat{\ell}}_\mm$ hosts a belt of an $\mm$ membrane at its boundary such that multiplying all the lineon symmetries in the region sweeps an $\mm$ loop around the region.

As an aside, in Appendix~\ref{app:gaugedSPT} we provide the stabilizers for the gauged SPT in the $\ZZ_N$ case, and show that when $N$ is prime, the resulting model is always stacks of $\ZZ_N$ Toric codes for any choice of non-trivial SPT.

\subsubsection{Boundary anomaly of SPT}
We may restrict the symmetry action to the boundary in order to analyze the mixed anomaly between the planar subsystem symmetry (which terminates as a line subsystem symmetry) and the 1-form symmetry. Because of the rigidity of the subsystem symmetry, different boundary planes can give rise to different symmetry actions on the boundary. We choose the $(001)$ for simplicity of the discussion. We also note that there are different boundary terminations one can choose. However, some choices may result in certain symmetry actions acting trivially in the boundary Hilbert space.\footnote{A similar phenomenon happens when choosing the smooth boundary for the 2D cluster state on the Lieb lattice, which is protected by a mix of $0$-form and $1$-form $\ZZ_2$ symmetries in 2+1D~\cite{verresen2022higgs}. In that case, one finds that the 1-form acts trivially, allowing a boundary with no edge modes.} Thus we must carefully choose which boundary truncation to take. For example, one can show that choosing the ``smooth" boundary (removing all plaquettes) results in a trivial 1-form symmetry action on the boundary, while choosing the ``rough" boundary (removing all vertices) makes the subsystem symmetry act trivially.

We choose the following boundary truncation, which results in non-trivial symmetry actions for both subsystem and 1-form symmetries: first pick the $xy$ plane $z=0$ on the direct lattice. Perform $CX_{v_2,v_1}$ on all boundary vertices, then truncate the stabilizers by throwing away
\begin{enumerate}
    \item every first qubit of each vertex strictly above the plane, and 
    \item every second qubit of each vertex and every qubit of each plaquette on and above the plane.
\end{enumerate}
The resulting truncated stabilizers on the boundary are
\begin{equation}
\begin{aligned}
\label{eq:SPTlineon1form}
    A^{\SPT,\text{trunc}}_{v,yz} &= \raisebox{-0.5\height}{\includegraphics{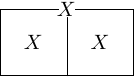}},\\
A^{\SPT,\text{trunc}}_{v,xz}&=\raisebox{-0.5\height}{\includegraphics{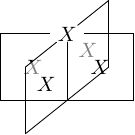}},\\
     A^{\SPT,\text{trunc}}_{v,xy} &=\raisebox{-0.5\height}{\includegraphics{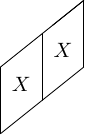}},\\
    B^{\SPT,\text{trunc}}_{p_{yz}} &=\raisebox{-0.5\height}{\includegraphics{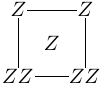}}, \\
    B^{\SPT,\text{trunc}}_{p_{xz}} &=\raisebox{-0.5\height}{\includegraphics{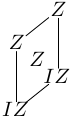}}, \\
    B^{\SPT,\text{trunc}}_{p_{xy}} &=\raisebox{-0.5\height}{\includegraphics{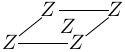}}.
\end{aligned}
\end{equation}

The symmetry restricted to the boundary are realized as follows. First, the planar subsystem symmetry restricts the line subsystem symmetries

  \begin{equation}
\begin{aligned}
\hat{\eta}^{\hat{\ell}}_\mm(\Sigma_{yz}^{(i_0)}) &\xrightarrow{\text{restrict}} \prod_{v \subset \Gamma_{y}^{(i_0)}} X_v,\\
\hat{\eta}^{\hat{\ell}}_\mm(\Sigma_{xz}^{(j_0)}) &\xrightarrow{\text{restrict}}  \prod_{v \subset \Gamma_{x}^{(j_0)}} A^{\SPT,\text{trunc}}_{v,yz},\\
\hat \eta^{(1)}_\ee({\Sigma}_{yz}^{(i)}) &\xrightarrow{\text{restrict}} \prod_{p_{yz} \in \hat{\Gamma}_{y}^{(i)}} Z_{p_{yz}},\\
    \hat \eta^{(1)}_\ee({\Sigma}_{xz}^{(j)}) &\xrightarrow{\text{restrict}} \prod_{p_{xz} \in \hat{\Gamma}_{x}^{(j)}} Z_{p_{xz}},\\
\end{aligned}
\end{equation}
where $\Gamma_{y}^{(i_0)}$ is a rigid line along the $y$ direction with $x$ coordinate $i_0$ on the boundary, and $ \hat{\Gamma}_{y}^{(i)}$ are all the $yz$ plaquettes directly below the boundary of the lattice with $x$-coordinate $i$.

We may use the above truncated boundary stabilizers to construct gapped boundaries as in Refs.~\onlinecite{bulmashboundary,Aitchison2024}. We have the following options:
\begin{enumerate}
    \item If we use $A^{\SPT,\text{trunc}}_{v,xz}$, then we realize the 2+1D Toric Code on the boundary, which spontaneously breaks the 1-form symmetry.
    \item If we use $B^{\SPT,\text{trunc}}_{p_{xy}}$, then we realize the 2+1D plaquette Ising model, which spontaneously breaks the lineon subsystem symmetry
\end{enumerate}

\subsection{Gauging $G^{\ell}_\ee$ and $G^{(2)}_\ee$: Trivial phase with group extension}

Starting from the Hamiltonian $H_{\text{TC},\mm}$, we may instead choose to gauge the symmetry generated by the Wilson lines of the $\ee$ particle of the Toric Code. To facilitate this gauging, it is helpful to perform a basis transformation using Eq.~\eqref{eq:U_SSET} on both the wavefunction and the symmetry action. In this basis, the subsystem symmetry $\hat G^{\hat{\ell}}_\mm$ now acts as:

  \begin{equation}
\begin{aligned}
\hat{\eta}^{\hat{\ell}}_\mm(\Sigma_{yz}) &\xmapsto{U_\text{SSET}} \prod_{v \subset \Sigma_{yz}} XI_{v},\\
\hat{\eta}^{\hat{\ell}}_\mm(\Sigma_{xz}) &\xmapsto{U_\text{SSET}}\prod_{v \subset \Sigma_{xz}} IX_{v},\\
\hat{\eta}^{\hat{\ell}}_\mm(\Sigma_{xy}) &\xmapsto{U_\text{SSET}} \prod_{v \subset \Sigma_{xy}} XX_{v} \prod_{e \supset v} X_e.
\end{aligned}
\label{eq:lineonicsubsystem}
\end{equation}

Under this basis transformation, the stabilizers now take the form of the Toric Code as in Eq.~\eqref{eq:TC1}, which manifestly has a 2-form symmetry $G^{(2)}_\ee$ given by Eq.~\eqref{eq:G2e}. We can therefore now gauge it using $\DD^{G_\ee^{(2)}}$ Eq.~\eqref{eq:GaugeG2e}. Since we already have two qubits on vertices, we let $\DD^{G_\ee^{(2)}}$ map the edges to a new third qubit on each vertex.  We rename the operators under this mapping as follows $\mathcal O^{\TC,\mm} \xmapsto{ \DD^{G^{(2)}_\ee} U_\text{SSET}} \mathcal O^{0,\mm}$ The resulting Hamiltonian is
\begin{align}
 H_{0,\mm} = -\sum_v (A^{0,\mm}_{v,yz}+A^{0,\mm}_{v,xz}+A^{0,\mm}_{v,xy}), 
    \label{eq:H0m}
\end{align}
where
  \begin{equation}
\begin{aligned}
    A^{0,\mm}_{v,yz} &= XII_v,\\
   A^{0,\mm}_{v,xz} &=  IXI_v ,\\
    A^{0,\mm}_{v,xy} &=  XXX_v.
\end{aligned}
\end{equation}
Here, we have used the fact that the plaquette operators which are contractible 2-form Wilson loops are mapped to the identity. Indeed, it is apparent that the ground state of the above Hamiltonian is a product state.

Let us investigate what happens to the symmetries under this gauging. First, the subsystem symmetries $\hat G^{\hat{\ell}}_\mm$ are mapped to
  \begin{equation}
\begin{aligned}
\label{eq:3foliated}
\hat{\eta}^{\hat{\ell}}_\mm(\Sigma_{yz}) &\xmapsto{ \DD^{G^{(2)}_\ee} U_\text{SSET}}\prod_{v \in \Sigma_{yz}} XII_v,\\
\hat{\eta}^{\hat{\ell}}_\mm(\Sigma_{xz}) &\xmapsto{ \DD^{G^{(2)}_\ee} U_\text{SSET}} \prod_{v \in \Sigma_{xz}} IXI_v,\\
\hat{\eta}^{\hat{\ell}}_\mm(\Sigma_{xy}) &\xmapsto{ \DD^{G^{(2)}_\ee} U_\text{SSET}} \prod_{v \in \Sigma_{xy}} XXX_v.
\end{aligned}
\end{equation}
Moreover, we also have a dual 0-form symmetry $\hat G^{(0)}_\mm$ acting as 
\begin{align}
    \hat \eta^{(0)}_\mm =\prod_v IIX_v.
\end{align}
Notice that the dual 0-form symmetry is just the product of all the subsystem symmetries in all three directions. This signifies that $\hat G^{\hat{\ell}}_\mm$ has been extended by $\hat G^{(0)}_\mm$. The resulting symmetry acts on each foliation separately, and is therefore is a 3-foliated 0-form symmetry $\hat G^{(0),\text{fol}}_\mm$. This relation corresponds to the group extension Eq.~\eqref{eq:ext1}, which says that after modding out the diagonal 0-form symmetry from the foliated 0-form symmetry, the remaining symmetry is a subsystem symmetry whose charges are lineons. This relation can also be expressed via the relation between the gauging maps
\begin{align}
    \DD^{\hat G^{(0),\text{fol}}_\mm} =  \DD^{\hat G^{\hat{\ell}}_\mm}  U_\text{SSET}^\dagger   \DD^{\hat G^{(0)}_\mm}.
\end{align}
Taking the dagger of this equation gives the dual group extension Eq.~\eqref{eq:ext2}
\begin{align}
    \DD^{G^{(1),\text{fol}}_\ee} =    \DD^{G^{(2)}_\ee} U_\text{SSET}  \DD^{G^{\ell}_\ee},
\end{align}
which says that after modding out the lineon Wilson operators from the foliated 1-form symmetry, the remaining symmetry is a topological 2-form symmetry.

\subsection{Gauging $G^{(1)}_\mm$: X-Cube}

Starting again from the foliated stacks of Toric Codes Eq.~\eqref{eq:TCstacks}, we instead gauge the 1-form symmetry $G^{(1)}_\mm$. Applying the map, $\DD^{G^{(1)}_\mm}$ in Eq.~\eqref{eq:GaugeG1m}, the resulting model has one qubit on each edge and one qubit on each plaquette.  We rename the operators under this mapping as follows $\mathcal O^\fol \xmapsto{\mathbbm 1 \otimes \DD^{G^{(1)}_\mm}} \mathcal O^{\XC,\ee}$. The resulting Hamiltonian is
  \begin{equation}
\begin{aligned}
 H_{\XC,\ee} =& -\sum_v (A^{\XC,\ee}_{v,yz}+A^{\XC,\ee}_{v,xz}+A^{\XC,\ee}_{v,xy})\\
& - \sum_p B^{\XC,\ee}_{p} - \sum_c V^{\XC,\ee}_c,
    \label{eq:HXCe}
\end{aligned}
\end{equation}
where
  \begin{equation}
\begin{aligned}
    A^{\XC,\ee}_{v,yz}  &= \prod_{e_y,e_z \supset v } X_e   \prod_{p\supset e_y} X_p && =\raisebox{-0.5\height}{\includegraphics{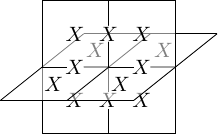}}, \\
    A^{\XC,\ee}_{v,xz} &=  \prod_{e_x,e_z \supset v } X_e   \prod_{p\supset e_z} X_p&& =\raisebox{-0.5\height}{\includegraphics{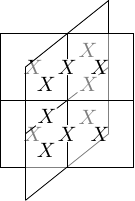}},\\
    A^{\XC,\ee}_{v,xy}&=   \prod_{e_x,e_y \supset v } X_e  \prod_{p\supset e_x} X_p&& =\raisebox{-0.5\height}{\includegraphics{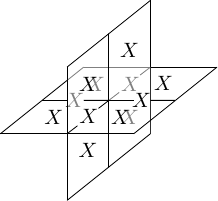}},\\
    B^{\XC,\ee}_{p_{yz}} &= Z_p \prod_{e_z \supset p} Z_e&&=\raisebox{-0.5\height}{\includegraphics{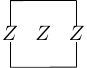}},  \\
    B^{\XC,\ee}_{p_{xz}} &= Z_p \prod_{e_x \supset p} Z_e&&=\raisebox{-0.5\height}{\includegraphics{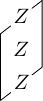}},\\
    B^{\XC,\ee}_{p_{xy}} &= Z_p \prod_{e_y \supset p} Z_e&&=\raisebox{-0.5\height}{\includegraphics{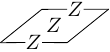}},\\
     V^{\XC,\ee}_c &=  \hat \eta^{(1)}_\ee(c) =\prod_{p \subset c} Z_p,
\label{eq:SEF_XC}
\end{aligned}
\end{equation}
where importantly, we have also enforced the local generators $\hat \eta^{(1)}_\ee(c)$ of the dual 1-form symmetry $\hat G^{(1)}_\ee$ in Eq.~\eqref{eq:G1ebasis} energetically to avoid an extensive degeneracy.

 To see that this model is indeed equivalent to the X-Cube model, we apply the following unitary
\begin{align}
    U_\text{SEF}=& \prod_{p_{yz}} \prod_{e_z \supset p_{yz}}CX_{e_z,p_{yz}} \times \prod_{p_{xz}} \prod_{e_x \supset p_{xz}}CX_{e_x,p_{xz}} \nonumber\\
    &\times \prod_{p_{xy}} \prod_{e_y \supset p_{xy}}CX_{e_y,p_{xy}}
\end{align}
The result is
\begin{subequations}
\begin{align}
    A^{\XC,\ee}_{v,yz} &\xmapsto{U_\text{SEF}  }\prod_{e_y,e_z \supset v } X_e = A_{v,yz} ,\\
     A^{\XC,\ee}_{v,xz} &\xmapsto{U_\text{SEF}} \prod_{e_x,e_z \supset v } X_e= A_{v,xz} ,\\
    A^{\XC,\ee}_{v,xy} &\xmapsto{U_\text{SEF} }  \prod_{e_x,e_y \supset v } X_e = A_{v,xy} ,\\
    B^{\XC,\ee}_{p} &\xmapsto{U_\text{SEF} } Z_p  ,\\
    V_c &\xmapsto{U_\text{SEF} }\prod_{e \subset c} Z_e \prod_{p \subset c} Z_p  = W_c \prod_{p \subset c} Z_p. 
\end{align}
\label{eq:Xcube}
\end{subequations}

So the $A_v$ terms are naturally the vertex terms of X-Cube. Moreover, by restricting to the subspace where  $B^{\XC,\ee}_{p}= Z_p=1$, we see that $V_c$ matches exactly the cube term $W_c$ of the X-Cube model in Eq.~\eqref{eq:XC0terms}.

The fact that the cube term arises from enforcing the dual 1-form symmetry energetically signifies that the fractons were not originally local excitations $H_\fol$. In fact, they should be thought of as an semi-infinite product of $m$ excitations in each foliation.

Consider the symmetry action of the 1-form on a region with boundary. For concreteness, we consider a rectangle in the $xy$ plane. In the ground state, we may use the fact that $B^{\XC,\ee}_{p_{xy}}=1$ to substitute $Z_p$ for $\prod_{e_y \supset p} Z_e$. Thus, the corners of this rectangle hosts lineons. If we further act with a one-form symmetry surrounding a region, (say a cube $c$), using the plaquette operators, we may similarly replace $\prod_{p \subset c} Z_p$ with $W_c$, which detects the fracton. Therefore, we conclude that the symmetry enrichment comes from the fact that the fracton is charged under the 1-form symmetry.

\subsubsection{Gauging $G^{\ell}_\ee$}

Starting from $ H_{\XC,\ee}$ in Eq.~\eqref{eq:HXCe}, we may further gauge $G^{\ell}_\ee$ using the mapping $D^{G^{\ell}_\ee}$. This gives rise to the SPT Hamiltonian in Eq.~\eqref{eq:HSPT}.

\subsection{Gauging $G^{(1)}_\mm$ and $G^{f}_\mm$: Trivial phase with group extension}
We perform a basis transformation on both $ H_{\XC,\ee}$ and the symmetry using $U_\text{SEF}$. The Hamiltonian is now effectively the X-Cube model, while the dual 1-form symmetry $\hat G^{(1)}_\ee$ now acts in an unusual manner. Starting from the basis for $\hat G^{(1)}_\ee$ in Eq.~\eqref{eq:G1ebasis}, applying $U_\text{SEF}$ results in 
\begin{align}
    \hat \eta^{(1)}_\ee(c) &\xmapsto{U_\text{SEF}} \prod_{p \subset c} Z_p \prod_{e \subset c} Z_e,\\
   \hat \eta^{(1)}_\ee({\Sigma}_{yz}) &\xmapsto{U_\text{SEF}}\prod_{p_{yz} \in \Sigma_{yz}} Z_{p_{yz}},\\
   \hat \eta^{(1)}_\ee({\Sigma}_{xz}) &\xmapsto{U_\text{SEF}} \prod_{p_{xz} \in \Sigma_{xz}} Z_{p_{xz}},\\
    \hat \eta^{(1)}_\ee({\Sigma}_{xy}) &\xmapsto{U_\text{SEF}} \prod_{p_{xy} \in \Sigma_{xy}} Z_{p_{xy}}.
\end{align}

Let us now gauge  $G^{f}_\mm$ via the map $\DD^{G^{f}_\mm}$ in Eq.~\eqref{eq:GaugeGfm}. We rename the operators under this mapping as follows $\mathcal O^{\XC,\ee} \xmapsto{\DD^{G^{f}_\mm} U_\text{SEF}} \mathcal O^{0,\ee}$. The resulting Hamiltonian is
  \begin{equation}
\begin{aligned}
 H_{0,\ee} =& - \sum_p B^{0,\ee}_{p} - \sum_c V^{0,\ee}_c,
    \label{eq:H0e}
\end{aligned}
\end{equation}
where
  \begin{equation}
\begin{aligned}
    B^{0,\ee}_{p} &= Z_p,\\
    V^{0,\ee}_c &= Z_c\prod_{p \subset c} Z_p, 
\end{aligned}
\end{equation}
and we have used the fact that $\mathcal A^{\XC,\ee}_v \xmapsto{\DD^{G^{f}_m} U_\text{SEF}} 1$.

Under this gauging, let us call the symmetry operators of $\hat{G}^{(1)}_\ee$ as ${\hat \eta^{(1)}_\ee}{'}$. In particular, the local symmetry action is
\begin{align}
{\hat \eta^{(1)}_\ee}{'}(c) \equiv Z_c\prod_{p \subset c} Z_p, 
\end{align}
while $\hat \eta^{(1)}_\ee({\Sigma}_{xz})$, $\hat \eta^{(1)}_\ee({\Sigma}_{yz})$, $\hat \eta^{(1)}_\ee({\Sigma}_{xy})$ are invariant under this mapping.
In addition, we also have a dual $\hat G^{\hat{f}}_\ee$ which acts according to Eq.~\eqref{eq:hatGhatfe}.

After gauging, we now find that the 1-form symmetry is extended by the planar subsystem symmetry. In particular, even in the subspace where the ``local symmetry action" of the 1-form $ Z_c\prod_{p \subset c} Z_p$ is set to one on all cubes, the 1-form symmetry is still not topological. For example, consider the product of $\eta^{(1)}_\ee({\Sigma}_{xy})$ along two adjacent $xy$ planes which sandwich the dual plane $\hat \Sigma_{xy}$. This product differs by
\begin{align}
\prod_{c\in \hat \Sigma_{xy}} \left[ Z_c\prod_{p \subset c} Z_p\right] = \prod_{c\in \hat \Sigma_{xy}}  Z_c = \hat \eta^{\hat f}_\ee (\hat \Sigma_{xy}),
\end{align}
which is precisely the subsystem planar symmetry defined on $\hat \Sigma_{xy}$. Thus, we see that more generally, homologous 1-form symmetries now differ by the product of all the subsystem symmetries in the intermediate layers. 

We can further elaborate on the nature of this extended symmetry by a basis transformation on both the Hamiltonian and the symmetries
\begin{align}
    U_{0,\ee}= \prod_c \prod_{p \subset c} CX_{p,c}.
\end{align}

In this basis, the terms in the Hamiltonian are
\begin{align}
    B^{0,\ee}_p &\xmapsto{U_{0,\ee}} Z_p,\\
    V^{0,\ee}_c &\xmapsto{U_{0,\ee}} Z_c .
\end{align}
and the symmetries now act as
\begin{align}
{\hat \eta^{(1)}_\ee}{'}(c) &\xmapsto{U_{0,\ee}} Z_c,\\
{\hat \eta^{(1)}_\ee}{'}({\Sigma}_{yz}) & \xmapsto{U_{0,\ee}} \prod_{p_{yz} \in \Sigma_{yz}} Z_{p_{yz}},\\
{\hat \eta^{(1)}_\ee}{'}({\Sigma}_{xz}) & \xmapsto{U_{0,\ee}} \prod_{p_{yz} \in \Sigma_{xz}} Z_{p_{yz}},\\
{\hat \eta^{(1)}_\ee}{'}({\Sigma}_{xy}) & \xmapsto{U_{0,\ee}} \prod_{p_{yz} \in \Sigma_{xy}} Z_{p_{yz}},\\
 \hat{\eta}^{\hat{f}}_\ee(\hat \Sigma_{yz,i})  & \xmapsto{U_{0,\ee}} \prod_{c \subset \hat \Sigma_{yz,i}} \left[Z_c \prod_{p \subset c} Z_p \right],\\
    \hat{\eta}^{\hat{f}}_\ee(\hat \Sigma_{xz,j}) & \xmapsto{U_{0,\ee}} \prod_{c \subset \hat \Sigma_{xz,j}} \left[Z_c \prod_{p \subset c} Z_p \right],\\
   \hat{\eta}^{\hat{f}}_\ee(\hat \Sigma_{xy,k})  & \xmapsto{U_{0,\ee}} \prod_{c \subset \hat \Sigma_{xy,k}}\left[Z_c \prod_{p \subset c} Z_p \right].
\end{align}
We may now restrict to the subspace where $Z_c=1$ on every cube. In this subspace, the planar subsystem symmetry reduces to a product of two foliated 0-form symmetries on adjacent planes. This, along with ${\hat{\eta}^{(1)}_\ee}{'}$ which generates a foliated 0-form symmetry on a single plane, generates all the foliated 0-form symmetries $\hat G^{(0),\text{fol}}_\ee$, which we may instead represent with the basis
\begin{align}
    \prod_{xy} Z_p, && \prod_{yz} Z_p, && \prod_{xz} Z_p.
\end{align}

To conclude, the foliated 0-form symmetry has a subgroup which is a fracton planar subsystem symmetry generated by the product of all consecutive planes. The remaining quotient group is a topological 1-form symmetry, and corresponds to the group extension Eq.~\eqref{eq:ext3}. This relation can also be expressed via the relation between the gauging maps
\begin{align}
       \DD^{\hat G^{(0),\text{fol}}_\ee} =  \DD^{\hat G^{(1)}_\ee}  U_\text{SEF}^\dagger   \DD^{\hat G^{\hat{f}}_\ee}.
\end{align}
Taking the dagger of this equation gives the dual group extension Eq.~\eqref{eq:ext4}
\begin{align}
    \DD^{G^{(1),\text{fol}}_\mm} =    \DD^{G^{f}_\mm} U_\text{SEF}  \DD^{G^{(1)}_\mm},
\end{align}
which says that after modding out the 1-form symmetries generated by the $p$-strings from the foliated 1-form symmetry, the remaining symmetry is generated by fracton Wilson operators.


\section{Gauging web in field theory}
\label{sec:cont-gaugingweb}

In this Section, we give a continuum field theory perspective of the gauging web. In addition to the conventions mentioned in Sec.~\ref{sec:symreview}, we use boldfaced letters for background fields and ordinary letters for dynamical fields. Furthermore, in foliated field theory, we use lower-case letters for bulk fields and upper-case letters for fields on the layers.

\subsection{Foliated stack}
The starting point is a 3-foliated stack of 2+1D $\mathbb Z_N$ gauge theories in 3+1D described by the Lagrangian
\ie\label{3dfollag}
\mathcal L_\fol = \sum_i\sum_{n_i=1}^{L_i} \frac{\ii N}{2\pi} A_\mm^{(i)} dA_\ee^{(i)} \delta(x_i - n_i \ell_i) dx_i~,
\fe
where $\ell_i$ is the spacing between the layers in the foliation orthogonal the $i$-th spatial direction and $L_i$ is the number of such layers. Here, $A_{\mm,\ee}^{(x)}=A_{\mm,\ee}^{(x)}(\tau,y,z;n_x)$, i.e., $n_x$ labels a layer orthogonal to the $x$ direction, and $y,z$ are the spatial coordinates along the layer. Similar comments apply to the fields on the layers orthogonal to the $y$ and $z$ directions.

Each layer has two $G=\mathbb Z_N$ 1-form symmetries generated by the Wilson line operators of $A^{(1)}_{\mm,\ee}$:
\ie
\eta_{\mm,yz}^{(1),\fol}(\Gamma^{n_x}) = \exp\left(\ii \oint_{\Gamma^{n_x}} A_\mm^{(x)}\right)~,
\fe
and its variants in the other directions, and similar operators with $\mm\rightarrow\ee$. Here, $\Gamma^{n_x}$ is a closed curve in the $n_x$-th layer. We refer to the total symmetry generated by these operators as a foliated 1-form symmetry, denoted as $G^{(1),\fol}_{\mm,\ee}$.

Let us identify two subgroups that we are interested in gauging.
\begin{enumerate}

\item The first is the subgroup of $G^{(1),\fol}_\ee$ generated by the lineon operators
\ie\label{lineonop}
\eta_\ee^\ell(\Gamma^{n_x,n_y}) := \eta_{\ee,yz}^{(1),\fol}(\Gamma^{n_x,n_y}) \eta_{\ee,xz}^{(1),\fol}(\Gamma^{n_x,n_y})^{-1}~,
\fe
and its variants in the other directions. Here, $\Gamma^{n_x,n_y}$ is a closed curve along the intersection of the $n_x$-th and $n_y$-th layers. This symmetry is denoted as $G_\ee^\ell$.

\item The second is the diagonal subgroup of $G^{(1),\fol}_\mm$ generated by the surface operators
\ie
\eta_\mm^{(1)}(\Sigma) := \prod_i \prod_{n_i=1}^{L_i} \eta_{\mm,jk}^{(1),\fol}(\Gamma^{n_i})~,
\fe
where $i,j,k$ are cyclic, $\Sigma$ is a closed surface, and $\Gamma^{n_i}$ is the intersection of $\Sigma$ with the $n_i$-th layer. When $\Sigma$ is open, the application of $\eta$ creates a defect on its boundary made of a string of $\mm$ anyons, commonly known as a $p$-string. As the notation suggests, this operator generates a global 1-form symmetry in 3+1D, so we denote it as $G_\mm^{(1)}$.

\end{enumerate}
Although the $\ee$ and $\mm$ anyons in each layer braid nontrivially, signalling a mixed 't Hooft anomaly between the foliated 1-form symmetries they generate, there is no mixed 't Hooft anomaly between $G_\ee^\ell$ and $G_\mm^{(1)}$,\footnote{Here is one way to see the absence of the mixed anomaly: $G_\ee^\ell$ is generated by the bound state of two $\ee$ anyons from orthogonal layers at their intersection, and it is easy to check that this combination braids trivially with the $p$-string that generates $G_\mm^{(1)}$. See Sec.~\ref{sec:decouple} for a brief discussion of $p$-string condensation.} so we can gauge these symmetries simultaneously.

Alternatively, we can first gauge one of the two symmetries, and then gauge the other. The two orders in which the symmetries are gauged lead to a rich gauging web involving the 3+1D $\mathbb Z_N$ gauge theory, the 3+1D foliated field theory of $\mathbb Z_N$ X-Cube model, and a new 3+1D SPT protected by a subsystem symmetry and a 1-form symmetry. This produces the lower half of the gauging web in Fig.~\ref{fig:gaugingweb}. To complete the web, one just has to repeat this discussion after exchanging $\mm$ and $\ee$.

\subsection{Gauging $G_\ee^\ell$}\label{sec:gaugelsym}
Before gauging $G_\ee^\ell$, we should first address what the gauge fields associated with this symmetry are. The idea behind constructing the gauge fields can be best understood in a simpler example. Consider a QFT $\mathcal T$ in $d$ spacetime dimensions with a $p$-form abelian $G$ symmetry. Say we want to gauge a subgroup $H\subset G$. One can directly couple the theory to $H$-valued $(p+1)$-form gauge fields. But there is an alternative approach. Consider the short exact sequence
\ie
1 \to H \to G \to K \to 1~,
\fe
where $K\cong G/H$. By Pontrjagin duality, there is a dual sequence,
\ie
1 \to \hat K \to \hat G \to \hat H \to 1~,
\fe
where $\hat K \subset \hat G$ and $\hat H \cong \hat G/\hat K$. Here, $\hat G$ denotes the Pontrjagin dual of $G$. It is well known that gauging a $p$-form $G$ symmetry give rise to a dual $(d-p-2)$-form $\hat G$ symmetry, and gauging the dual symmetry takes us back to the original theory with $p$-form $G$ symmetry. Therefore, instead of gauging $H$ directly, one can equivalently gauge $G$ and then gauge the subgroup $\hat K$ of the dual symmetry $\hat G$. The resulting theory is the same in both cases. Moreover, the symmetry at the end of both procedures is the $p$-form $K$ symmetry times the dual $(d-p-2)$-form $\hat H$ symmetry.

In an analogous way, consider the group extension given by Eq.~\eqref{eq:ext2}. Instead of gauging $G_\ee^\ell$ directly in the foliated stack Eq.~\eqref{3dfollag}, we could first gauge the full foliated 1-form symmetry $G_\ee^{(1),\fol}$, and then gauge a diagonal subgroup $\hat G_\mm^{(0)}$ of the dual foliated 0-form symmetry $\hat G_\mm^{(0),\fol}$. This is emphasized in the commutativity of the bottom-left triangle in Fig.~\ref{fig:gaugingweb}.

At the end of the first step, we land in a trivial theory, denoted as $\mathcal T_{0,\mm}$ in Fig.~\ref{fig:gaugingweb}:
\ie\label{3dfollag-gaugefol1-form}
\mathcal L_{0,\mm} = \sum_i\sum_{n_i=1}^{L_i} \frac{\ii N}{2\pi} &\Big[ A_\ee^{(i)} (dA_\mm^{(i)} - B_\mm^{(i)})
\\
&- \varPhi_\ee^{(i)} dB_\mm^{(i)} \Big]\delta(x_i - n_i \ell_i) dx_i~,
\fe
where $B_\mm^{(i)}$ is a 2-form gauge field on the layers, and $\varPhi_\ee^{(i)}$ is a Lagrange multiplier (compact scalar) on the layers that constrains $B_\mm^{(i)}$ to be a $\mathbb Z_N$ 2-form layer gauge field. The gauge symmetry acts as
\ie
&A_\ee^{(i)} \sim A_\ee^{(i)} + d\alpha_\ee^{(i)}~,\quad \varPhi_\ee^{(i)} \sim \varPhi_\ee^{(i)} + \alpha_\ee^{(i)}~,
\\
&A_\mm^{(i)} \sim A_\mm^{(i)} + \beta_\mm^{(i)}~,\quad B_\mm^{(i)} \sim B_\mm^{(i)} + d\beta_\mm^{(i)}~.
\fe
This theory has a dual foliated 0-form symmetry $\hat G_\mm^{(0),\fol}$ generated by the Wilson surface operators $\hat \eta_{\mm,jk}^{(0),\fol}(\Sigma^{n_i}) = \exp(\ii \oint_{\Sigma^{n_i}} B_\mm^{(i)})$, where $\Sigma^{n_i}$ is a closed surface in the $n_i$-th layer. The diagonal subgroup $\hat G_\mm^{(0)}$ of this dual symmetry is generated by the operator
\ie
\hat \eta_\mm^{(0)}(M) := \prod_{i,j,k\atop \mathrm{cyclic}} \prod_{n_i=1}^{L_i} \hat \eta_{\mm,jk}^{(0),\fol}(\Sigma^{n_i})~,
\fe
where $M$ is a 3-cycle and $\Sigma^{n_i}$ is the intersection of $M$ with the $n_i$-th layer.

In the next step, we gauge $\hat G_\mm^{(0)}$ by coupling $\mathcal T_{0,\mm}$ to a $\mathbb Z_N$ 1-form bulk gauge field $a_\ee$:
\ie\label{3dfollag-gaugelsym}
&\mathcal L_{\TC,\mm} = \sum_i\sum_{n_i=1}^{L_i} \frac{\ii N}{2\pi} \Big[ A_\ee^{(i)} (dA_\mm^{(i)} - B_\mm^{(i)})
\\
&+ (d\varPhi_\ee^{(i)} - a_\ee)B_\mm^{(i)} \Big]\delta(x_i - n_i \ell_i) dx_i + \frac{\ii N}{2\pi} b_\mm da_\ee~,
\fe
where $b_\mm$ is a Lagrange multiplier (2-form bulk gauge field) that constrains $a_\ee$ to be a $\mathbb Z_N$ 1-form bulk gauge field. The gauge symmetry acts as\footnote{The gauge fields $(B_\mm^{(i)},b_\mm)$ together can be interpreted as the foliated-equivalent of the exotic tensor gauge fields $(b^\mm_{\tau ij}, b^\mm_{[ij]k})$, whereas the combination $(\varPhi_\ee^{(i)},a_\ee)$ is the foliated-equivalent of the exotic scalar field $\phi_\ee^{[ij]k}$. One can derive this correspondence using methods similar to the ones in Ref.~\onlinecite{Ohmori:2022rzz}.}
\ie
&A_\ee^{(i)} \sim A_\ee^{(i)} + d\alpha_\ee^{(i)}~,\quad a_\ee \sim a_\ee + d\alpha_\ee~,
\\
&\varPhi_\ee^{(i)} \sim \varPhi_\ee^{(i)} + \alpha_\ee^{(i)} + \alpha_\ee~,
\\
&A_\mm^{(i)} \sim A_\mm^{(i)} + \beta_\mm^{(i)}~,\quad B_\mm^{(i)} \sim B_\mm^{(i)} + d\beta_\mm^{(i)}~,
\\
&b_\mm \sim b_\mm + d\beta_\mm + \sum_i \sum_{n_i=1}^{L_i} \beta_\mm^{(i)}\delta(x_i - n_i \ell_i) dx_i~.
\fe

To ensure that we indeed gauged the lineon subsystem symmetry $G_\ee^\ell$, note that the equation of motion of $B_\mm^{(i)}$ implies that the line operator of $A_\ee^{(i)}$ is identified with the line operator of $a_\ee$. Therefore, the lineon operator $\eta_\ee^\ell(\Gamma^{n_i,n_j})$ in Eq.~\eqref{lineonop} acts trivially. Moreover, the nontriviality of the surface operator in the lower right corner of Eq.~\eqref{3dzncorr} implies that we gauged only the subgroup $G_\ee^\ell \subset G_\ee^{(1),\fol}$ and not more.

As the notation suggests, the Lagrangian Eq.~\eqref{3dfollag-gaugelsym} describes the ground states of the 3+1D $\mathbb Z_N$ Toric Code. The correspondence between the operators in the 3+1D $\mathbb Z_N$ gauge theory Eq.~\eqref{3dznlag} and those in Eq.~\eqref{3dfollag-gaugelsym} is given as follows:
\ie\label{3dzncorr}
& \exp\left(\ii \oint_\gamma a_\ee \right) \leftrightarrow \exp\left(\ii \oint_\gamma a_\ee\right)~,
\\
& \exp\left(\ii \oint_\Sigma b_\mm \right) \leftrightarrow \exp\left(\ii \oint_\Sigma b_\mm - \ii \sum_i \sum_{n_i=1}^{L_i} \oint_{\Gamma^{n_i}} A_\mm^{(i)} \right)~,
\fe
where $\gamma$ is a closed curve, $\Sigma$ is a closed surface, and $\Gamma^{n_i}$ is the intersection of $\Sigma$ with the $n_i$-th layer. The line operator in the first line generates the $\mathbb Z_N$ 2-form symmetry, denoted as $G_\ee^{(2)} \cong G_\ee^{(1),\fol} / G_\ee^\ell$,\footnote{One can see why this quotient is a 2-form symmetry from the ``lineon condensation'' perspective. An $\ee_{xy}$ anyon in an $xy$ layer can turn into an $\ee_{yz}$ anyon in a $yz$ layer by leaving behind a lineon formed by the bound state $\ee_{xy}\ee_{yz}^{-1}$ at the intersection of the two layers. This bound state is absorbed into the lineon condensate, and so an $\ee$ anyon becomes completely mobile in 3+1D. This corresponds to the $\ee$ particle of the 3+1D $\mathbb Z_N$ Toric Code that generates the 2-form symmetry. Alternatively, the 2-form symmetry is the ``dual'' symmetry obtained from gauging the 0-form symmetry $\hat G_\mm^{(0)}$ in 3+1D.} whereas the surface operator in the second line generates the $\mathbb Z_N$ 1-form symmetry $G_\mm^{(1)}$.

In addition to these two global symmetries, there is a dual lineon subsystem symmetry, denoted $\hat G_\mm^{\hat \ell}$, coming from gauging the lineon subsystem symmetry $G_\ee^\ell$. This dual symmetry is generated by the Wilson surface operators
\ie
\hat \eta_\mm^{\hat \ell}(\Sigma_{yz}^{n_x}) = \exp\left(\ii \oint_{\Sigma_{yz}^{n_x}} B_\mm^{(x)} \right)~,
\fe
and its variants in the other directions. Here, $\Sigma_{yz}^{n_x}$ is a closed surface in the $n_x$-th layer. They satisfy the constraint
\ie
\prod_{i,j,k \atop \mathrm{cyclic}} \prod_{n_i=1}^{L_i} \hat \eta^{\hat \ell}_\mm(\Sigma^{n_i}_{jk}) = 1~,
\fe
which follows from the equation of motion of $a_\ee$.

The 3+1D $\mathbb Z_N$ gauge theory is enriched by the dual symmetry $\hat G_\mm^{\hat \ell}$ as can be seen from the fact that the $\ee$ particle is charged under it.\footnote{The $\ee$ particle in the bulk is identified with the $\ee_{yz}$ anyon in the $yz$ layer by the equation of motion of $B_\mm^{(x)}$, and the latter is detected by the surface operator $\hat \eta_\mm^{\hat \ell}(\Sigma_{yz}^{n_x})$.} In other words, the Lagrangian Eq.~\eqref{3dfollag-gaugelsym} describes a subsystem symmetry-enriched 3+1D $\mathbb Z_N$ gauge theory.

\subsubsection{Gauging $\hat G_\mm^{\hat \ell}$}

Gauging the dual lineon subsystem symmetry $\hat G_\mm^{\hat \ell}$ takes us back to the foliated stack Eq.~\eqref{3dfollag}. To see this, we couple the theory $\mathcal T_{\TC,\mm}$ to the foliated version of the ``lineon gauge fields'':
\ie
\mathcal L_\fol &= \sum_i\sum_{n_i=1}^{L_i} \frac{\ii N}{2\pi} \Big[ A_\ee^{(i)} (dA_\mm^{(i)} - B_\mm^{(i)}) + A'^{(i)}_\mm (dA'^{(i)}_\ee - b'_\ee)
\\
&+ (d\varPhi_\ee^{(i)} - a_\ee - A'^{(i)}_\ee)B_\mm^{(i)} \Big]\delta(x_i - n_i \ell_i) dx_i
\\
&+ \frac{\ii N}{2\pi} \left[b_\mm (da_\ee + b'_\ee) + a'_\mm db'_\ee \right]~,
\fe
where the combination $(A'^{(i)}_\ee,b'_\ee)$ serves as the foliated version of the lineon gauge fields $(a'^{k(ij)}_{\ee,\tau},a'^{ij}_\ee)$, and the combination $(A'^{(i)}_\mm,a'_\mm)$ serve as the foliated version of the Lagrange multipliers $(a'^\mm_\tau,a'^\mm_{ij})$ that constrain $(A'^{(i)}_\ee,b'_\ee)$ to be $\mathbb Z_N$ gauge fields. We encounter these combinations of gauge fields in Sec.~\ref{sec:gauge1-form}. The gauge symmetry acts as
\ie
\!&A_\ee^{(i)} \sim A_\ee^{(i)} + d\alpha_\ee^{(i)}~,\quad a_\ee \sim a_\ee - \beta'_\ee~,
\\
&\varPhi_\ee^{(i)} \sim \varPhi_\ee^{(i)} + \alpha_\ee^{(i)} + \alpha'^{(i)}_\ee~,
\\
&A_\ee'^{(i)} \sim A_\ee'^{(i)} + d\alpha_\ee'^{(i)} + \beta'_\ee~,\quad b'_\ee \sim b'_\ee + d\beta'_\ee~,
\\
&A_\mm^{(i)} \sim A_\mm^{(i)} + \beta_\mm^{(i)}~,\quad B_\mm^{(i)} \sim B_\mm^{(i)} + d\beta_\mm^{(i)}~,
\\
&b_\mm \sim b_\mm + d\beta_\mm + \sum_i \sum_{n_i=1}^{L_i} \beta_\mm^{(i)}\delta(x_i - n_i \ell_i) dx_i~,
\\
&A_\mm'^{(i)} \sim A_\mm'^{(i)} + d\alpha_\mm'^{(i)} + \beta_\mm^{(i)}~,
\\
&a_\mm' \sim a_\mm' - \beta_\mm + \sum_i\sum_{n_i=1}^{L_i} \alpha_\mm'^{(i)} \delta(x_i - n_i \ell_i) dx_i~.
\fe
It is not too hard to verify that the only nontrivial gauge invariant operators in this theory are the Wilson lines on the layers given by
\ie
\!&\exp\left( \ii \oint_{\Gamma^{n_i}} A^{(i)}_\ee \right) = \exp\left( -\ii \oint_{\Gamma^{n_i}} (A'^{(i)}_\ee + a_\ee) \right)~,
\\
&\exp\left( \ii \oint_{\Gamma^{n_i}} (A^{(i)}_\mm - A'^{(i)}_\mm) \right)~,
\fe
where the equality in the first line follows from the equation of motion of $B^{(i)}_\mm$. This matches precisely with the operator content of the foliated stack Eq.~\eqref{3dfollag}.

\subsubsection{Gauging $G_\ee^{(2)}$}

On the other hand, if we gauge the 2-form symmetry $G_\ee^{(2)}$, we end up in the trivial theory $\mathcal T_{0,\mm}$. This is because gauging $G_\ee^{(2)}$ in Eq.~\eqref{3dfollag-gaugelsym} is equivalent to gauging the foliated 1-form symmetry $G_\ee^{(1),\fol}$ in the foliated stack Eq.~\eqref{3dfollag}. This can be seen explicitly by coupling the theory Eq.~\eqref{3dfollag-gaugelsym} to a $\mathbb Z_N$ 3-form gauge field, but we do not discuss the details of this here.

\subsubsection{Gauging $G_\mm^{(1)}$}

Finally, we can also gauge the 1-form symmetry $G_\mm^{(1)}$. We defer the discussion of this gauging to Sec.~\ref{sec:contSPT}.

\subsection{Gauging $G_\mm^{(1)}$}\label{sec:gauge1-form}
We can gauge $G_\mm^{(1)}$ by coupling the foliated stack Eq.~\eqref{3dfollag} to a $\mathbb Z_N$ 2-form gauge field $b_\ee$ in the 3+1D bulk. The resulting Lagrangian is
\ie\label{3dfollag-gauge1-form}
\mathcal L_{\XC,\ee} &= \sum_i\sum_{n_i=1}^{L_i} \frac{\ii N}{2\pi}A_\mm^{(i)} (dA_\ee^{(i)} - b_\ee) \delta(x_i - n_i \ell_i) dx_i
\\
& + \frac{\ii N}{2\pi} a_\mm db_\ee~,
\fe
where $a_\mm$ is a Lagrange multiplier (1-form bulk gauge field) that constrains $b_\ee$ to be a $\mathbb Z_N$ 2-form bulk gauge field. The gauge symmetry acts as
\ie
&A_\ee^{(i)} \sim A_\ee^{(i)} + d\alpha_\ee^{(i)} + \beta_\ee~,\quad b_\ee \sim b_\ee + d\beta_\ee~,
\\
&A_\mm^{(i)} \sim A_\mm^{(i)} + d \alpha_\mm^{(i)}~,
\\
&a_\mm \sim a_\mm + d \alpha_\mm + \sum_i\sum_{n_i=1}^{L_i} \alpha_\mm^{(i)} \delta(x_i - n_i \ell_i) dx_i~.
\fe
One can recognise Eq.~\eqref{3dfollag-gauge1-form} as the Lagrangian for the folitated field theory description of the $\mathbb Z_N$ X-Cube model~\cite{Slagle21,SlagleSMN,HsinSlagle21}. The correspondence between the exotic field theory Eq.~\eqref{3dxclag} and the foliated field theory Eq.~\eqref{3dfollag-gauge1-form} is spelled out in~\cite{Ohmori:2022rzz}. The connection between gauging $G_\mm^{(1)}$ and $p$-string condensation was discussed recently in a companion paper~\cite{GPTW25}.

\begin{figure}
    \centering
    \includegraphics[scale=0.3]{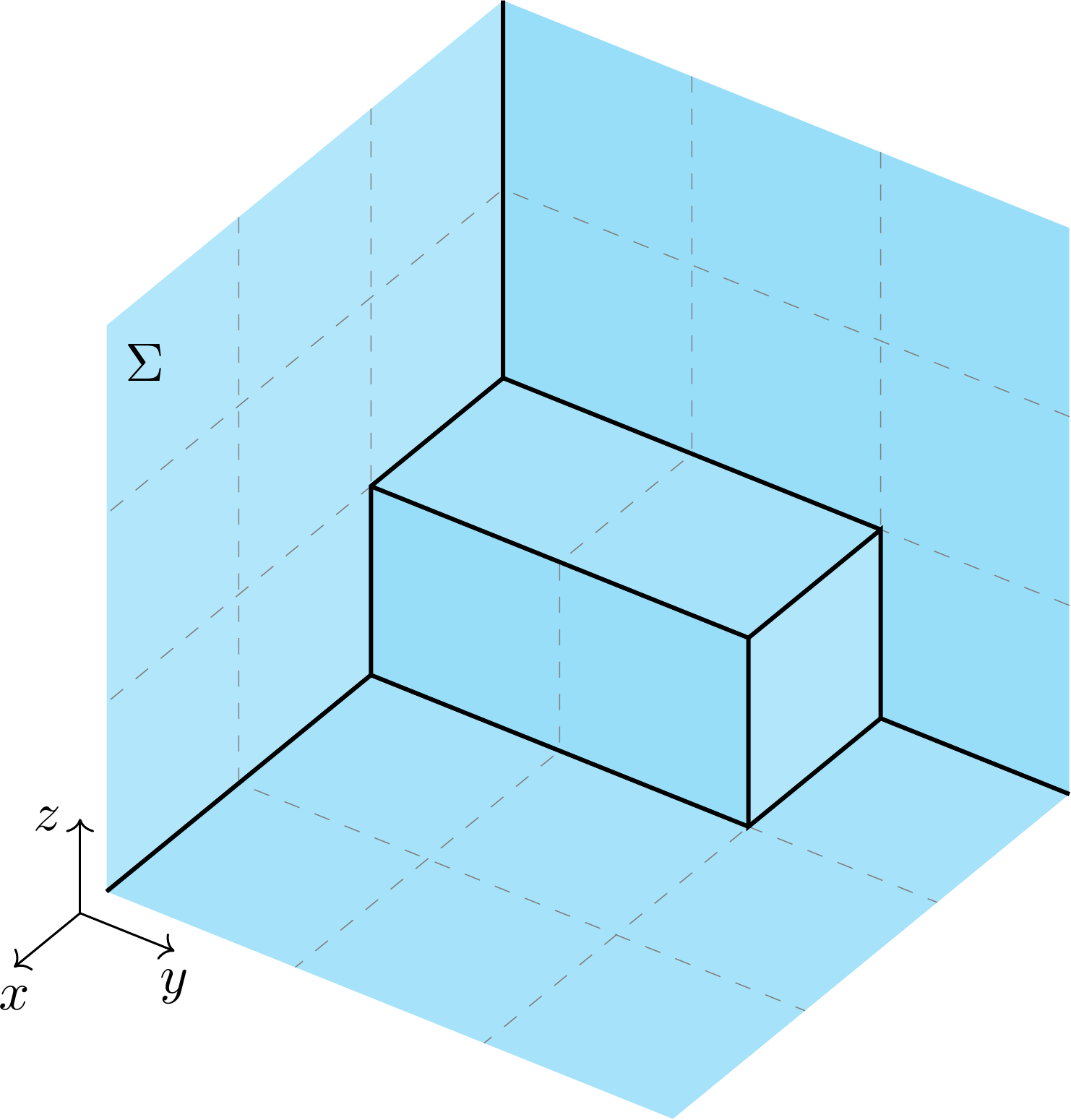}
    \caption{The cyan surface $\Sigma$ is along the layers (not shown) and its cage, $\mathrm{cage}(\Sigma)$, is made of black solid lines along intersections of orthogonal layers. The gray dashed lines represent the intersections of $\Sigma$ with the orthogonal layers.}
    \label{fig:surf-cage}
\end{figure}

In addition to the two types of subsystem symmetries, $G_\mm^f$ and $G_\ee^\ell$, the gauged theory has a dual 1-form symmetry, denoted as $\hat G^{(1)}_\ee$, generated by the Wilson surface operator $\hat \eta_\ee^{(1)}(\Sigma) = \exp(\ii \oint_\Sigma b_\ee)$. In other words, the foliated field theory Lagrangian Eq.~\eqref{3dfollag-gauge1-form} describes a 1-form symmetry-enriched $\mathbb Z_N$ X-Cube model. The enrichment can be seen from the fact that the fracton, described by the Wilson line defect $\exp(\ii \oint d\tau~a_{\mm,\tau})$, is charged under the dual 1-form symmetry. Indeed, using the equation of motion of $A^{(i)}_\mm$, we find that, whenever $\Sigma$ is along the layers, the surface operator $\hat \eta_\ee^{(1)}(\Sigma)$ can be identified with the lineon cage operator,
\ie
\prod_{\Gamma\in \mathrm{cage}(\Sigma)} \eta_\ee^\ell(\Gamma)~,
\fe
which detects a fracton inside the cage. Here, $\mathrm{cage}(\Sigma)$ is the cage associated with the surface $\Sigma$ and $\Gamma$'s are the segments of the cage (see Fig.~\ref{fig:surf-cage}).

\subsubsection{Gauging $\hat G^{(1)}_\ee$}
Gauging the dual 1-form symmetry $\hat G^{(1)}_\ee$ takes us back to the foliated stack Eq.~\eqref{3dfollag}. This can be seen explicitly by coupling the theory Eq.~\eqref{3dfollag-gauge1-form} to a $\mathbb Z_N$ 2-form gauge field, but we do not discuss the details of this here.

\subsubsection{Gauging $G_\mm^f$}
On the other hand, if we gauge the fracton subsystem symmetry $G_\mm^f \cong G_\mm^{(1),\fol}/G_\mm^{(1)}$, we end up in a trivial theory, denoted as $\mathcal T_{0,\ee}$ in Fig.~\ref{fig:gaugingweb}. This can be seen by coupling the theory Eq.~\eqref{3dfollag-gauge1-form} to $\mathbb Z_N$ gauge field combination $(B_\ee^{(i)},c_\ee)$:
\ie
\mathcal L_{0,\ee} &= \sum_i\sum_{n_i=1}^{L_i} \frac{\ii N}{2\pi} \Big[ A_\mm^{(i)} (dA_\ee^{(i)} - b_\ee - B^{(i)}_\ee)
\\
&+ \varPhi^{(i)}_\mm (dB_\ee^{(i)} - c_\ee) \Big] \delta(x_i - n_i \ell_i) dx_i
\\
& + \frac{\ii N}{2\pi} \Big[ a_\mm (db_\ee + c_\ee) + \phi_\mm dc_\ee \Big]~, 
\fe
where $(\varPhi^{(i)}_\mm,\phi_\mm)$ are a combination of Lagrange multipliers (both compact scalars) that constrain $(B_\ee^{(i)},c_\ee)$ to be $\mathbb Z_N$ gauge fields. The gauge symmetry acts as\footnote{The gauge fields $(B_\ee^{(i)},c_\ee)$ together can be interpreted as the foliated-equivalent of the exotic tensor gauge fields $(b^{ij}_{\ee,\tau}, b_\ee)$, whereas the combination $(\varPhi_\mm^{(i)},\phi_\mm)$ is the foliated-equivalent of the exotic scalar field $\phi^\mm$. One can derive this correspondence using methods similar to the ones in Ref.~\onlinecite{Ohmori:2022rzz}.}
\ie
&A_\ee^{(i)} \sim A_\ee^{(i)} + d\alpha_\ee^{(i)} + \beta_\ee + \beta^{(i)}_\ee~,\quad b_\ee \sim b_\ee + d\beta_\ee + \gamma_\ee~,
\\
&B_\ee^{(i)} \sim B_\ee^{(i)} + d \beta_\ee^{(i)} + \gamma_\ee~, \quad c_\ee \sim c_\ee + d\gamma_\ee~,
\\
&A_\mm^{(i)} \sim A_\mm^{(i)} + d \alpha_\mm^{(i)}~, \quad \phi_\mm \sim \phi_\mm + \alpha_m~,
\\
&\varPhi^{(i)}_\mm \sim \varPhi^{(i)}_\mm + \alpha^{(i)}_\mm~,
\\
&a_\mm \sim a_\mm + d \alpha_\mm + \sum_i\sum_{n_i=1}^{L_i} \alpha_\mm^{(i)} \delta(x_i - n_i \ell_i) dx_i~.
\fe
It is easy to see that there are no nontrivial gauge invariant operators in this theory.

Alternatively, gauging $G_\mm^f$ in the foliated field theory of X-Cube Eq.~\eqref{3dfollag-gauge1-form} is equivalent to gauging $G_\mm^{(1),\fol}$ in the foliated stack Eq.~\eqref{3dfollag} in two steps: first gauge $G_\mm^{(1)}$ and then gauge $G_\mm^f$. The latter interpretation clearly results in the trivial theory $\mathcal T_{0,\ee}$. Moreover, gauging a $\mathbb Z_N$ 1-form symmetry in 2+1D yields a dual $\mathbb Z_N$ 0-form symmetry, so $\mathcal T_{0,\ee}$ has a dual foliated 0-form symmetry $\hat G_\ee^{(0),\fol}$.

A subgroup of $\hat G_\ee^{(0),\fol}$ is the dual fracton subsystem symmetry $\hat G_\ee^{\hat f}$, which is the dual symmetry coming from gauging the fracton subsystem symmetry $G_\mm^f$. Indeed, gauging $\hat G_\ee^{\hat f}$ in $\mathcal T_{0,\ee}$ takes us back to the X-Cube foliated field theory Eq.~\eqref{3dfollag-gauge1-form}. We do not go into the details here.

\subsubsection{Gauging $G_\ee^\ell$}
Finally, we can also gauge the lineon subsystem symmetry $G_\ee^\ell$. This is explored in Sec.~\ref{sec:contSPT}.

\subsection{Gauging $G_\ee^\ell$ and $G_\mm^{(1)}$}\label{sec:contSPT}

Since there is no mixed 't Hooft anomaly between $G_\ee^\ell$ and $G_\mm^{(1)}$ in the foliated stack Eq.~\eqref{3dfollag}, we can gauge both of these symmetries simultaneously. The resulting Lagrangian is
\ie
\mathcal L_\SPT &= \sum_i\sum_{n_i=1}^{L_i} \frac{\ii N}{2\pi} \Big[ A_\ee^{(i)}dA_\mm^{(i)} - A_\ee^{(i)} B_\mm^{(i)} - b_\ee A^{(i)}_\mm
\\
&+ (d\varPhi_\ee^{(i)} - a_\ee)B_\mm^{(i)} \Big]\delta(x_i - n_i \ell_i) dx_i
\\
&+ \frac{\ii N}{2\pi} \left[ b_\mm da_\ee  + a_\mm db_\ee + b_\mm b_\ee \right]~,
\fe
where the gauge symmetry acts as
\ie
&A_\ee^{(i)} \sim A_\ee^{(i)} + d\alpha_\ee^{(i)} + \beta_\ee~,\quad b_\ee \sim b_\ee + d\beta_\ee~,
\\
&\varPhi_\ee^{(i)} \sim \varPhi_\ee^{(i)} + \alpha_\ee^{(i)}~,\quad a_\ee \sim a_\ee - \hat\beta_\ee~,
\\
&A_\mm^{(i)} \sim A_\mm^{(i)} + \beta_\mm^{(i)}~,\quad B_\mm^{(i)} \sim B_\mm^{(i)} + d\beta_\mm^{(i)}~,
\\
&a_\mm \sim a_\mm - \beta_\mm + \sum_i\sum_{n_i=1}^{L_i} \alpha_\mm^{(i)} \delta(x_i - n_i \ell_i) dx_i~,
\\
&b_\mm \sim b_\mm + d\beta_\mm + \sum_i \sum_{n_i=1}^{L_i} \beta_\mm^{(i)}\delta(x_i - n_i \ell_i) dx_i~.
\fe
There are no nontrivial gauge invariant operators in this theory, so it describes a trivially gapped phase. In fact, it describes an SPT phase protected by the dual symmetries $\hat G^{(1)}_\ee$ and $\hat G^{\hat \ell}_\mm$. To see this, let us compute the response theory by coupling it to background gauge fields of the dual symmetries. The Lagrangian is
\begin{widetext}
\ie
\mathcal L_\SPT[(\bb A'^{(i)}_\ee, b'_\ee); \bb b'_\mm] &= \sum_i\sum_{n_i=1}^{L_i} \frac{\ii N}{2\pi} \Big[ A_\ee^{(i)}dA_\mm^{(i)} - A_\ee^{(i)} B_\mm^{(i)} - b_\ee A^{(i)}_\mm
\\
&+ (d\varPhi_\ee^{(i)} - a_\ee - \bb A'^{(i)}_\ee)B_\mm^{(i)} + A'^{(i)}_\mm (d\bb A'^{(i)}_\ee - b'_\ee) \Big]\delta(x_i - n_i \ell_i) dx_i
\\
&+ \frac{\ii N}{2\pi} \left[ b_\mm (da_\ee + b'_\ee)  + b_\ee (da_\mm + \bb b'_\mm) + b_\mm b_\ee + a'_\mm db'_\ee + a'_\ee d\bb b'_\mm \right]~,
\fe
\end{widetext}
where the combination $(\bb A'^{(i)}_\ee,b'_\ee)$ is the foliated version of the background\footnote{As mentioned at the beginning of Sec.~\ref{sec:cont-gaugingweb}, we use boldfaced letters for background gauge fields. We do not use boldfaced letter for $b'_\ee$ because it is a dynamical field that turns the gauge field $\bb A'^{(i)}_\ee$ for the dual foliated 0-form symmetry $\hat G^{(0),\fol}_\mm$ into a gauge field for $\hat G^{\hat \ell}_\mm$.} ``lineon'' gauge fields for $\hat G^{\hat \ell}_\mm$ and $(A'^{(i)}_\mm,a'_\mm)$ are Lagrange multipliers (``fracton'' gauge fields) that constrain them to be $\mathbb Z_N$ gauge fields, whereas $\bb b'_\mm$ is the background 2-form gauge field for $\hat G^{(1)}_\ee$ and $a'_\ee$ is the Lagrange multiplier (1-form gauge field) that constrains it to be a $\mathbb Z_N$ gauge field. The gauge symmetry acts as
\begin{widetext}
\ie
\!&A_\ee^{(i)} \sim A_\ee^{(i)} + d\alpha_\ee^{(i)} + \beta_\ee~,\qquad &&A_\mm^{(i)} \sim A_\mm^{(i)} + d\alpha^{(i)}_\mm + \beta_\mm^{(i)}~,
\\
&b_\ee \sim b_\ee + d\beta_\ee~,&&a_\mm \sim a_\mm - \beta_\mm - \bb \beta'_\mm + \sum_i\sum_{n_i=1}^{L_i} \alpha_\mm^{(i)} \delta(x_i - n_i \ell_i) dx_i~,
\\
&\varPhi_\ee^{(i)} \sim \varPhi_\ee^{(i)} + \alpha_\ee^{(i)} + \bb \alpha_\ee'^{(i)}~,&&B_\mm^{(i)} \sim B_\mm^{(i)} + d\beta_\mm^{(i)}~,
\\
&a_\ee \sim a_\ee - \beta_\ee - \beta'_\ee~,&&b_\mm \sim b_\mm + d\beta_\mm + \sum_i \sum_{n_i=1}^{L_i} \beta_\mm^{(i)}\delta(x_i - n_i \ell_i) dx_i~,
\\
&\bb A'^{(i)}_\ee \sim \bb A'^{(i)}_\ee + d\bb \alpha_\ee'^{(i)} + \beta'_\ee~,&& A_\mm'^{(i)} \sim A_\mm'^{(i)} + d \alpha_\mm'^{(i)} + \beta_\mm^{(i)}~,
\\
&b'_\ee \sim b'_\ee + d\beta'_\ee~,&&a'_\mm \sim a'_\mm + d \alpha'_\mm - \beta_\mm + \sum_i\sum_{n_i=1}^{L_i} \alpha_\mm'^{(i)} \delta(x_i - n_i \ell_i) dx_i~,
\\
&a'_\ee \sim a'_\ee + d\alpha'_\ee - \beta_\ee~,&& \bb b'_\mm \sim \bb b'_\mm + d\bb \beta'_\mm~.
\fe
\end{widetext}

We can simplify this Lagrangian considerably by integrating out some of the dynamical fields. For instance, upon integrating out $B^{(i)}_\mm$ on the layers and $b_\mm$ in the bulk, and then replacing
\ie
A'^{(i)}_\mm &\to A'^{(i)}_\mm + A^{(i)}_\mm~,
\\
a'_\mm &\to a'_\mm + a_\mm~,
\\
a'_\ee &\to a'_\ee + a_\ee~,
\fe
we get a much simpler Lagrangian involving only the primed gauge fields. The resulting Lagrangian is given by setting $p=1$ in the following more general Lagrangian labelled by an integer $p\mod N$:
\ie\label{3dlagspt-p}
\!&\mathcal L_{\SPT_p}[(\bb A^{(i)}_\ee, b_\ee); \bb b_\mm]
\\
&~= \sum_i\sum_{n_i=1}^{L_i} \frac{\ii N}{2\pi}  A^{(i)}_\mm (d\bb A^{(i)}_\ee - b_\ee) \delta(x_i - n_i \ell_i) dx_i
\\
&~ + \frac{\ii N}{2\pi} \left[ a_\mm db_\ee + a_\ee d\bb b_\mm - pb_\ee \bb b_\mm \right]~,
\fe
where we dropped the primes on the gauge fields for brevity. The gauge symmetry acts as
\ie\label{3dlagspt-p-gaugesym}
\!&\bb A^{(i)}_\ee \sim \bb A^{(i)}_\ee + d\bb \alpha_\ee^{(i)} + \beta_\ee~,\quad b_\ee \sim b_\ee + d\beta_\ee~,
\\
&a_\ee \sim a_\ee + d\alpha_\ee + p\beta_\ee~,
\\
&A_\mm^{(i)} \sim A_\mm^{(i)} + d \alpha_\mm^{(i)}~,\quad \bb b_\mm \sim \bb b_\mm + d\bb \beta_\mm~,
\\
&a_\mm \sim a_\mm + d \alpha_\mm + p\bb \beta_\mm + \sum_i\sum_{n_i=1}^{L_i} \alpha_\mm^{(i)} \delta(x_i - n_i \ell_i) dx_i~.
\fe
As the notation suggests, this describes a distinct SPT phase labelled by $p=0,1,\ldots,N-1$ and protected by the 1-form symmetry $\hat G^{(1)}_\ee$ and the subsystem symmetry $\hat G^{\hat \ell}_\mm$.

One can probe the nontriviality of these SPTs by gauging the symmetries protecting them, i.e., by promoting the background gauge fields to dynamical gauge fields ($\bb A^{(i)}_\ee \to A^{(i)}_\ee$ and $\bb b_\mm \to b_\mm$).
\begin{enumerate}
\item When $p=0$, one can recognize the second line and the first term of the third line of Eq.~\eqref{3dlagspt-p} together correspond to the foliated field theory of $\mathbb Z_N$ X-Cube model Eq.~\eqref{3dfollag-gauge1-form}, whereas the remaining term corresponds to the 3+1D $\mathbb Z_N$ gauge theory. Relatedly, there is a fully mobile particle-like excitation described by the Wilson line $\exp(i\oint_\gamma a_\ee)$. This is the expected result when we gauge $\hat G^{\hat \ell}_\mm$ and $\hat G^{(1)}_\ee$ in a trivial theory, so this corresponds to the trivial SPT.

\item When $p=1$, which is the case of interest to us, the resulting Lagrangian describes a foliated stack of 2+1D $\mathbb Z_N$ gauge theories Eq.~\eqref{3dfollag}. This can be seen by integrating out $b_\mm$, which sets $b_\ee = da_\ee$, and then replacing $A^{(i)}_\ee \to A^{(i)}_\ee + a_\ee$, which yields Eq.~\eqref{3dfollag}. Relatedly, this theory does not have fully mobile particle-like excitations. Therefore, this corresponds to a nontrivial SPT of $\hat G^{\hat \ell}_\mm$ and $\hat G^{(1)}_\ee$.
\end{enumerate}
Similarly, other nonzero values of $p$ correspond to distinct nontrivial SPTs protected by $\hat G^{\hat \ell}_\mm$ and $\hat G^{(1)}_\ee$. We explicitly write down the lattice model of the gauged SPT$_p$ in Appendix \ref{app:gaugedSPT}.

One way to see that the SPTs labelled by distinct $p \mod N$ are indeed distinct is by comparing the gauge invariant observables of the gauged SPT for different values of $p$. For instance, we can detect the value of $p \mod N$ using the correlation function described below. Consider the following gauge invariant open surface operators:
\ie
\!&V_\mm(\Sigma):=\exp\left( \ii \oint_\gamma a_\mm + \ii \sum_i \sum_{n_i=1}^{L_i} \int_{\Gamma^{n_i}} A_\mm^{(i)}(n_i) - \ii p \oint_\Sigma b_\mm \right)~,
\\
&V_\ee(\tilde \Sigma) := \exp\left( \ii \oint_{\tilde \gamma} a_\ee - \ii p \oint_{\tilde \Sigma} b_\ee \right)~,
\fe
where $\Sigma$ and $\tilde \Sigma$ are two open surfaces in the 3+1d spacetime with boundaries $\gamma$ and $\tilde \gamma$, respectively, 
and $\Gamma^{n_i}$ is the intersection of $\Sigma$ with the $n_i$-th layer (it is an open curve with endpoints on $\gamma$). The correlation function of these two operators is given by
\ie
\langle V_\mm(\Sigma) V_\ee(\tilde \Sigma) \rangle = e^{-2\pi \ii p \#(\Sigma,\tilde \Sigma)/N} \langle V_\mm(\Sigma)\rangle \langle V_\ee(\tilde \Sigma) \rangle~,
\fe
where $\#(\Sigma,\tilde \Sigma)$ is the intersection number of the surfaces $\Sigma$ and $\tilde \Sigma$. This shows that the SPTs labelled by distinct $p$ are different from each other.

\subsubsection{Boundary of $\SPT_p$}
One of the hallmarks of an SPT phase is that it is not invariant under background gauge transformations in the presence of a boundary. Consider the variation of the Lagrangian Eq.~\eqref{3dlagspt-p} of $\SPT_p$ under the gauge transformation Eq.~\eqref{3dlagspt-p-gaugesym}:
\ie\label{3dlagspt-p-gaugevariation}
\!&\delta \mathcal L_{\SPT_p} = d\bigg(\sum_i\sum_{n_i=1}^{L_i} \frac{\ii N}{2\pi} \alpha^{(i)}_\mm (d\bb A^{(i)}_\ee -  b_\ee) \delta(x_i - n_i \ell_i) dx_i
\\
&+ \frac{\ii N}{2\pi} \left[  \alpha_\mm d b_\ee + \alpha_\ee d\bb b_\mm - p ( \beta_\ee \bb b_\mm +  b_\ee \bb \beta_\mm +  \beta_\ee d \bb \beta_\mm ) \right] \bigg)~.
\fe
When there is no boundary, the variation vanishes because it is a total derivative. In contrast, in the presence of a boundary, we get a nontrivial variation localised along on the boundary. We can impose the Dirichlet boundary condition for the Lagrange multipliers $(A^{(i)}_\mm, a_\mm)$ and $a_\ee$, which sets their gauge parameters $(\alpha^{(i)}_\mm, \alpha_\mm)$ and $\alpha_\ee$ to zero on the boundary.\footnote{\label{ftnt:boundary-cond}This can be done by adding the terms
\ie
\!&\sum_i \sum_{n_i=1}^{L_i} \frac{\ii N}{2\pi} \varPhi^{(i)}_{\dd,\mm} (d\bb A^{(i)}_\ee -  b_e) \delta(x_i - n_i \ell_i) dx_i
\\
&+ \frac{\ii N}{2\pi} \big[  \phi_{\dd,\mm} d b_\ee + \phi_{\dd,\ee} d\bb b_\mm \big]
\fe
on the boundary with gauge symmetry $\varPhi^{(i)}_{\dd,\mm} \sim \varPhi^{(i)}_{\dd,\mm} - \alpha^{(i)}_\mm$, $ \phi_{\dd,\mm} \sim  \phi_{\dd,\mm} -  \alpha_\mm$, and $\phi_{\dd,\ee} \sim \phi_{\dd,\ee} - \alpha_\ee$. Here, the subscript $\dd$ indicates boundary fields.} What is left are the terms proportional to $p$, which depend only on the background gauge fields and their gauge parameters on the boundary.

The non-invariance along the boundary is cancelled by an anomalous boundary theory via the \emph{anomaly inflow} mechanism~\cite{Callan:1984sa}. We now explore several gapped boundary theories that realize a $\hat G^{\hat \ell}_\mm \times \hat G^{(1)}_\ee$ symmetry with a mixed 't Hooft anomaly that is cancelled by the above SPT. For simplicity, we assume that the boundary is along a fixed $z$ plane.

The starting point for both examples is a decoupled 2-foliated stack of 1+1D $\mathbb Z_N$ gauge theories in 2+1D, described by the Lagrangian:
\ie\label{2dfollag}
\mathcal L_{\dd,\fol} = \sum_i \sum_{n_i=1}^{L_i} \frac{\ii N}{2\pi} A^{(i)}_\dd d\varPhi^{(i)}_\dd \delta(x_i - n_i \ell_i) dx_i~,
\fe
where $i$ runs over $x,y$, and the subscript $\dd$ indicates boundary fields. This theory has a foliated 0-form symmetry $G^{(0),\fol}$ generated by the Wilson line operators $\exp\big(\ii\oint_{\Gamma^{n_i}_\dd} A^{(i)}_\dd\big)$, where $\Gamma^{n_i}_\partial$ is a closed curve in the $n_i$-th layer on the boundary, and a foliated 1-form symmetry generated by the local operators $\exp(\ii \varPhi^{(i)}_\dd)$. We are interested in the theories obtained by gauging the following two subgroups of the foliated 0-form symmetry.\footnote{One can also consider gauging similar subgroups of the foliated 1-form symmetries. Together, all the resulting theories fit into a 2+1D gauging web, analogous to the one in Fig.~\ref{fig:gaugingweb}, relating the plaquette Ising model, the Toric Code, and the Ising model. We do not go into the details here and point the interested readers to Ref.~\onlinecite{domhigherform}.}

\paragraph{2+1D plaquette Ising model.} The first is the global 0-form symmetry $G^{(0)} \subset G^{(0),\fol}$ associated with the diagonal subgroup generated by the following operator on a closed surface $\Sigma_\dd$:
\ie
\prod_i \prod_{n_i=1}^{L_i} \exp\left( \ii \oint_{\Gamma^{n_i}_\dd} A^{(i)}_\dd \right)~,
\fe
where $\Gamma^{n_i}_\dd$ is the intersection of $\Sigma_\dd$ with the $n_i$-th layer on the boundary. This symmetry can be gauged by coupling the theory Eq.~\eqref{2dfollag} to a 1-form gauge field $a_\dd$:
\ie\label{2dfollag-pim}
\mathcal L_{\dd,\PIM} &= \sum_i \sum_{n_i=1}^{L_i} \frac{\ii N}{2\pi} A^{(i)}_\dd (d\varPhi^{(i)}_\dd - a_\dd) \delta(x_i - n_i \ell_i) dx_i
\\
&\quad+ \frac{\ii N}{2\pi} \tilde a_\dd da_\dd~,
\fe
where $\tilde a_\dd$ is a Lagrange multiplier (1-form gauge field) that constrains $a_\dd$ to be a $\mathbb Z_N$ gauge field. The gauge symmetry acts
\ie
&\varPhi^{(i)}_\dd \sim \varPhi^{(i)}_\dd + 2\pi N^{(i)}_\dd(n_i) + \alpha_\dd~,\quad a_\dd \sim a_\dd + d\alpha_\dd~,
\\
&A^{(i)}_\dd \sim A^{(i)}_\dd + d\alpha^{(i)}_\dd~,
\\
&\tilde a_\dd \sim \tilde a_\dd + d\tilde \alpha_\dd - \sum_i \sum_{n_i=1}^{L_i} \alpha^{(i)}_\dd \delta(x_i-n_i\ell_i) dx_i~.
\fe
In fact, this theory describes the ferromagnetic phase of the 2+1D $\mathbb Z_N$ plaquette Ising model~\cite{Ohmori:2022rzz}, where the connection to the exotic field theory of Ref.~\onlinecite{Seiberg:2020bhn} is also explored. This is the 2+1D analogue of the discussion in Sec.~\ref{sec:gauge1-form}.

The gauged theory Eq.~\eqref{2dfollag-pim} has dual 1-form symmetry $\hat G^{(1)}$ generated by $\exp(\ii\oint_{\gamma_\dd} a_\dd)$, where $\gamma_\dd$ is a closed curve on the boundary, so it describes a 1-form symmetry-enriched plaquette Ising model. This is in addition to the two remaining subsystem symmetries: $G^f \cong G^{(0),\fol}/G^{(0)}$ generated by the Wilson line operators $\exp\big(\ii\oint_{\Gamma^{n_i}_\dd} A^{(i)}_\dd\big)$ and $G^p \subset G^{(1),\fol}$ generated by the local (point-like) operators $\exp(\ii \varPhi^{(x)}_\dd(n_x)) \exp(\ii \varPhi^{(y)}_\dd(n_y))^{-1}$ at the intersection of the $n_x$-th and $n_y$-th layers. The line operators satisfy the constraints
\ie
\prod_{i\ne j}\prod_{n_i=1}^{L_i} \exp\left(\ii\oint dx_j~ A^{(i)}_{\dd,j} \right) = 1~,
\fe
which follows from the equation of motion of $a_\dd$.

The 1-form symmetry $\hat G^{(1)}$ and the fracton subsystem symmetry $G^f$ have a mixed 't Hooft anomaly. To see this, let us couple the theory Eq.~\eqref{2dfollag-pim} to the corresponding background gauge fields:
\begin{widetext}
\ie\label{2dfollag-pim-background}
\mathcal L_{\dd,\PIM}[(\bb A^{(i)}_\ee, b_\ee);\bb b_\mm] &= \sum_i \sum_{n_i=1}^{L_i} \frac{\ii N}{2\pi} \Big[ A^{(i)}_\dd (d\varPhi^{(i)}_\dd - a_\dd - \bb A^{(i)}_\ee) + \varPhi^{(i)}_{\dd,\mm} (d\bb A^{(i)}_\ee -  b_\ee) \Big] \delta(x_i - n_i \ell_i) dx_i
\\
&+ \frac{\ii N}{2\pi} \big[ \tilde a_\dd (da_\dd +  b_\ee) + a_\dd \bb b_\mm +  \phi_{\dd,\mm} d b_\ee + \phi_{\dd,\ee} d\bb b_\mm \big]~,
\fe
\end{widetext}
where one can recognize $(\bb A^{(i)}_\ee, b_\ee)$ as the foliated version of the background fracton gauge fields for $G^f$, and $\bb b_\mm$ as the background 2-form gauge field for $\hat G^{(1)}$, whereas the dynamical fields $(\varPhi^{(i)}_{\dd,\mm}, \phi_{\dd,\mm})$ and $\phi_{\dd,\ee}$ are the corresponding Lagrange multipliers that constrain them to be $\mathbb Z_N$ gauge fields on the boundary.\footnote{These Lagrange multipliers are the same fields that appear in Footnote~\ref{ftnt:boundary-cond}.} We do not use the subscript $\dd$ on the background gauge fields since they are the restrictions of the bulk background gauge fields to the boundary. The full gauge symmetry acts as
\begin{widetext}
\ie
\!&\varPhi^{(i)}_\dd \sim \varPhi^{(i)}_\dd + 2\pi N^{(i)}_\dd(n_i) + \alpha_\dd + \bb \alpha^{(i)}_\ee~,\quad && A^{(i)}_\dd \sim A^{(i)}_\dd + d\alpha^{(i)}_\dd~,
\\
&a_\dd \sim a_\dd + d\alpha_\dd -  \beta_\ee~, && \tilde a_\dd \sim \tilde a_\dd + d\tilde \alpha_\dd - \bb \beta_\mm - \sum_i \sum_{n_i=1}^{L_i} \alpha^{(i)}_\dd \delta(x_i-n_i\ell_i) dx_i~,
\\
&\bb A^{(i)}_\ee \sim \bb A^{(i)}_\ee + d\bb \alpha^{(i)}_\ee +  \beta_\ee~,&&\varPhi^{(i)}_{\dd,\mm} \sim \varPhi^{(i)}_{\dd,\mm} - \alpha^{(i)}_\dd~,
\\
& b_\ee \sim  b_\ee + d \beta_\ee~,&& \phi_{\dd,\mm} \sim  \phi_{\dd,\mm} - \tilde \alpha_\dd~,
\\
&\phi_{\dd,\ee}\sim \phi_{\dd,\ee} - \alpha_\dd~,&&\bb b_\mm \sim \bb b_\mm + d\bb \beta_\mm~,
\fe
\end{widetext}
where, once again, fields without the subscript $\dd$ are restrictions of the bulk fields to the boundary. It is easy to check that the boundary Lagrangian Eq.~\eqref{2dfollag-pim-background} is not invariant under this gauge symmetry. In fact, the variation precisely cancels the boundary variation of $\SPT_{p=1}$ in Eq.~\eqref{3dlagspt-p-gaugevariation}.

More generally, one can take $p$ copies of the 2+1D plaquette Ising model enriched by the 1-form symmetry. Then, the diagonal 1-form symmetry and the diagonal fracton subsystem symmetry have a mixed 't Hooft anomaly that is cancelled by $\SPT_p$.

\paragraph{2+1D Toric Code.} The second is the subsystem symmetry $G^{\hat f} \subset G^{(0),\fol}$ associated with the subgroup generated by the operators
\ie
\exp\left( \ii \oint dy~A^{(x)}_{\dd,y}(n_x+1) \right) \exp\left( \ii \oint dy~A^{(x)}_{\dd,y}(n_x) \right)^{-1}~,
\fe
where $n_x =1,\ldots,L_x$, and similar operators in the other direction. Similar to the discussion in Sec.~\ref{sec:gaugelsym}, we can gauge this symmetry by first gauging the full foliated 0-form symmetry and then gauging the diagonal subgroup of the dual foliated 0-form symmetry, $\hat G^{(0)} \subset \hat G^{(0),\fol}$. The resulting Lagrangian is
\ie\label{2dfollag-tc}
\mathcal L_{\dd,\TC} &= \sum_i \sum_{n_i=1}^{L_i} \frac{\ii N}{2\pi} \Big[ A^{(i)}_\dd (d\varPhi^{(i)}_\dd - \tilde A^{(i)}_\dd) 
\\
&\quad - \tilde A^{(i)}_\dd (d\tilde \varPhi^{(i)}_\dd - a_\dd) \Big]\delta(x_i - n_i \ell_i) dx_i~,
\\
&\quad +\frac{\ii N}{2\pi} \tilde a_\dd d a_\dd~,
\fe
and the gauge symmetry acts as
\ie
&\varPhi^{(i)}_\dd \sim \varPhi^{(i)}_\dd + \tilde \alpha^{(i)}_\dd~,\quad A^{(i)}_\dd \sim A^{(i)}_\dd + d\alpha^{(i)}_\dd~,
\\
&\tilde \varPhi^{(i)}_\dd \sim \tilde \varPhi^{(i)}_\dd + \alpha^{(i)}_\dd + \alpha_\dd~,\quad a_\dd \sim a_\dd + d\alpha_\dd~,
\\
&\tilde A^{(i)}_\dd \sim \tilde A^{(i)}_\dd + d\tilde \alpha^{(i)}_\dd~,
\\
&\tilde a_\dd \sim \tilde a_\dd + d\tilde \alpha_\dd + \sum_i \sum_{n_i=1}^{L_i} \tilde \alpha^{(i)}_\dd \delta(x_i-n_i\ell_i) dx_i~.
\fe
This theory describes the low energy phase of the 2+1D $\mathbb Z_N$ Toric Code. Indeed, the nontrivial gauge invariant operators are
\ie
\!&\exp\left(\ii \oint_{\gamma_\dd} a_\dd\right)~,
\\
&\exp\left( \ii \oint_{\tilde \gamma_\dd} \tilde a_\dd - \ii \sum_i \sum_{n_i=1}^{L_i} \varPhi^{(i)}_\dd|_{p^{n_i}_\dd} \right)~,
\fe
where $\gamma_\dd,\tilde \gamma_\dd$ are closed curves on the boundary, and $p^{n_i}_\dd$ is the point of intersection of $\tilde \gamma_\dd$ with the $n_i$-th layer on the boundary. These operators are in one-one correspondence with the $\ee$ and $\mm$ anyons of the Toric Code, so they generate two 1-form symmetries: $G^{(1)} \cong G^{(0),\fol}/G^{\hat f}$ generated by the first line and $\tilde G^{(1)} \subset G^{(1),\fol}$ generated by the second line.

In addition, the gauged theory Eq.~\eqref{2dfollag-tc} has a dual fracton subsystem symmetry, $\hat G^f$ generated by the Wilson line operators
\ie
\exp\left( \ii \oint dy~\tilde A^{(x)}_{\dd,y}(n_x) \right)~,
\fe
and similar operators in the other direction. They satisfy the constraint
\ie
\prod_{i\ne j} \prod_{n_i=1}^{L_i}\exp\left( \ii \oint dx_j~\tilde A^{(i)}_{\dd,j}(n_i) \right) = 1~,
\fe
which follows from the equation of motion of $a_\dd$. The 2+1D $\mathbb Z_N$ Toric Code is enriched by this subsystem symmetry.

The 1-form symmetry $G^{(1)}$ and the fracton subsystem symmetry $\hat G^f$ have a mixed 't Hooft anomaly. To see this, we couple the theory Eq.~\eqref{2dfollag-tc} to the respective gauge fields:
\begin{widetext}
\ie\label{2dfollag-tc-background}
\mathcal L_{\dd,\TC}[(\bb A^{(i)}_\ee, b_\ee);\bb b_\mm] &= \sum_i \sum_{n_i=1}^{L_i} \frac{\ii N}{2\pi} \Big[ A^{(i)}_\dd (d\varPhi^{(i)}_\dd - \tilde A^{(i)}_\dd)  - \tilde A^{(i)}_\dd (d\tilde \varPhi^{(i)}_\dd - a_\dd - \bb A^{(i)}_\ee) + \varPhi^{(i)}_{\dd,\mm} (d\bb A^{(i)}_\ee - b_\ee) \Big]\delta(x_i - n_i \ell_i) dx_i~,
\\
&\quad +\frac{\ii N}{2\pi} \big[ \tilde a_\dd (da_\dd +  b_\ee) + a_\dd \bb b_\mm + \phi_{\dd,\mm} d b_\ee + \phi_{\dd,\ee} d\bb b_\mm \big]~,
\fe
where the gauge symmetry acts as
\ie
\!&A^{(i)}_\dd \sim A^{(i)}_\dd + d\alpha^{(i)}_\dd~,&& \varPhi^{(i)}_\dd \sim \varPhi^{(i)}_\dd + \tilde \alpha^{(i)}_\dd~,
\\
&\tilde \varPhi^{(i)}_\dd \sim \tilde \varPhi^{(i)}_\dd + \alpha^{(i)}_\dd + \alpha_\dd + \bb \alpha^{(i)}_\ee~,\quad && \tilde A^{(i)}_\dd \sim \tilde A^{(i)}_\dd + d\tilde \alpha^{(i)}_\dd~,
\\
&a_\dd \sim a_\dd + d\alpha_\dd -  \beta_\ee~, && \tilde a_\dd \sim \tilde a_\dd + d\tilde \alpha_\dd - \bb \beta_\mm + \sum_i \sum_{n_i=1}^{L_i} \tilde \alpha^{(i)}_\dd \delta(x_i-n_i\ell_i) dx_i~,
\\
&\bb A^{(i)}_\ee \sim \bb A^{(i)}_\ee + d\bb \alpha^{(i)}_\ee + \beta_\ee~,&&\varPhi^{(i)}_{\dd,\mm} \sim \varPhi^{(i)}_{\dd,\mm} + \tilde \alpha^{(i)}_\dd~,
\\
& b_\ee \sim  b_\ee + d \beta_\ee~,&& \phi_{\dd,\mm} \sim  \phi_{\dd,\mm} - \tilde \alpha_\dd~,
\\
&\phi_{\dd,\ee}\sim \phi_{\dd,\ee} - \alpha_\dd~,&&\bb b_\mm \sim \bb b_\mm + d\bb \beta_\mm~.
\fe
\end{widetext}
Once again, it is easy to verify that the boundary Lagrangian Eq.~\eqref{2dfollag-tc-background} has a nontrivial variation under this gauge symmetry which cancels the boundary variation of $\SPT_{p=1}$ in Eq.~\eqref{3dlagspt-p-gaugevariation}.

More generally, one can take $p$ copies of the 2+1D Toric Code enriched by the fracton subsystem symmetry. Then, a diagonal 1-form symmetry and the diagonal fracton subsystem symmetry have a mixed 't Hooft anomaly that is cancelled by $\SPT_p$.


\section{Outlook}
\label{sec:outlook}

In this work, we have unveiled a rich gauging web relating topological phases (including SPTs and topological orders) and fractons phases, obtained by gauging subgroups of the total symmetry of a stack of 2+1D Abelian gauge theories in 3+1D. In the process, we have uncovered exotic topological phenomena including subsystem symmetry fractionalization in a topological order, twisted gauge theories that involve higher-form and subsystem symmetries, and unconventional extensions of higher form symmetries by subsystem symmetries. Our work thus points to vast generalizations of gauging webs (typically based on conventional global symmetries) that can appear once the geometric nature of subsystem symmetries is properly accounted for. Exploring such gauging webs promises to uncover new models and new connections between known models, involving a mix of higher form and subsystem symmetries. This raises a number of directions to be addressed in future work. 

A natural next step is to explore the gauging web of other fracton models that admit a p-string condensation construction, such as the Four Color Cube model of Ref.~\onlinecite{han}. In future work, we will report how the Checkerboard model (which is equivalent to two copies of the X-Cube model~\cite{foliatedcb}) can be constructed by gauging a topological 1-form symmetry in the X-Cube model. Moreover, the gauging web we have explored can be further extended by considering twisted gauging maps in addition to the standard gauging maps we have used. Building on Ref.~\onlinecite{GPTW25}, it would be interesting to understand whether there are obstructions to non-Abelian $p$-string condensation that can be understood as stemming from certain mixed anomalies between subsystem and higher-form symmetries that prevent gauging certain symmetries.

Along similar lines, while we have focused entirely on planar subsystem symmetries here, one can also consider fractal subsystem symmetries and ask whether new models and relations between them can be obtained by constructing analogous gauging webs. An interesting observation in this regard is that any gauging duality of a spin Hamiltonian whose terms are products of only Pauli-$X$ or Pauli-$Z$ terms (such as the 1+1D Ising model) can be lifted to a higher dimensional model with fractal symmetry via fractalization~\cite{Devakul2020b}. The fractal models constructed in this way inherit the dualities of the original model, once appropriate spatial symmetries are incorporated, providing a plausible route to obtaining gauging webs involving fractal subsystem symmetries.

Looking towards a more general formalism that captures gauging and dualities between topological and fracton models, a natural candidate would be a ``Symmetry QFT" perspective, where gapped boundaries of an appropriate 4+1D QFT capture the symmetry algebras of subsystem and higher-form symmetries (see Refs.~\onlinecite{shao2023review,sakurareview} for a review of the established Symmetry TQFT paradigm). An obvious choice for the bulk 4+1D model corresponds to a stack of 3+1D cluster states~\cite{Raussendorf2005} in 4+1D. Our gauging web implies that the X-Cube, 3+1D Toric Code can also arise as gapped boundary of this 4+1D bulk. It would be interesting to work out these boundary conditions explicitly.

We also note that the 3+1D SPT we have uncovered in this paper is protected by a 1-form symmetry as well as a planar subsystem symmetry. It is natural to ask whether this SPT remains non-trivial when the 1-form symmetry is restricted to a planar subsystem symmetry. Finally, an especially intriguing question is developing an algebraic framework for capturing the non-trivial symmetry extensions we have found in this work: incorporating the role of translation symmetry explicitly or considering the full chain complex (see e.g. Refs.~\onlinecite{williamson,kubica}) capturing the symmetries may provide a viable route forward.


\stoptoc
\begin{acknowledgements}
We are especially thankful to Nati Seiberg for stimulating discussions and for his enthusiastic encouragement along this unexpected journey, which required frequent breaks for second breakfast. N.T.~is grateful to Xie Chen and Mike Hermele for helpful discussions. This research was supported in part by grant NSF PHY-2309135 to the Kavli Institute for Theoretical Physics (KITP). Part of this work was conducted while P.G.~and A.P.~were visiting the Okinawa Institute of Science and Technology (OIST) through the Theoretical Sciences Visiting Program (TSVP). P.G.~was supported by the Physics Department of Princeton University and the Simons Collaboration on Global Categorical Symmetries. This material is based upon work supported by the Sivian Fund and the Paul Dirac Fund at the Institute for Advanced Study and the U.S. Department of Energy, Office of Science, Office of High Energy Physics under Award Number DE-SC0009988 (A.P.). N.T.~was supported by the Walter Burke Institute for Theoretical
Physics at Caltech. D.W.~is supported by the Australian Research Council
Discovery Early Career Research Award (DE220100625). 
The authors of this paper were ordered alphabetically.
\end{acknowledgements}


\resumetoc

\clearpage
\appendix


\section{Lattice model for gauged $\text{SPT}_p$}
\label{app:gaugedSPT}

In this appendix, we provide the lattice model for the $G=\ZZ_N$ gauged $\text{SPT}_p$. We will use the clock and shift operators satisfying  $Z^N = 1, X^N = 1, ZX = e^{2\pi i/N}XZ$.
 \begin{equation}
\begin{aligned}
A^\fol_{v,yz} &=  \raisebox{-0.5\height}{\includegraphics{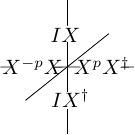}},& B^\text{fol}_{p_{yz}} &=  \raisebox{-0.5\height}{\includegraphics{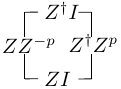}},\\
 A^\fol_{v,xz} &=  \raisebox{-0.5\height}{\includegraphics{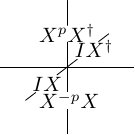}},& B^\text{fol}_{p_{xz}} &=  \raisebox{-0.5\height}{\includegraphics{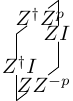}},\\  
A^\fol_{v,xy} &=  \raisebox{-0.5\height}{\includegraphics{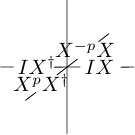}},& B^\text{fol}_{p_{xy}} &=  \raisebox{-0.5\height}{\includegraphics{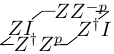}},\\
A_v &=  \raisebox{-0.5\height}{\includegraphics{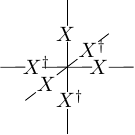}} \otimes \mathbbm 1, &
W_c &= \mathbbm 1 \otimes  \raisebox{-0.5\height}{\includegraphics{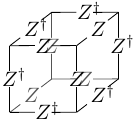}}.
\end{aligned}
\label{eq:TCstacksZN}
\end{equation}
The last two stabilizers $A_v$ and $W_c$ are the gauge constraints arising from gauging the 1-form symmetry and the lineon planar symmetry, respectively.

For $p=0$ the two sets of qudits decouple. $B^\text{fol}_p$ and $A_v$ realizes the Toric Code on the first qudit, while $A^\text{fol}_p$ and $W_c$ realizes the X-Cube model.

Let us show that if $N$ and $p$ are coprime, this model reduces to a 3-foliated stack of 2+1D Toric Codes. First, note that  $(A^\text{fol}_{v,yz}A^\text{fol}_{v,xz}A^\text{fol}_{v,xy})^{1/p} =A_v $ where $1/p$ denotes the integer that satisfies $(1/p) \times p = 1 {\text{ (mod }N)}$. This is well defined because $N$ and $p$ are coprime.
Similarly, the product of $B^\text{fol}_{p}$ around a cube generates $W_c$. Therefore, we can discard the final two stabilizers. For the special case of $N=2$ and $p=1$, this reduces to the stabilizers shown in  Eq.~\eqref{eq:TCstacks}.

We now perform a local unitary
\begin{align}
   U = \prod_e (V(p) \otimes \mathbbm 1)_e \times (CX_{e})^{1/p} ,
\end{align}
where $V(p)$ is the qudit automorphism $V(p) = \sum_n \ket{n}\bra{pn}$ which acts as
\begin{align}
    X^p &\xmapsto{V(p)} X &  Z &\xmapsto{V(p)} Z^p
\end{align}
and the qudit $CX$ gate acts on the two qudits as
\begin{equation}
\begin{aligned}
    XI &\xmapsto{CX} XX, & IX &\xmapsto{CX} IX,\\
    ZI &\xmapsto{CX} ZI, & IZ &\xmapsto{CX} Z^\dagger Z.
\end{aligned}
\end{equation}
This unitary is the generalization of Eq.~\eqref{eq:decoupledbasistransform} in the qubit case. Explicitly, its action is
\begin{equation}
\begin{aligned}
    X^pX^\dagger &\xmapsto{U} XI , & IX &\xmapsto{U} IX ,\\
    ZI &\xmapsto{U} Z^pI , & Z^\dagger Z^{p} &\xmapsto{U} IZ^p .
\end{aligned}
\end{equation}
Thus, the stabilizers map as
  \begin{equation}
\begin{aligned}
A^\fol_{v,yz} &\xmapsto{U} \raisebox{-0.5\height}{\includegraphics{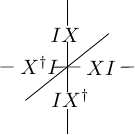}},&B^\text{fol}_{p_{yz}} &\xmapsto{U}  \left(\raisebox{-0.5\height}{\includegraphics{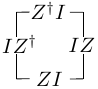}}\right)^p,\\
A^\fol_{v,xz} &\xmapsto{U}  \raisebox{-0.5\height}{\includegraphics{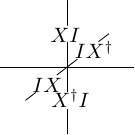}},&B^\text{fol}_{p_{xz}} &\xmapsto{U}  \left(\raisebox{-0.5\height}{\includegraphics{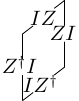}}\right)^p,\\  
A^\fol_{v,xy} &\xmapsto{U}  \raisebox{-0.5\height}{\includegraphics{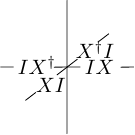}},&B^\text{fol}_{p_{xy}} &\xmapsto{U}   \left(\raisebox{-0.5\height}{\includegraphics{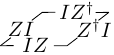}}\right)^p,
\end{aligned}
\end{equation}
which describes the stabilizers of decoupled Toric Codes.

We remark that the fact that the gauged SPTs for distinct indices $p$ give rise to the same topological order after gauging is not surprising. These models couple to the background gauge field of the dual symmetry differently, and therefore one can use the coefficient of this coupling to distinguish the SPTs. A similar phenomenon happens in $\ZZ_N \times \ZZ_N$ twisted gauge theory in 2+1D with a type-II cocycle when N is prime. Labelling the index of this cocycle by $p$, for any non-zero $p$, the resulting group extension
\begin{align}
    1 \to \ZZ_N \to \ZZ_{N^2} \to \ZZ_N \to 1
\end{align}
then always gives rise to a $\ZZ_{N^2}$ group. Therefore, the topological order is always equivalent to a $\ZZ_{N^2}$ Toric Code.


\stoptoc
\bibliography{library}

\begin{thebibliography}{81}%
\makeatletter
\providecommand \@ifxundefined [1]{%
 \@ifx{#1\undefined}
}%
\providecommand \@ifnum [1]{%
 \ifnum #1\expandafter \@firstoftwo
 \else \expandafter \@secondoftwo
 \fi
}%
\providecommand \@ifx [1]{%
 \ifx #1\expandafter \@firstoftwo
 \else \expandafter \@secondoftwo
 \fi
}%
\providecommand \natexlab [1]{#1}%
\providecommand \enquote  [1]{``#1''}%
\providecommand \bibnamefont  [1]{#1}%
\providecommand \bibfnamefont [1]{#1}%
\providecommand \citenamefont [1]{#1}%
\providecommand \href@noop [0]{\@secondoftwo}%
\providecommand \href [0]{\begingroup \@sanitize@url \@href}%
\providecommand \@href[1]{\@@startlink{#1}\@@href}%
\providecommand \@@href[1]{\endgroup#1\@@endlink}%
\providecommand \@sanitize@url [0]{\catcode `\\12\catcode `\$12\catcode
  `\&12\catcode `\#12\catcode `\^12\catcode `\_12\catcode `\%12\relax}%
\providecommand \@@startlink[1]{}%
\providecommand \@@endlink[0]{}%
\providecommand \url  [0]{\begingroup\@sanitize@url \@url }%
\providecommand \@url [1]{\endgroup\@href {#1}{\urlprefix }}%
\providecommand \urlprefix  [0]{URL }%
\providecommand \Eprint [0]{\href }%
\providecommand \doibase [0]{https://doi.org/}%
\providecommand \selectlanguage [0]{\@gobble}%
\providecommand \bibinfo  [0]{\@secondoftwo}%
\providecommand \bibfield  [0]{\@secondoftwo}%
\providecommand \translation [1]{[#1]}%
\providecommand \BibitemOpen [0]{}%
\providecommand \bibitemStop [0]{}%
\providecommand \bibitemNoStop [0]{.\EOS\space}%
\providecommand \EOS [0]{\spacefactor3000\relax}%
\providecommand \BibitemShut  [1]{\csname bibitem#1\endcsname}%
\let\auto@bib@innerbib\@empty
\bibitem [{\citenamefont {{Nandkishore}}\ and\ \citenamefont
  {{Hermele}}(2019)}]{fractonreview}%
  \BibitemOpen
  \bibfield  {author} {\bibinfo {author} {\bibfnamefont {R.~M.}\ \bibnamefont
  {{Nandkishore}}}\ and\ \bibinfo {author} {\bibfnamefont {M.}~\bibnamefont
  {{Hermele}}},\ }\bibfield  {title} {\bibinfo {title} {{Fractons}},\ }\href
  {https://doi.org/10.1146/annurev-conmatphys-031218-013604} {\bibfield
  {journal} {\bibinfo  {journal} {Annual Review of Condensed Matter Physics}\
  }\textbf {\bibinfo {volume} {10}},\ \bibinfo {pages} {295} (\bibinfo {year}
  {2019})}\BibitemShut {NoStop}%
\bibitem [{\citenamefont {{Pretko}}\ \emph {et~al.}(2020)\citenamefont
  {{Pretko}}, \citenamefont {{Chen}},\ and\ \citenamefont
  {{You}}}]{fractonreview2}%
  \BibitemOpen
  \bibfield  {author} {\bibinfo {author} {\bibfnamefont {M.}~\bibnamefont
  {{Pretko}}}, \bibinfo {author} {\bibfnamefont {X.}~\bibnamefont {{Chen}}},\
  and\ \bibinfo {author} {\bibfnamefont {Y.}~\bibnamefont {{You}}},\ }\bibfield
   {title} {\bibinfo {title} {{Fracton phases of matter}},\ }\href
  {https://doi.org/10.1142/S0217751X20300033} {\bibfield  {journal} {\bibinfo
  {journal} {International Journal of Modern Physics A}\ }\textbf {\bibinfo
  {volume} {35}},\ \bibinfo {eid} {2030003} (\bibinfo {year}
  {2020})}\BibitemShut {NoStop}%
\bibitem [{\citenamefont {Haah}(2011)}]{haah}%
  \BibitemOpen
  \bibfield  {author} {\bibinfo {author} {\bibfnamefont {J.}~\bibnamefont
  {Haah}},\ }\bibfield  {title} {\bibinfo {title} {Local stabilizer codes in
  three dimensions without string logical operators},\ }\href
  {https://doi.org/10.1103/PhysRevA.83.042330} {\bibfield  {journal} {\bibinfo
  {journal} {Phys. Rev. A}\ }\textbf {\bibinfo {volume} {83}},\ \bibinfo
  {pages} {042330} (\bibinfo {year} {2011})}\BibitemShut {NoStop}%
\bibitem [{\citenamefont {Bravyi}\ and\ \citenamefont {Haah}(2013)}]{haah2}%
  \BibitemOpen
  \bibfield  {author} {\bibinfo {author} {\bibfnamefont {S.}~\bibnamefont
  {Bravyi}}\ and\ \bibinfo {author} {\bibfnamefont {J.}~\bibnamefont {Haah}},\
  }\bibfield  {title} {\bibinfo {title} {Quantum self-correction in the 3d
  cubic code model},\ }\href {https://doi.org/10.1103/PhysRevLett.111.200501}
  {\bibfield  {journal} {\bibinfo  {journal} {Phys. Rev. Lett.}\ }\textbf
  {\bibinfo {volume} {111}},\ \bibinfo {pages} {200501} (\bibinfo {year}
  {2013})}\BibitemShut {NoStop}%
\bibitem [{\citenamefont {Vijay}\ \emph {et~al.}(2015)\citenamefont {Vijay},
  \citenamefont {Haah},\ and\ \citenamefont {Fu}}]{fracton1}%
  \BibitemOpen
  \bibfield  {author} {\bibinfo {author} {\bibfnamefont {S.}~\bibnamefont
  {Vijay}}, \bibinfo {author} {\bibfnamefont {J.}~\bibnamefont {Haah}},\ and\
  \bibinfo {author} {\bibfnamefont {L.}~\bibnamefont {Fu}},\ }\bibfield
  {title} {\bibinfo {title} {A new kind of topological quantum order: A
  dimensional hierarchy of quasiparticles built from stationary excitations},\
  }\href {https://doi.org/10.1103/PhysRevB.92.235136} {\bibfield  {journal}
  {\bibinfo  {journal} {Phys. Rev. B}\ }\textbf {\bibinfo {volume} {92}},\
  \bibinfo {pages} {235136} (\bibinfo {year} {2015})}\BibitemShut {NoStop}%
\bibitem [{\citenamefont {Vijay}\ \emph {et~al.}(2016)\citenamefont {Vijay},
  \citenamefont {Haah},\ and\ \citenamefont {Fu}}]{fracton2}%
  \BibitemOpen
  \bibfield  {author} {\bibinfo {author} {\bibfnamefont {S.}~\bibnamefont
  {Vijay}}, \bibinfo {author} {\bibfnamefont {J.}~\bibnamefont {Haah}},\ and\
  \bibinfo {author} {\bibfnamefont {L.}~\bibnamefont {Fu}},\ }\bibfield
  {title} {\bibinfo {title} {Fracton topological order, generalized lattice
  gauge theory, and duality},\ }\href
  {https://doi.org/10.1103/PhysRevB.94.235157} {\bibfield  {journal} {\bibinfo
  {journal} {Phys. Rev. B}\ }\textbf {\bibinfo {volume} {94}},\ \bibinfo
  {pages} {235157} (\bibinfo {year} {2016})}\BibitemShut {NoStop}%
\bibitem [{\citenamefont {Williamson}(2016)}]{williamson}%
  \BibitemOpen
  \bibfield  {author} {\bibinfo {author} {\bibfnamefont {D.~J.}\ \bibnamefont
  {Williamson}},\ }\bibfield  {title} {\bibinfo {title} {Fractal symmetries:
  Ungauging the cubic code},\ }\href
  {https://doi.org/10.1103/PhysRevB.94.155128} {\bibfield  {journal} {\bibinfo
  {journal} {Phys. Rev. B}\ }\textbf {\bibinfo {volume} {94}},\ \bibinfo
  {pages} {155128} (\bibinfo {year} {2016})}\BibitemShut {NoStop}%
\bibitem [{\citenamefont {Shirley}\ \emph {et~al.}(2018)\citenamefont
  {Shirley}, \citenamefont {Slagle}, \citenamefont {Wang},\ and\ \citenamefont
  {Chen}}]{shirleygeneral}%
  \BibitemOpen
  \bibfield  {author} {\bibinfo {author} {\bibfnamefont {W.}~\bibnamefont
  {Shirley}}, \bibinfo {author} {\bibfnamefont {K.}~\bibnamefont {Slagle}},
  \bibinfo {author} {\bibfnamefont {Z.}~\bibnamefont {Wang}},\ and\ \bibinfo
  {author} {\bibfnamefont {X.}~\bibnamefont {Chen}},\ }\bibfield  {title}
  {\bibinfo {title} {Fracton models on general three-dimensional manifolds},\
  }\href {https://doi.org/10.1103/PhysRevX.8.031051} {\bibfield  {journal}
  {\bibinfo  {journal} {Phys. Rev. X}\ }\textbf {\bibinfo {volume} {8}},\
  \bibinfo {pages} {031051} (\bibinfo {year} {2018})}\BibitemShut {NoStop}%
\bibitem [{\citenamefont {Slagle}\ and\ \citenamefont {Kim}(2018)}]{slagle3}%
  \BibitemOpen
  \bibfield  {author} {\bibinfo {author} {\bibfnamefont {K.}~\bibnamefont
  {Slagle}}\ and\ \bibinfo {author} {\bibfnamefont {Y.~B.}\ \bibnamefont
  {Kim}},\ }\bibfield  {title} {\bibinfo {title} {X-cube model on generic
  lattices: Fracton phases and geometric order},\ }\href
  {https://doi.org/10.1103/PhysRevB.97.165106} {\bibfield  {journal} {\bibinfo
  {journal} {Phys. Rev. B}\ }\textbf {\bibinfo {volume} {97}},\ \bibinfo
  {pages} {165106} (\bibinfo {year} {2018})}\BibitemShut {NoStop}%
\bibitem [{\citenamefont {{Slagle}}\ \emph
  {et~al.}(2019{\natexlab{a}})\citenamefont {{Slagle}}, \citenamefont
  {{Prem}},\ and\ \citenamefont {{Pretko}}}]{symmetric}%
  \BibitemOpen
  \bibfield  {author} {\bibinfo {author} {\bibfnamefont {K.}~\bibnamefont
  {{Slagle}}}, \bibinfo {author} {\bibfnamefont {A.}~\bibnamefont {{Prem}}},\
  and\ \bibinfo {author} {\bibfnamefont {M.}~\bibnamefont {{Pretko}}},\
  }\bibfield  {title} {\bibinfo {title} {{Symmetric tensor gauge theories on
  curved spaces}},\ }\href {https://doi.org/10.1016/j.aop.2019.167910}
  {\bibfield  {journal} {\bibinfo  {journal} {Annals of Physics}\ }\textbf
  {\bibinfo {volume} {410}},\ \bibinfo {eid} {167910} (\bibinfo {year}
  {2019}{\natexlab{a}})}\BibitemShut {NoStop}%
\bibitem [{\citenamefont {Slagle}\ and\ \citenamefont
  {Kim}(2017)}]{SlagleXcubeQFT}%
  \BibitemOpen
  \bibfield  {author} {\bibinfo {author} {\bibfnamefont {K.}~\bibnamefont
  {Slagle}}\ and\ \bibinfo {author} {\bibfnamefont {Y.~B.}\ \bibnamefont
  {Kim}},\ }\bibfield  {title} {\bibinfo {title} {Quantum field theory of
  x-cube fracton topological order and robust degeneracy from geometry},\
  }\href {https://doi.org/10.1103/PhysRevB.96.195139} {\bibfield  {journal}
  {\bibinfo  {journal} {Phys. Rev. B}\ }\textbf {\bibinfo {volume} {96}},\
  \bibinfo {pages} {195139} (\bibinfo {year} {2017})}\BibitemShut {NoStop}%
\bibitem [{\citenamefont {Slagle}(2021)}]{Slagle21}%
  \BibitemOpen
  \bibfield  {author} {\bibinfo {author} {\bibfnamefont {K.}~\bibnamefont
  {Slagle}},\ }\bibfield  {title} {\bibinfo {title} {Foliated quantum field
  theory of fracton order},\ }\href
  {https://doi.org/10.1103/PhysRevLett.126.101603} {\bibfield  {journal}
  {\bibinfo  {journal} {Phys. Rev. Lett.}\ }\textbf {\bibinfo {volume} {126}},\
  \bibinfo {pages} {101603} (\bibinfo {year} {2021})}\BibitemShut {NoStop}%
\bibitem [{\citenamefont {Seiberg}(2020)}]{SeibergSymmetry}%
  \BibitemOpen
  \bibfield  {author} {\bibinfo {author} {\bibfnamefont {N.}~\bibnamefont
  {Seiberg}},\ }\bibfield  {title} {\bibinfo {title} {{Field Theories With a
  Vector Global Symmetry}},\ }\href
  {https://doi.org/10.21468/SciPostPhys.8.4.050} {\bibfield  {journal}
  {\bibinfo  {journal} {SciPost Phys.}\ }\textbf {\bibinfo {volume} {8}},\
  \bibinfo {pages} {50} (\bibinfo {year} {2020})}\BibitemShut {NoStop}%
\bibitem [{\citenamefont {Seiberg}\ and\ \citenamefont
  {Shao}(2021)}]{Seiberg:2020bhn}%
  \BibitemOpen
  \bibfield  {author} {\bibinfo {author} {\bibfnamefont {N.}~\bibnamefont
  {Seiberg}}\ and\ \bibinfo {author} {\bibfnamefont {S.-H.}\ \bibnamefont
  {Shao}},\ }\bibfield  {title} {\bibinfo {title} {{Exotic Symmetries, Duality,
  and Fractons in 2+1-Dimensional Quantum Field Theory}},\ }\href
  {https://doi.org/10.21468/SciPostPhys.10.2.027} {\bibfield  {journal}
  {\bibinfo  {journal} {SciPost Phys.}\ }\textbf {\bibinfo {volume} {10}},\
  \bibinfo {pages} {027} (\bibinfo {year} {2021})},\ \Eprint
  {https://arxiv.org/abs/2003.10466} {arXiv:2003.10466 [cond-mat.str-el]}
  \BibitemShut {NoStop}%
\bibitem [{\citenamefont {{Seiberg}}\ and\ \citenamefont
  {{Shao}}(2021)}]{seiberg2021zn}%
  \BibitemOpen
  \bibfield  {author} {\bibinfo {author} {\bibfnamefont {N.}~\bibnamefont
  {{Seiberg}}}\ and\ \bibinfo {author} {\bibfnamefont {S.-H.}\ \bibnamefont
  {{Shao}}},\ }\bibfield  {title} {\bibinfo {title} {{Exotic $\mathbb{Z}_N$
  symmetries, duality, and fractons in 3+1-dimensional quantum field theory}},\
  }\href {https://doi.org/10.21468/SciPostPhys.10.1.003} {\bibfield  {journal}
  {\bibinfo  {journal} {SciPost Physics}\ }\textbf {\bibinfo {volume} {10}},\
  \bibinfo {eid} {003} (\bibinfo {year} {2021})}\BibitemShut {NoStop}%
\bibitem [{\citenamefont {{Gorantla}}\ \emph {et~al.}(2021)\citenamefont
  {{Gorantla}}, \citenamefont {{Lam}}, \citenamefont {{Seiberg}},\ and\
  \citenamefont {{Shao}}}]{gorantla2021villain}%
  \BibitemOpen
  \bibfield  {author} {\bibinfo {author} {\bibfnamefont {P.}~\bibnamefont
  {{Gorantla}}}, \bibinfo {author} {\bibfnamefont {H.~T.}\ \bibnamefont
  {{Lam}}}, \bibinfo {author} {\bibfnamefont {N.}~\bibnamefont {{Seiberg}}},\
  and\ \bibinfo {author} {\bibfnamefont {S.-H.}\ \bibnamefont {{Shao}}},\
  }\bibfield  {title} {\bibinfo {title} {{A modified Villain formulation of
  fractons and other exotic theories}},\ }\href
  {https://doi.org/10.1063/5.0060808} {\bibfield  {journal} {\bibinfo
  {journal} {Journal of Mathematical Physics}\ }\textbf {\bibinfo {volume}
  {62}},\ \bibinfo {eid} {102301} (\bibinfo {year} {2021})}\BibitemShut
  {NoStop}%
\bibitem [{\citenamefont {{Slagle}}\ \emph
  {et~al.}(2019{\natexlab{b}})\citenamefont {{Slagle}}, \citenamefont
  {{Aasen}},\ and\ \citenamefont {{Williamson}}}]{SlagleSMN}%
  \BibitemOpen
  \bibfield  {author} {\bibinfo {author} {\bibfnamefont {K.}~\bibnamefont
  {{Slagle}}}, \bibinfo {author} {\bibfnamefont {D.}~\bibnamefont {{Aasen}}},\
  and\ \bibinfo {author} {\bibfnamefont {D.}~\bibnamefont {{Williamson}}},\
  }\bibfield  {title} {\bibinfo {title} {Foliated field theory and
  string-membrane-net condensation picture of fracton order},\ }\href
  {https://doi.org/10.21468/SciPostPhys.6.4.043} {\bibfield  {journal}
  {\bibinfo  {journal} {SciPost Physics}\ }\textbf {\bibinfo {volume} {6}},\
  \bibinfo {eid} {043} (\bibinfo {year} {2019}{\natexlab{b}})}\BibitemShut
  {NoStop}%
\bibitem [{\citenamefont {Wen}(2020)}]{Wen2020}%
  \BibitemOpen
  \bibfield  {author} {\bibinfo {author} {\bibfnamefont {X.-G.}\ \bibnamefont
  {Wen}},\ }\bibfield  {title} {\bibinfo {title} {Systematic construction of
  gapped nonliquid states},\ }\href
  {https://doi.org/10.1103/PhysRevResearch.2.033300} {\bibfield  {journal}
  {\bibinfo  {journal} {Phys. Rev. Res.}\ }\textbf {\bibinfo {volume} {2}},\
  \bibinfo {pages} {033300} (\bibinfo {year} {2020})}\BibitemShut {NoStop}%
\bibitem [{\citenamefont {{Aasen}}\ \emph {et~al.}(2020)\citenamefont
  {{Aasen}}, \citenamefont {{Bulmash}}, \citenamefont {{Prem}}, \citenamefont
  {{Slagle}},\ and\ \citenamefont {{Williamson}}}]{defectnetworks}%
  \BibitemOpen
  \bibfield  {author} {\bibinfo {author} {\bibfnamefont {D.}~\bibnamefont
  {{Aasen}}}, \bibinfo {author} {\bibfnamefont {D.}~\bibnamefont {{Bulmash}}},
  \bibinfo {author} {\bibfnamefont {A.}~\bibnamefont {{Prem}}}, \bibinfo
  {author} {\bibfnamefont {K.}~\bibnamefont {{Slagle}}},\ and\ \bibinfo
  {author} {\bibfnamefont {D.~J.}\ \bibnamefont {{Williamson}}},\ }\bibfield
  {title} {\bibinfo {title} {{Topological defect networks for fractons of all
  types}},\ }\href {https://doi.org/10.1103/PhysRevResearch.2.043165}
  {\bibfield  {journal} {\bibinfo  {journal} {Physical Review Research}\
  }\textbf {\bibinfo {volume} {2}},\ \bibinfo {eid} {043165} (\bibinfo {year}
  {2020})},\ \Eprint {https://arxiv.org/abs/2002.05166} {arXiv:2002.05166
  [cond-mat.str-el]} \BibitemShut {NoStop}%
\bibitem [{\citenamefont {Wang}(2022)}]{Juven2020}%
  \BibitemOpen
  \bibfield  {author} {\bibinfo {author} {\bibfnamefont {J.}~\bibnamefont
  {Wang}},\ }\bibfield  {title} {\bibinfo {title} {Nonliquid cellular states:
  Gluing gauge-higher-symmetry-breaking versus gauge-higher-symmetry-extension
  interfacial defects},\ }\href
  {https://doi.org/10.1103/PhysRevResearch.4.023258} {\bibfield  {journal}
  {\bibinfo  {journal} {Phys. Rev. Res.}\ }\textbf {\bibinfo {volume} {4}},\
  \bibinfo {pages} {023258} (\bibinfo {year} {2022})}\BibitemShut {NoStop}%
\bibitem [{\citenamefont {Song}\ \emph {et~al.}(2023)\citenamefont {Song},
  \citenamefont {Dua}, \citenamefont {Shirley},\ and\ \citenamefont
  {Williamson}}]{Song2023}%
  \BibitemOpen
  \bibfield  {author} {\bibinfo {author} {\bibfnamefont {Z.}~\bibnamefont
  {Song}}, \bibinfo {author} {\bibfnamefont {A.}~\bibnamefont {Dua}}, \bibinfo
  {author} {\bibfnamefont {W.}~\bibnamefont {Shirley}},\ and\ \bibinfo {author}
  {\bibfnamefont {D.~J.}\ \bibnamefont {Williamson}},\ }\bibfield  {title}
  {\bibinfo {title} {{Topological Defect Network Representations of Fracton
  Stabilizer Codes}},\ }\href
  {https://doi.org/10.1103/PRXQUANTUM.4.010304/FIGURES/8/MEDIUM} {\bibfield
  {journal} {\bibinfo  {journal} {PRX Quantum}\ }\textbf {\bibinfo {volume}
  {4}},\ \bibinfo {pages} {010304} (\bibinfo {year} {2023})},\ \Eprint
  {https://arxiv.org/abs/2112.14717} {arXiv:2112.14717} \BibitemShut {NoStop}%
\bibitem [{\citenamefont {Yoshida}(2013)}]{yoshida}%
  \BibitemOpen
  \bibfield  {author} {\bibinfo {author} {\bibfnamefont {B.}~\bibnamefont
  {Yoshida}},\ }\bibfield  {title} {\bibinfo {title} {Exotic topological order
  in fractal spin liquids},\ }\href
  {https://doi.org/10.1103/PhysRevB.88.125122} {\bibfield  {journal} {\bibinfo
  {journal} {Phys. Rev. B}\ }\textbf {\bibinfo {volume} {88}},\ \bibinfo
  {pages} {125122} (\bibinfo {year} {2013})}\BibitemShut {NoStop}%
\bibitem [{\citenamefont {Song}\ \emph {et~al.}(2019)\citenamefont {Song},
  \citenamefont {Prem}, \citenamefont {Huang},\ and\ \citenamefont
  {Martin-Delgado}}]{twisted}%
  \BibitemOpen
  \bibfield  {author} {\bibinfo {author} {\bibfnamefont {H.}~\bibnamefont
  {Song}}, \bibinfo {author} {\bibfnamefont {A.}~\bibnamefont {Prem}}, \bibinfo
  {author} {\bibfnamefont {S.-J.}\ \bibnamefont {Huang}},\ and\ \bibinfo
  {author} {\bibfnamefont {M.~A.}\ \bibnamefont {Martin-Delgado}},\ }\bibfield
  {title} {\bibinfo {title} {Twisted fracton models in three dimensions},\
  }\href {https://doi.org/10.1103/PhysRevB.99.155118} {\bibfield  {journal}
  {\bibinfo  {journal} {Phys. Rev. B}\ }\textbf {\bibinfo {volume} {99}},\
  \bibinfo {pages} {155118} (\bibinfo {year} {2019})}\BibitemShut {NoStop}%
\bibitem [{\citenamefont {Prem}\ \emph {et~al.}(2019)\citenamefont {Prem},
  \citenamefont {Huang}, \citenamefont {Song},\ and\ \citenamefont
  {Hermele}}]{cagenet}%
  \BibitemOpen
  \bibfield  {author} {\bibinfo {author} {\bibfnamefont {A.}~\bibnamefont
  {Prem}}, \bibinfo {author} {\bibfnamefont {S.-J.}\ \bibnamefont {Huang}},
  \bibinfo {author} {\bibfnamefont {H.}~\bibnamefont {Song}},\ and\ \bibinfo
  {author} {\bibfnamefont {M.}~\bibnamefont {Hermele}},\ }\bibfield  {title}
  {\bibinfo {title} {Cage-net fracton models},\ }\href
  {https://doi.org/10.1103/PhysRevX.9.021010} {\bibfield  {journal} {\bibinfo
  {journal} {Phys. Rev. X}\ }\textbf {\bibinfo {volume} {9}},\ \bibinfo {pages}
  {021010} (\bibinfo {year} {2019})}\BibitemShut {NoStop}%
\bibitem [{\citenamefont {Prem}\ and\ \citenamefont
  {Williamson}(2019)}]{premgauging}%
  \BibitemOpen
  \bibfield  {author} {\bibinfo {author} {\bibfnamefont {A.}~\bibnamefont
  {Prem}}\ and\ \bibinfo {author} {\bibfnamefont {D.~J.}\ \bibnamefont
  {Williamson}},\ }\bibfield  {title} {\bibinfo {title} {{Gauging permutation
  symmetries as a route to non-Abelian fractons}},\ }\href
  {https://doi.org/10.21468/SciPostPhys.7.5.068} {\bibfield  {journal}
  {\bibinfo  {journal} {SciPost Phys.}\ }\textbf {\bibinfo {volume} {7}},\
  \bibinfo {pages} {68} (\bibinfo {year} {2019})}\BibitemShut {NoStop}%
\bibitem [{\citenamefont {Bulmash}\ and\ \citenamefont
  {Barkeshli}(2019)}]{bulmashgauging}%
  \BibitemOpen
  \bibfield  {author} {\bibinfo {author} {\bibfnamefont {D.}~\bibnamefont
  {Bulmash}}\ and\ \bibinfo {author} {\bibfnamefont {M.}~\bibnamefont
  {Barkeshli}},\ }\bibfield  {title} {\bibinfo {title} {Gauging fractons:
  Immobile non-abelian quasiparticles, fractals, and position-dependent
  degeneracies},\ }\href {https://doi.org/10.1103/PhysRevB.100.155146}
  {\bibfield  {journal} {\bibinfo  {journal} {Phys. Rev. B}\ }\textbf {\bibinfo
  {volume} {100}},\ \bibinfo {pages} {155146} (\bibinfo {year}
  {2019})}\BibitemShut {NoStop}%
\bibitem [{\citenamefont {Tantivasadakarn}\ and\ \citenamefont
  {Vijay}(2020{\natexlab{a}})}]{Tantivasadakarnsearching20}%
  \BibitemOpen
  \bibfield  {author} {\bibinfo {author} {\bibfnamefont {N.}~\bibnamefont
  {Tantivasadakarn}}\ and\ \bibinfo {author} {\bibfnamefont {S.}~\bibnamefont
  {Vijay}},\ }\bibfield  {title} {\bibinfo {title} {Searching for fracton
  orders via symmetry defect condensation},\ }\href
  {https://doi.org/10.1103/PhysRevB.101.165143} {\bibfield  {journal} {\bibinfo
   {journal} {Phys. Rev. B}\ }\textbf {\bibinfo {volume} {101}},\ \bibinfo
  {pages} {165143} (\bibinfo {year} {2020}{\natexlab{a}})}\BibitemShut
  {NoStop}%
\bibitem [{\citenamefont {You}\ \emph {et~al.}(2020)\citenamefont {You},
  \citenamefont {Devakul}, \citenamefont {Sondhi},\ and\ \citenamefont
  {Burnell}}]{FractonCSBF20}%
  \BibitemOpen
  \bibfield  {author} {\bibinfo {author} {\bibfnamefont {Y.}~\bibnamefont
  {You}}, \bibinfo {author} {\bibfnamefont {T.}~\bibnamefont {Devakul}},
  \bibinfo {author} {\bibfnamefont {S.~L.}\ \bibnamefont {Sondhi}},\ and\
  \bibinfo {author} {\bibfnamefont {F.~J.}\ \bibnamefont {Burnell}},\
  }\bibfield  {title} {\bibinfo {title} {Fractonic chern-simons and bf
  theories},\ }\href {https://doi.org/10.1103/PhysRevResearch.2.023249}
  {\bibfield  {journal} {\bibinfo  {journal} {Phys. Rev. Res.}\ }\textbf
  {\bibinfo {volume} {2}},\ \bibinfo {pages} {023249} (\bibinfo {year}
  {2020})}\BibitemShut {NoStop}%
\bibitem [{\citenamefont {Tantivasadakarn}(2020)}]{JWfracton20}%
  \BibitemOpen
  \bibfield  {author} {\bibinfo {author} {\bibfnamefont {N.}~\bibnamefont
  {Tantivasadakarn}},\ }\bibfield  {title} {\bibinfo {title} {Jordan-wigner
  dualities for translation-invariant hamiltonians in any dimension: Emergent
  fermions in fracton topological order},\ }\href
  {https://doi.org/10.1103/PhysRevResearch.2.023353} {\bibfield  {journal}
  {\bibinfo  {journal} {Phys. Rev. Res.}\ }\textbf {\bibinfo {volume} {2}},\
  \bibinfo {pages} {023353} (\bibinfo {year} {2020})}\BibitemShut {NoStop}%
\bibitem [{\citenamefont {Shirley}(2020)}]{Shirley20}%
  \BibitemOpen
  \bibfield  {author} {\bibinfo {author} {\bibfnamefont {W.}~\bibnamefont
  {Shirley}},\ }\href@noop {} {\bibinfo {title} {Fractonic order and emergent
  fermionic gauge theory}} (\bibinfo {year} {2020}),\ \Eprint
  {https://arxiv.org/abs/2002.12026} {arXiv:2002.12026 [cond-mat.str-el]}
  \BibitemShut {NoStop}%
\bibitem [{\citenamefont {Tantivasadakarn}\ \emph
  {et~al.}(2021{\natexlab{a}})\citenamefont {Tantivasadakarn}, \citenamefont
  {Ji},\ and\ \citenamefont {Vijay}}]{TJV1}%
  \BibitemOpen
  \bibfield  {author} {\bibinfo {author} {\bibfnamefont {N.}~\bibnamefont
  {Tantivasadakarn}}, \bibinfo {author} {\bibfnamefont {W.}~\bibnamefont
  {Ji}},\ and\ \bibinfo {author} {\bibfnamefont {S.}~\bibnamefont {Vijay}},\
  }\bibfield  {title} {\bibinfo {title} {Hybrid fracton phases: Parent orders
  for liquid and nonliquid quantum phases},\ }\href
  {https://doi.org/10.1103/PhysRevB.103.245136} {\bibfield  {journal} {\bibinfo
   {journal} {Phys. Rev. B}\ }\textbf {\bibinfo {volume} {103}},\ \bibinfo
  {pages} {245136} (\bibinfo {year} {2021}{\natexlab{a}})}\BibitemShut
  {NoStop}%
\bibitem [{\citenamefont {Tantivasadakarn}\ \emph
  {et~al.}(2021{\natexlab{b}})\citenamefont {Tantivasadakarn}, \citenamefont
  {Ji},\ and\ \citenamefont {Vijay}}]{TJV2}%
  \BibitemOpen
  \bibfield  {author} {\bibinfo {author} {\bibfnamefont {N.}~\bibnamefont
  {Tantivasadakarn}}, \bibinfo {author} {\bibfnamefont {W.}~\bibnamefont
  {Ji}},\ and\ \bibinfo {author} {\bibfnamefont {S.}~\bibnamefont {Vijay}},\
  }\bibfield  {title} {\bibinfo {title} {Non-abelian hybrid fracton orders},\
  }\href {https://doi.org/10.1103/PhysRevB.104.115117} {\bibfield  {journal}
  {\bibinfo  {journal} {Phys. Rev. B}\ }\textbf {\bibinfo {volume} {104}},\
  \bibinfo {pages} {115117} (\bibinfo {year} {2021}{\natexlab{b}})}\BibitemShut
  {NoStop}%
\bibitem [{\citenamefont {Hsin}\ and\ \citenamefont
  {Slagle}(2021)}]{HsinSlagle21}%
  \BibitemOpen
  \bibfield  {author} {\bibinfo {author} {\bibfnamefont {P.-S.}\ \bibnamefont
  {Hsin}}\ and\ \bibinfo {author} {\bibfnamefont {K.}~\bibnamefont {Slagle}},\
  }\bibfield  {title} {\bibinfo {title} {{Comments on foliated gauge theories
  and dualities in 3+1d}},\ }\href
  {https://doi.org/10.21468/SciPostPhys.11.2.032} {\bibfield  {journal}
  {\bibinfo  {journal} {SciPost Phys.}\ }\textbf {\bibinfo {volume} {11}},\
  \bibinfo {pages} {032} (\bibinfo {year} {2021})}\BibitemShut {NoStop}%
\bibitem [{\citenamefont {{Gorantla}}\ \emph {et~al.}(2023)\citenamefont
  {{Gorantla}}, \citenamefont {{Lam}}, \citenamefont {{Seiberg}},\ and\
  \citenamefont {{Shao}}}]{gorantla2023graphs}%
  \BibitemOpen
  \bibfield  {author} {\bibinfo {author} {\bibfnamefont {P.}~\bibnamefont
  {{Gorantla}}}, \bibinfo {author} {\bibfnamefont {H.~T.}\ \bibnamefont
  {{Lam}}}, \bibinfo {author} {\bibfnamefont {N.}~\bibnamefont {{Seiberg}}},\
  and\ \bibinfo {author} {\bibfnamefont {S.-H.}\ \bibnamefont {{Shao}}},\
  }\bibfield  {title} {\bibinfo {title} {{Gapped lineon and fracton models on
  graphs}},\ }\href {https://doi.org/10.1103/PhysRevB.107.125121} {\bibfield
  {journal} {\bibinfo  {journal} {\prb}\ }\textbf {\bibinfo {volume} {107}},\
  \bibinfo {eid} {125121} (\bibinfo {year} {2023})}\BibitemShut {NoStop}%
\bibitem [{\citenamefont {Boesl}\ \emph {et~al.}(2025)\citenamefont {Boesl},
  \citenamefont {Liu}, \citenamefont {Xu}, \citenamefont {Pollmann},\ and\
  \citenamefont {Knap}}]{boesl2025}%
  \BibitemOpen
  \bibfield  {author} {\bibinfo {author} {\bibfnamefont {J.}~\bibnamefont
  {Boesl}}, \bibinfo {author} {\bibfnamefont {Y.-J.}\ \bibnamefont {Liu}},
  \bibinfo {author} {\bibfnamefont {W.-T.}\ \bibnamefont {Xu}}, \bibinfo
  {author} {\bibfnamefont {F.}~\bibnamefont {Pollmann}},\ and\ \bibinfo
  {author} {\bibfnamefont {M.}~\bibnamefont {Knap}},\ }\bibfield  {title}
  {\bibinfo {title} {Quantum phase transitions between symmetry-enriched
  fracton phases},\ }\href@noop {} {\bibfield  {journal} {\bibinfo  {journal}
  {arXiv preprint arXiv:2501.18688}\ } (\bibinfo {year} {2025})}\BibitemShut
  {NoStop}%
\bibitem [{\citenamefont {{Dua}}\ \emph {et~al.}(2019)\citenamefont {{Dua}},
  \citenamefont {{Kim}}, \citenamefont {{Cheng}},\ and\ \citenamefont
  {{Williamson}}}]{DuaSorting}%
  \BibitemOpen
  \bibfield  {author} {\bibinfo {author} {\bibfnamefont {A.}~\bibnamefont
  {{Dua}}}, \bibinfo {author} {\bibfnamefont {I.~H.}\ \bibnamefont {{Kim}}},
  \bibinfo {author} {\bibfnamefont {M.}~\bibnamefont {{Cheng}}},\ and\ \bibinfo
  {author} {\bibfnamefont {D.~J.}\ \bibnamefont {{Williamson}}},\ }\bibfield
  {title} {\bibinfo {title} {{Sorting topological stabilizer models in three
  dimensions}},\ }\href {https://doi.org/10.1103/PhysRevB.100.155137}
  {\bibfield  {journal} {\bibinfo  {journal} {\prb}\ }\textbf {\bibinfo
  {volume} {100}},\ \bibinfo {eid} {155137} (\bibinfo {year}
  {2019})}\BibitemShut {NoStop}%
\bibitem [{\citenamefont {You}\ \emph {et~al.}(2018)\citenamefont {You},
  \citenamefont {Devakul}, \citenamefont {Burnell},\ and\ \citenamefont
  {Sondhi}}]{yizhi1}%
  \BibitemOpen
  \bibfield  {author} {\bibinfo {author} {\bibfnamefont {Y.}~\bibnamefont
  {You}}, \bibinfo {author} {\bibfnamefont {T.}~\bibnamefont {Devakul}},
  \bibinfo {author} {\bibfnamefont {F.~J.}\ \bibnamefont {Burnell}},\ and\
  \bibinfo {author} {\bibfnamefont {S.~L.}\ \bibnamefont {Sondhi}},\ }\bibfield
   {title} {\bibinfo {title} {Subsystem symmetry protected topological order},\
  }\href {https://doi.org/10.1103/PhysRevB.98.035112} {\bibfield  {journal}
  {\bibinfo  {journal} {Phys. Rev. B}\ }\textbf {\bibinfo {volume} {98}},\
  \bibinfo {pages} {035112} (\bibinfo {year} {2018})}\BibitemShut {NoStop}%
\bibitem [{\citenamefont {Devakul}\ \emph {et~al.}(2018)\citenamefont
  {Devakul}, \citenamefont {Williamson},\ and\ \citenamefont
  {You}}]{strongsspt}%
  \BibitemOpen
  \bibfield  {author} {\bibinfo {author} {\bibfnamefont {T.}~\bibnamefont
  {Devakul}}, \bibinfo {author} {\bibfnamefont {D.~J.}\ \bibnamefont
  {Williamson}},\ and\ \bibinfo {author} {\bibfnamefont {Y.}~\bibnamefont
  {You}},\ }\bibfield  {title} {\bibinfo {title} {Classification of subsystem
  symmetry-protected topological phases},\ }\href
  {https://doi.org/10.1103/PhysRevB.98.235121} {\bibfield  {journal} {\bibinfo
  {journal} {Phys. Rev. B}\ }\textbf {\bibinfo {volume} {98}},\ \bibinfo
  {pages} {235121} (\bibinfo {year} {2018})}\BibitemShut {NoStop}%
\bibitem [{\citenamefont {Williamson}\ \emph
  {et~al.}(2019{\natexlab{a}})\citenamefont {Williamson}, \citenamefont {Dua},\
  and\ \citenamefont {Cheng}}]{spurious}%
  \BibitemOpen
  \bibfield  {author} {\bibinfo {author} {\bibfnamefont {D.~J.}\ \bibnamefont
  {Williamson}}, \bibinfo {author} {\bibfnamefont {A.}~\bibnamefont {Dua}},\
  and\ \bibinfo {author} {\bibfnamefont {M.}~\bibnamefont {Cheng}},\ }\bibfield
   {title} {\bibinfo {title} {Spurious topological entanglement entropy from
  subsystem symmetries},\ }\href
  {https://doi.org/10.1103/PhysRevLett.122.140506} {\bibfield  {journal}
  {\bibinfo  {journal} {Phys. Rev. Lett.}\ }\textbf {\bibinfo {volume} {122}},\
  \bibinfo {pages} {140506} (\bibinfo {year} {2019}{\natexlab{a}})}\BibitemShut
  {NoStop}%
\bibitem [{\citenamefont {Williamson}\ \emph
  {et~al.}(2019{\natexlab{b}})\citenamefont {Williamson}, \citenamefont {Bi},\
  and\ \citenamefont {Cheng}}]{williamsonSET}%
  \BibitemOpen
  \bibfield  {author} {\bibinfo {author} {\bibfnamefont {D.~J.}\ \bibnamefont
  {Williamson}}, \bibinfo {author} {\bibfnamefont {Z.}~\bibnamefont {Bi}},\
  and\ \bibinfo {author} {\bibfnamefont {M.}~\bibnamefont {Cheng}},\ }\bibfield
   {title} {\bibinfo {title} {{Fractonic matter in symmetry-enriched U(1) gauge
  theory}},\ }\href {https://doi.org/10.1103/PhysRevB.100.125150} {\bibfield
  {journal} {\bibinfo  {journal} {Phys. Rev. B}\ }\textbf {\bibinfo {volume}
  {100}},\ \bibinfo {pages} {125150} (\bibinfo {year}
  {2019}{\natexlab{b}})}\BibitemShut {NoStop}%
\bibitem [{\citenamefont {{Shirley}}\ \emph {et~al.}(2019)\citenamefont
  {{Shirley}}, \citenamefont {{Slagle}},\ and\ \citenamefont
  {{Chen}}}]{shirleygauging}%
  \BibitemOpen
  \bibfield  {author} {\bibinfo {author} {\bibfnamefont {W.}~\bibnamefont
  {{Shirley}}}, \bibinfo {author} {\bibfnamefont {K.}~\bibnamefont
  {{Slagle}}},\ and\ \bibinfo {author} {\bibfnamefont {X.}~\bibnamefont
  {{Chen}}},\ }\bibfield  {title} {\bibinfo {title} {{Foliated fracton order
  from gauging subsystem symmetries}},\ }\href
  {https://doi.org/10.21468/SciPostPhys.6.4.041} {\bibfield  {journal}
  {\bibinfo  {journal} {SciPost Physics}\ }\textbf {\bibinfo {volume} {6}},\
  \bibinfo {eid} {041} (\bibinfo {year} {2019})}\BibitemShut {NoStop}%
\bibitem [{\citenamefont {Stephen}\ \emph {et~al.}(2020)\citenamefont
  {Stephen}, \citenamefont {Garre-Rubio}, \citenamefont {Dua},\ and\
  \citenamefont {Williamson}}]{Stephen2020}%
  \BibitemOpen
  \bibfield  {author} {\bibinfo {author} {\bibfnamefont {D.~T.}\ \bibnamefont
  {Stephen}}, \bibinfo {author} {\bibfnamefont {J.}~\bibnamefont
  {Garre-Rubio}}, \bibinfo {author} {\bibfnamefont {A.}~\bibnamefont {Dua}},\
  and\ \bibinfo {author} {\bibfnamefont {D.~J.}\ \bibnamefont {Williamson}},\
  }\bibfield  {title} {\bibinfo {title} {{Subsystem symmetry enriched
  topological order in three dimensions}},\ }\bibfield  {journal} {\bibinfo
  {journal} {Physical Review Research}\ }\textbf {\bibinfo {volume} {2}},\
  \href {https://doi.org/10.1103/physrevresearch.2.033331}
  {10.1103/physrevresearch.2.033331} (\bibinfo {year} {2020}),\ \Eprint
  {https://arxiv.org/abs/2004.04181} {arXiv:2004.04181} \BibitemShut {NoStop}%
\bibitem [{\citenamefont {Shirley}\ \emph {et~al.}(2023)\citenamefont
  {Shirley}, \citenamefont {Liu},\ and\ \citenamefont {Dua}}]{Shirley23}%
  \BibitemOpen
  \bibfield  {author} {\bibinfo {author} {\bibfnamefont {W.}~\bibnamefont
  {Shirley}}, \bibinfo {author} {\bibfnamefont {X.}~\bibnamefont {Liu}},\ and\
  \bibinfo {author} {\bibfnamefont {A.}~\bibnamefont {Dua}},\ }\bibfield
  {title} {\bibinfo {title} {Emergent fermionic gauge theory and foliated
  fracton order in the chamon model},\ }\href
  {https://doi.org/10.1103/PhysRevB.107.035136} {\bibfield  {journal} {\bibinfo
   {journal} {Phys. Rev. B}\ }\textbf {\bibinfo {volume} {107}},\ \bibinfo
  {pages} {035136} (\bibinfo {year} {2023})}\BibitemShut {NoStop}%
\bibitem [{\citenamefont {{Vijay}}(2017)}]{sagar}%
  \BibitemOpen
  \bibfield  {author} {\bibinfo {author} {\bibfnamefont {S.}~\bibnamefont
  {{Vijay}}},\ }\bibfield  {title} {\bibinfo {title} {{Isotropic Layer
  Construction and Phase Diagram for Fracton Topological Phases}},\ }\bibfield
  {journal} {\bibinfo  {journal} {arXiv e-prints}\ }\href
  {https://doi.org/10.48550/arXiv.1701.00762} {10.48550/arXiv.1701.00762}
  (\bibinfo {year} {2017}),\ \Eprint {https://arxiv.org/abs/1701.00762}
  {arXiv:1701.00762 [cond-mat.str-el]} \BibitemShut {NoStop}%
\bibitem [{\citenamefont {Ma}\ \emph {et~al.}(2017)\citenamefont {Ma},
  \citenamefont {Lake}, \citenamefont {Chen},\ and\ \citenamefont
  {Hermele}}]{han}%
  \BibitemOpen
  \bibfield  {author} {\bibinfo {author} {\bibfnamefont {H.}~\bibnamefont
  {Ma}}, \bibinfo {author} {\bibfnamefont {E.}~\bibnamefont {Lake}}, \bibinfo
  {author} {\bibfnamefont {X.}~\bibnamefont {Chen}},\ and\ \bibinfo {author}
  {\bibfnamefont {M.}~\bibnamefont {Hermele}},\ }\bibfield  {title} {\bibinfo
  {title} {Fracton topological order via coupled layers},\ }\href
  {https://doi.org/10.1103/PhysRevB.95.245126} {\bibfield  {journal} {\bibinfo
  {journal} {Phys. Rev. B}\ }\textbf {\bibinfo {volume} {95}},\ \bibinfo
  {pages} {245126} (\bibinfo {year} {2017})}\BibitemShut {NoStop}%
\bibitem [{\citenamefont {Hal\'asz}\ \emph {et~al.}(2017)\citenamefont
  {Hal\'asz}, \citenamefont {Hsieh},\ and\ \citenamefont {Balents}}]{balents}%
  \BibitemOpen
  \bibfield  {author} {\bibinfo {author} {\bibfnamefont {G.~B.}\ \bibnamefont
  {Hal\'asz}}, \bibinfo {author} {\bibfnamefont {T.~H.}\ \bibnamefont
  {Hsieh}},\ and\ \bibinfo {author} {\bibfnamefont {L.}~\bibnamefont
  {Balents}},\ }\bibfield  {title} {\bibinfo {title} {Fracton topological
  phases from strongly coupled spin chains},\ }\href
  {https://doi.org/10.1103/PhysRevLett.119.257202} {\bibfield  {journal}
  {\bibinfo  {journal} {Phys. Rev. Lett.}\ }\textbf {\bibinfo {volume} {119}},\
  \bibinfo {pages} {257202} (\bibinfo {year} {2017})}\BibitemShut {NoStop}%
\bibitem [{\citenamefont {Williamson}\ and\ \citenamefont
  {Cheng}(2023)}]{designer}%
  \BibitemOpen
  \bibfield  {author} {\bibinfo {author} {\bibfnamefont {D.~J.}\ \bibnamefont
  {Williamson}}\ and\ \bibinfo {author} {\bibfnamefont {M.}~\bibnamefont
  {Cheng}},\ }\bibfield  {title} {\bibinfo {title} {Designer non-abelian
  fractons from topological layers},\ }\href
  {https://doi.org/10.1103/PhysRevB.107.035103} {\bibfield  {journal} {\bibinfo
   {journal} {Phys. Rev. B}\ }\textbf {\bibinfo {volume} {107}},\ \bibinfo
  {pages} {035103} (\bibinfo {year} {2023})}\BibitemShut {NoStop}%
\bibitem [{\citenamefont {Williamson}\ and\ \citenamefont
  {Devakul}(2021)}]{Williamson2020a}%
  \BibitemOpen
  \bibfield  {author} {\bibinfo {author} {\bibfnamefont {D.~J.}\ \bibnamefont
  {Williamson}}\ and\ \bibinfo {author} {\bibfnamefont {T.}~\bibnamefont
  {Devakul}},\ }\bibfield  {title} {\bibinfo {title} {{Type-II fractons from
  coupled spin chains and layers}},\ }\bibfield  {journal} {\bibinfo  {journal}
  {Physical Review B}\ }\textbf {\bibinfo {volume} {103}},\ \href
  {https://doi.org/10.1103/PhysRevB.103.155140} {10.1103/PhysRevB.103.155140}
  (\bibinfo {year} {2021}),\ \Eprint {https://arxiv.org/abs/2007.07894}
  {arXiv:2007.07894} \BibitemShut {NoStop}%
\bibitem [{\citenamefont {Sullivan}\ \emph {et~al.}(2021)\citenamefont
  {Sullivan}, \citenamefont {Iadecola},\ and\ \citenamefont
  {Williamson}}]{SullivanPlanarpstring}%
  \BibitemOpen
  \bibfield  {author} {\bibinfo {author} {\bibfnamefont {J.}~\bibnamefont
  {Sullivan}}, \bibinfo {author} {\bibfnamefont {T.}~\bibnamefont {Iadecola}},\
  and\ \bibinfo {author} {\bibfnamefont {D.~J.}\ \bibnamefont {Williamson}},\
  }\bibfield  {title} {\bibinfo {title} {Planar p-string condensation: Chiral
  fracton phases from fractional quantum hall layers and beyond},\ }\href
  {https://doi.org/10.1103/PhysRevB.103.205301} {\bibfield  {journal} {\bibinfo
   {journal} {Phys. Rev. B}\ }\textbf {\bibinfo {volume} {103}},\ \bibinfo
  {pages} {205301} (\bibinfo {year} {2021})}\BibitemShut {NoStop}%
\bibitem [{\citenamefont {Ebisu}\ \emph
  {et~al.}(2024{\natexlab{a}})\citenamefont {Ebisu}, \citenamefont {Honda},\
  and\ \citenamefont {Nakanishi}}]{Ebisu:2023idd}%
  \BibitemOpen
  \bibfield  {author} {\bibinfo {author} {\bibfnamefont {H.}~\bibnamefont
  {Ebisu}}, \bibinfo {author} {\bibfnamefont {M.}~\bibnamefont {Honda}},\ and\
  \bibinfo {author} {\bibfnamefont {T.}~\bibnamefont {Nakanishi}},\ }\bibfield
  {title} {\bibinfo {title} {{Foliated field theories and multipole
  symmetries}},\ }\href {https://doi.org/10.1103/PhysRevB.109.165112}
  {\bibfield  {journal} {\bibinfo  {journal} {Phys. Rev. B}\ }\textbf {\bibinfo
  {volume} {109}},\ \bibinfo {pages} {165112} (\bibinfo {year}
  {2024}{\natexlab{a}})},\ \Eprint {https://arxiv.org/abs/2310.06701}
  {arXiv:2310.06701 [cond-mat.str-el]} \BibitemShut {NoStop}%
\bibitem [{\citenamefont {Ebisu}\ \emph
  {et~al.}(2024{\natexlab{b}})\citenamefont {Ebisu}, \citenamefont {Honda},\
  and\ \citenamefont {Nakanishi}}]{Ebisu:2024cke}%
  \BibitemOpen
  \bibfield  {author} {\bibinfo {author} {\bibfnamefont {H.}~\bibnamefont
  {Ebisu}}, \bibinfo {author} {\bibfnamefont {M.}~\bibnamefont {Honda}},\ and\
  \bibinfo {author} {\bibfnamefont {T.}~\bibnamefont {Nakanishi}},\ }\bibfield
  {title} {\bibinfo {title} {{Multipole and fracton topological order via
  gauging foliated symmetry protected topological phases}},\ }\href
  {https://doi.org/10.1103/PhysRevResearch.6.023166} {\bibfield  {journal}
  {\bibinfo  {journal} {Phys. Rev. Res.}\ }\textbf {\bibinfo {volume} {6}},\
  \bibinfo {pages} {023166} (\bibinfo {year} {2024}{\natexlab{b}})},\ \Eprint
  {https://arxiv.org/abs/2401.10677} {arXiv:2401.10677 [cond-mat.str-el]}
  \BibitemShut {NoStop}%
\bibitem [{\citenamefont {Gorantla}\ \emph {et~al.}(2025)\citenamefont
  {Gorantla}, \citenamefont {Prem}, \citenamefont {Tantivasadakarn},\ and\
  \citenamefont {Williamson}}]{GPTW25}%
  \BibitemOpen
  \bibfield  {author} {\bibinfo {author} {\bibfnamefont {P.}~\bibnamefont
  {Gorantla}}, \bibinfo {author} {\bibfnamefont {A.}~\bibnamefont {Prem}},
  \bibinfo {author} {\bibfnamefont {N.}~\bibnamefont {Tantivasadakarn}},\ and\
  \bibinfo {author} {\bibfnamefont {D.~J.}\ \bibnamefont {Williamson}},\
  }\bibfield  {title} {\bibinfo {title} {String membrane nets from higher-form
  gauging: An alternate route to $p$-string condensation},\ }\href
  {https://doi.org/10.1103/qq9n-16hk} {\bibfield  {journal} {\bibinfo
  {journal} {Phys. Rev. B}\ }\textbf {\bibinfo {volume} {112}},\ \bibinfo
  {pages} {125124} (\bibinfo {year} {2025})}\BibitemShut {NoStop}%
\bibitem [{\citenamefont {Vafa}(1989)}]{Vafa:1989ih}%
  \BibitemOpen
  \bibfield  {author} {\bibinfo {author} {\bibfnamefont {C.}~\bibnamefont
  {Vafa}},\ }\bibfield  {title} {\bibinfo {title} {{Quantum Symmetries of
  String Vacua}},\ }\href {https://doi.org/10.1142/S0217732389001842}
  {\bibfield  {journal} {\bibinfo  {journal} {Mod. Phys. Lett. A}\ }\textbf
  {\bibinfo {volume} {4}},\ \bibinfo {pages} {1615} (\bibinfo {year}
  {1989})}\BibitemShut {NoStop}%
\bibitem [{\citenamefont {Tachikawa}(2020)}]{Tachikawa20}%
  \BibitemOpen
  \bibfield  {author} {\bibinfo {author} {\bibfnamefont {Y.}~\bibnamefont
  {Tachikawa}},\ }\bibfield  {title} {\bibinfo {title} {{On gauging finite
  subgroups}},\ }\href {https://doi.org/10.21468/SciPostPhys.8.1.015}
  {\bibfield  {journal} {\bibinfo  {journal} {SciPost Phys.}\ }\textbf
  {\bibinfo {volume} {8}},\ \bibinfo {pages} {015} (\bibinfo {year}
  {2020})}\BibitemShut {NoStop}%
\bibitem [{\citenamefont {Gaiotto}\ and\ \citenamefont
  {Kulp}(2021)}]{GaiottoKulp21}%
  \BibitemOpen
  \bibfield  {author} {\bibinfo {author} {\bibfnamefont {D.}~\bibnamefont
  {Gaiotto}}\ and\ \bibinfo {author} {\bibfnamefont {J.}~\bibnamefont {Kulp}},\
  }\bibfield  {title} {\bibinfo {title} {Orbifold groupoids},\ }\href
  {https://doi.org/10.1007/JHEP02(2021)132} {\bibfield  {journal} {\bibinfo
  {journal} {Journal of High Energy Physics}\ }\textbf {\bibinfo {volume}
  {2021}},\ \bibinfo {pages} {132} (\bibinfo {year} {2021})}\BibitemShut
  {NoStop}%
\bibitem [{\citenamefont {Zhu}\ \emph {et~al.}(2023)\citenamefont {Zhu},
  \citenamefont {Chen}, \citenamefont {Ye},\ and\ \citenamefont
  {Trebst}}]{zhu23}%
  \BibitemOpen
  \bibfield  {author} {\bibinfo {author} {\bibfnamefont {G.-Y.}\ \bibnamefont
  {Zhu}}, \bibinfo {author} {\bibfnamefont {J.-Y.}\ \bibnamefont {Chen}},
  \bibinfo {author} {\bibfnamefont {P.}~\bibnamefont {Ye}},\ and\ \bibinfo
  {author} {\bibfnamefont {S.}~\bibnamefont {Trebst}},\ }\bibfield  {title}
  {\bibinfo {title} {Topological fracton quantum phase transitions by tuning
  exact tensor network states},\ }\href
  {https://doi.org/10.1103/PhysRevLett.130.216704} {\bibfield  {journal}
  {\bibinfo  {journal} {Phys. Rev. Lett.}\ }\textbf {\bibinfo {volume} {130}},\
  \bibinfo {pages} {216704} (\bibinfo {year} {2023})}\BibitemShut {NoStop}%
\bibitem [{\citenamefont {Lake}\ and\ \citenamefont
  {Hermele}(2021)}]{lake2021sub}%
  \BibitemOpen
  \bibfield  {author} {\bibinfo {author} {\bibfnamefont {E.}~\bibnamefont
  {Lake}}\ and\ \bibinfo {author} {\bibfnamefont {M.}~\bibnamefont {Hermele}},\
  }\bibfield  {title} {\bibinfo {title} {Subdimensional criticality:
  Condensation of lineons and planons in the x-cube model},\ }\href
  {https://doi.org/10.1103/PhysRevB.104.165121} {\bibfield  {journal} {\bibinfo
   {journal} {Phys. Rev. B}\ }\textbf {\bibinfo {volume} {104}},\ \bibinfo
  {pages} {165121} (\bibinfo {year} {2021})}\BibitemShut {NoStop}%
\bibitem [{\citenamefont {Rayhaun}\ and\ \citenamefont
  {Williamson}(2023)}]{domhigherform}%
  \BibitemOpen
  \bibfield  {author} {\bibinfo {author} {\bibfnamefont {B.~C.}\ \bibnamefont
  {Rayhaun}}\ and\ \bibinfo {author} {\bibfnamefont {D.~J.}\ \bibnamefont
  {Williamson}},\ }\bibfield  {title} {\bibinfo {title} {{Higher-form subsystem
  symmetry breaking: Subdimensional criticality and fracton phase
  transitions}},\ }\href {https://doi.org/10.21468/SciPostPhys.15.1.017}
  {\bibfield  {journal} {\bibinfo  {journal} {SciPost Phys.}\ }\textbf
  {\bibinfo {volume} {15}},\ \bibinfo {pages} {017} (\bibinfo {year}
  {2023})}\BibitemShut {NoStop}%
\bibitem [{\citenamefont {Stephen}\ \emph {et~al.}(2022)\citenamefont
  {Stephen}, \citenamefont {Dua}, \citenamefont {Garre-Rubio}, \citenamefont
  {Williamson},\ and\ \citenamefont {Hermele}}]{Stephen22}%
  \BibitemOpen
  \bibfield  {author} {\bibinfo {author} {\bibfnamefont {D.~T.}\ \bibnamefont
  {Stephen}}, \bibinfo {author} {\bibfnamefont {A.}~\bibnamefont {Dua}},
  \bibinfo {author} {\bibfnamefont {J.}~\bibnamefont {Garre-Rubio}}, \bibinfo
  {author} {\bibfnamefont {D.~J.}\ \bibnamefont {Williamson}},\ and\ \bibinfo
  {author} {\bibfnamefont {M.}~\bibnamefont {Hermele}},\ }\bibfield  {title}
  {\bibinfo {title} {Fractionalization of subsystem symmetries in two
  dimensions},\ }\href {https://doi.org/10.1103/PhysRevB.106.085104} {\bibfield
   {journal} {\bibinfo  {journal} {Phys. Rev. B}\ }\textbf {\bibinfo {volume}
  {106}},\ \bibinfo {pages} {085104} (\bibinfo {year} {2022})}\BibitemShut
  {NoStop}%
\bibitem [{\citenamefont {Hsin}\ \emph {et~al.}(2025)\citenamefont {Hsin},
  \citenamefont {Stephen}, \citenamefont {Dua},\ and\ \citenamefont
  {Williamson}}]{hsin2025fft}%
  \BibitemOpen
  \bibfield  {author} {\bibinfo {author} {\bibfnamefont {P.-S.}\ \bibnamefont
  {Hsin}}, \bibinfo {author} {\bibfnamefont {D.~T.}\ \bibnamefont {Stephen}},
  \bibinfo {author} {\bibfnamefont {A.}~\bibnamefont {Dua}},\ and\ \bibinfo
  {author} {\bibfnamefont {D.~J.}\ \bibnamefont {Williamson}},\ }\bibfield
  {title} {\bibinfo {title} {{Subsystem symmetry fractionalization and foliated
  field theory}},\ }\href {https://doi.org/10.21468/SciPostPhys.18.5.147}
  {\bibfield  {journal} {\bibinfo  {journal} {SciPost Phys.}\ }\textbf
  {\bibinfo {volume} {18}},\ \bibinfo {pages} {147} (\bibinfo {year}
  {2025})}\BibitemShut {NoStop}%
\bibitem [{\citenamefont {Cobanera}\ \emph {et~al.}(2011)\citenamefont
  {Cobanera}, \citenamefont {Ortiz},\ and\ \citenamefont
  {Nussinov}}]{Cobanera11}%
  \BibitemOpen
  \bibfield  {author} {\bibinfo {author} {\bibfnamefont {E.}~\bibnamefont
  {Cobanera}}, \bibinfo {author} {\bibfnamefont {G.}~\bibnamefont {Ortiz}},\
  and\ \bibinfo {author} {\bibfnamefont {Z.}~\bibnamefont {Nussinov}},\
  }\bibfield  {title} {\bibinfo {title} {{The Bond-Algebraic Approach to
  Dualities}},\ }\href {https://doi.org/10.1080/00018732.2011.619814}
  {\bibfield  {journal} {\bibinfo  {journal} {Adv. Phys.}\ }\textbf {\bibinfo
  {volume} {60}},\ \bibinfo {pages} {679} (\bibinfo {year} {2011})},\ \Eprint
  {https://arxiv.org/abs/1103.2776} {arXiv:1103.2776 [cond-mat.stat-mech]}
  \BibitemShut {NoStop}%
\bibitem [{\citenamefont {Levin}\ and\ \citenamefont
  {Gu}(2012)}]{levin2012braiding}%
  \BibitemOpen
  \bibfield  {author} {\bibinfo {author} {\bibfnamefont {M.}~\bibnamefont
  {Levin}}\ and\ \bibinfo {author} {\bibfnamefont {Z.~C.}\ \bibnamefont {Gu}},\
  }\bibfield  {title} {\bibinfo {title} {{Braiding statistics approach to
  symmetry-protected topological phases}},\ }\href
  {https://doi.org/10.1103/PhysRevB.86.115109} {\bibfield  {journal} {\bibinfo
  {journal} {Physical Review B - Condensed Matter and Materials Physics}\
  }\textbf {\bibinfo {volume} {86}},\ \bibinfo {pages} {115109} (\bibinfo
  {year} {2012})},\ \Eprint {https://arxiv.org/abs/1202.3120} {arXiv:1202.3120}
  \BibitemShut {NoStop}%
\bibitem [{\citenamefont {Haegeman}\ \emph {et~al.}(2015)\citenamefont
  {Haegeman}, \citenamefont {{Van Acoleyen}}, \citenamefont {Schuch},
  \citenamefont {{Ignacio Cirac}},\ and\ \citenamefont
  {Verstraete}}]{HaegemanGaugingpaper}%
  \BibitemOpen
  \bibfield  {author} {\bibinfo {author} {\bibfnamefont {J.}~\bibnamefont
  {Haegeman}}, \bibinfo {author} {\bibfnamefont {K.}~\bibnamefont {{Van
  Acoleyen}}}, \bibinfo {author} {\bibfnamefont {N.}~\bibnamefont {Schuch}},
  \bibinfo {author} {\bibfnamefont {J.}~\bibnamefont {{Ignacio Cirac}}},\ and\
  \bibinfo {author} {\bibfnamefont {F.}~\bibnamefont {Verstraete}},\ }\bibfield
   {title} {\bibinfo {title} {{Gauging quantum states: From global to local
  symmetries in many-body systems}},\ }\href
  {https://doi.org/10.1103/PhysRevX.5.011024} {\bibfield  {journal} {\bibinfo
  {journal} {Physical Review X}\ }\textbf {\bibinfo {volume} {5}},\ \bibinfo
  {pages} {11024} (\bibinfo {year} {2015})},\ \Eprint
  {https://arxiv.org/abs/1407.1025} {arXiv:1407.1025} \BibitemShut {NoStop}%
\bibitem [{\citenamefont {Yoshida}(2016)}]{Yoshida15}%
  \BibitemOpen
  \bibfield  {author} {\bibinfo {author} {\bibfnamefont {B.}~\bibnamefont
  {Yoshida}},\ }\bibfield  {title} {\bibinfo {title} {{Topological phases with
  generalized global symmetries}},\ }\href
  {https://doi.org/10.1103/PhysRevB.93.155131} {\bibfield  {journal} {\bibinfo
  {journal} {Phys. Rev. B}\ }\textbf {\bibinfo {volume} {93}},\ \bibinfo
  {pages} {155131} (\bibinfo {year} {2016})},\ \Eprint
  {https://arxiv.org/abs/1508.03468} {arXiv:1508.03468 [cond-mat.str-el]}
  \BibitemShut {NoStop}%
\bibitem [{\citenamefont {{Kubica}}\ and\ \citenamefont
  {{Yoshida}}(2018)}]{kubica}%
  \BibitemOpen
  \bibfield  {author} {\bibinfo {author} {\bibfnamefont {A.}~\bibnamefont
  {{Kubica}}}\ and\ \bibinfo {author} {\bibfnamefont {B.}~\bibnamefont
  {{Yoshida}}},\ }\bibfield  {title} {\bibinfo {title} {{Ungauging quantum
  error-correcting codes}},\ }\bibfield  {journal} {\bibinfo  {journal} {arXiv
  e-prints}\ }\href {https://doi.org/10.48550/arXiv.1805.01836}
  {10.48550/arXiv.1805.01836} (\bibinfo {year} {2018}),\ \Eprint
  {https://arxiv.org/abs/1805.01836} {arXiv:1805.01836 [quant-ph]} \BibitemShut
  {NoStop}%
\bibitem [{\citenamefont {Radi\v{c}evi\'c}(2019)}]{Radicevic19}%
  \BibitemOpen
  \bibfield  {author} {\bibinfo {author} {\bibfnamefont {D.}~\bibnamefont
  {Radi\v{c}evi\'c}},\ }\bibfield  {title} {\bibinfo {title} {{Systematic
  Constructions of Fracton Theories}},\ }\href@noop {} {\  (\bibinfo {year}
  {2019})},\ \Eprint {https://arxiv.org/abs/1910.06336} {arXiv:1910.06336
  [cond-mat.str-el]} \BibitemShut {NoStop}%
\bibitem [{\citenamefont {Tantivasadakarn}\ and\ \citenamefont
  {Vijay}(2020{\natexlab{b}})}]{TantivasadakarnSearching}%
  \BibitemOpen
  \bibfield  {author} {\bibinfo {author} {\bibfnamefont {N.}~\bibnamefont
  {Tantivasadakarn}}\ and\ \bibinfo {author} {\bibfnamefont {S.}~\bibnamefont
  {Vijay}},\ }\bibfield  {title} {\bibinfo {title} {Searching for fracton
  orders via symmetry defect condensation},\ }\href
  {https://doi.org/10.1103/PhysRevB.101.165143} {\bibfield  {journal} {\bibinfo
   {journal} {Phys. Rev. B}\ }\textbf {\bibinfo {volume} {101}},\ \bibinfo
  {pages} {165143} (\bibinfo {year} {2020}{\natexlab{b}})}\BibitemShut
  {NoStop}%
\bibitem [{\citenamefont {Tantivasadakarn}\ \emph
  {et~al.}(2021{\natexlab{c}})\citenamefont {Tantivasadakarn}, \citenamefont
  {Thorngren}, \citenamefont {Vishwanath},\ and\ \citenamefont
  {Verresen}}]{Tantivasadakarn2021}%
  \BibitemOpen
  \bibfield  {author} {\bibinfo {author} {\bibfnamefont {N.}~\bibnamefont
  {Tantivasadakarn}}, \bibinfo {author} {\bibfnamefont {R.}~\bibnamefont
  {Thorngren}}, \bibinfo {author} {\bibfnamefont {A.}~\bibnamefont
  {Vishwanath}},\ and\ \bibinfo {author} {\bibfnamefont {R.}~\bibnamefont
  {Verresen}},\ }\bibfield  {title} {\bibinfo {title} {{Long-range entanglement
  from measuring symmetry-protected topological phases}},\ }\bibfield
  {journal} {\bibinfo  {journal} {Physical Review X}\ }\textbf {\bibinfo
  {volume} {14}},\ \href {https://doi.org/10.1103/PhysRevX.14.021040}
  {10.1103/PhysRevX.14.021040} (\bibinfo {year} {2021}{\natexlab{c}}),\ \Eprint
  {https://arxiv.org/abs/2112.01519v3} {arXiv:2112.01519v3} \BibitemShut
  {NoStop}%
\bibitem [{\citenamefont {Tantivasadakarn}\ \emph {et~al.}(2023)\citenamefont
  {Tantivasadakarn}, \citenamefont {Vishwanath},\ and\ \citenamefont
  {Verresen}}]{Tantivasadakarn2022}%
  \BibitemOpen
  \bibfield  {author} {\bibinfo {author} {\bibfnamefont {N.}~\bibnamefont
  {Tantivasadakarn}}, \bibinfo {author} {\bibfnamefont {A.}~\bibnamefont
  {Vishwanath}},\ and\ \bibinfo {author} {\bibfnamefont {R.}~\bibnamefont
  {Verresen}},\ }\bibfield  {title} {\bibinfo {title} {Hierarchy of topological
  order from finite-depth unitaries, measurement, and feedforward},\ }\href
  {https://doi.org/10.1103/PRXQuantum.4.020339} {\bibfield  {journal} {\bibinfo
   {journal} {PRX Quantum}\ }\textbf {\bibinfo {volume} {4}},\ \bibinfo {pages}
  {020339} (\bibinfo {year} {2023})}\BibitemShut {NoStop}%
\bibitem [{\citenamefont {Dolev}\ \emph {et~al.}(2022)\citenamefont {Dolev},
  \citenamefont {Calvera}, \citenamefont {Cree},\ and\ \citenamefont
  {Williamson}}]{Dolev21}%
  \BibitemOpen
  \bibfield  {author} {\bibinfo {author} {\bibfnamefont {K.}~\bibnamefont
  {Dolev}}, \bibinfo {author} {\bibfnamefont {V.}~\bibnamefont {Calvera}},
  \bibinfo {author} {\bibfnamefont {S.~S.}\ \bibnamefont {Cree}},\ and\
  \bibinfo {author} {\bibfnamefont {D.~J.}\ \bibnamefont {Williamson}},\
  }\bibfield  {title} {\bibinfo {title} {{Gauging the bulk: generalized gauging
  maps and holographic codes}},\ }\href
  {https://doi.org/10.1007/JHEP05(2022)158} {\bibfield  {journal} {\bibinfo
  {journal} {JHEP}\ }\textbf {\bibinfo {volume} {05}},\ \bibinfo {pages}
  {158}},\ \Eprint {https://arxiv.org/abs/2108.11402} {arXiv:2108.11402
  [quant-ph]} \BibitemShut {NoStop}%
\bibitem [{\citenamefont {Rakovszky}\ and\ \citenamefont
  {Khemani}(2023)}]{Rakovszky23}%
  \BibitemOpen
  \bibfield  {author} {\bibinfo {author} {\bibfnamefont {T.}~\bibnamefont
  {Rakovszky}}\ and\ \bibinfo {author} {\bibfnamefont {V.}~\bibnamefont
  {Khemani}},\ }\bibfield  {title} {\bibinfo {title} {{The Physics of (good)
  LDPC Codes I. Gauging and dualities}},\ }\href@noop {} {\  (\bibinfo {year}
  {2023})},\ \Eprint {https://arxiv.org/abs/2310.16032} {arXiv:2310.16032
  [quant-ph]} \BibitemShut {NoStop}%
\bibitem [{\citenamefont {Verresen}\ \emph {et~al.}(2022)\citenamefont
  {Verresen}, \citenamefont {Borla}, \citenamefont {Vishwanath}, \citenamefont
  {Moroz},\ and\ \citenamefont {Thorngren}}]{verresen2022higgs}%
  \BibitemOpen
  \bibfield  {author} {\bibinfo {author} {\bibfnamefont {R.}~\bibnamefont
  {Verresen}}, \bibinfo {author} {\bibfnamefont {U.}~\bibnamefont {Borla}},
  \bibinfo {author} {\bibfnamefont {A.}~\bibnamefont {Vishwanath}}, \bibinfo
  {author} {\bibfnamefont {S.}~\bibnamefont {Moroz}},\ and\ \bibinfo {author}
  {\bibfnamefont {R.}~\bibnamefont {Thorngren}},\ }\bibfield  {title} {\bibinfo
  {title} {Higgs condensates are symmetry-protected topological phases: I.
  discrete symmetries},\ }\href@noop {} {\bibfield  {journal} {\bibinfo
  {journal} {arXiv preprint arXiv:2211.01376}\ } (\bibinfo {year}
  {2022})}\BibitemShut {NoStop}%
\bibitem [{\citenamefont {{Bulmash}}\ and\ \citenamefont
  {{Iadecola}}(2019)}]{bulmashboundary}%
  \BibitemOpen
  \bibfield  {author} {\bibinfo {author} {\bibfnamefont {D.}~\bibnamefont
  {{Bulmash}}}\ and\ \bibinfo {author} {\bibfnamefont {T.}~\bibnamefont
  {{Iadecola}}},\ }\bibfield  {title} {\bibinfo {title} {{Braiding and Gapped
  Boundaries in Fracton Topological Phases}},\ }\href
  {https://doi.org/10.1103/PhysRevB.99.125132} {\bibfield  {journal} {\bibinfo
  {journal} {Phys. Rev. B}\ }\textbf {\bibinfo {volume} {99}},\ \bibinfo
  {pages} {125132} (\bibinfo {year} {2019})}\BibitemShut {NoStop}%
\bibitem [{\citenamefont {Aitchison}\ \emph {et~al.}(2024)\citenamefont
  {Aitchison}, \citenamefont {Bulmash}, \citenamefont {Dua}, \citenamefont
  {Doherty},\ and\ \citenamefont {Williamson}}]{Aitchison2024}%
  \BibitemOpen
  \bibfield  {author} {\bibinfo {author} {\bibfnamefont {C.~T.}\ \bibnamefont
  {Aitchison}}, \bibinfo {author} {\bibfnamefont {D.}~\bibnamefont {Bulmash}},
  \bibinfo {author} {\bibfnamefont {A.}~\bibnamefont {Dua}}, \bibinfo {author}
  {\bibfnamefont {A.~C.}\ \bibnamefont {Doherty}},\ and\ \bibinfo {author}
  {\bibfnamefont {D.~J.}\ \bibnamefont {Williamson}},\ }\bibfield  {title}
  {\bibinfo {title} {{Boundaries and defects in the cubic code}},\ }\bibfield
  {journal} {\bibinfo  {journal} {Physical Review B}\ }\textbf {\bibinfo
  {volume} {109}},\ \href {https://doi.org/10.1103/PhysRevB.109.205125}
  {10.1103/PhysRevB.109.205125} (\bibinfo {year} {2024}),\ \Eprint
  {https://arxiv.org/abs/2308.00138} {arXiv:2308.00138} \BibitemShut {NoStop}%
\bibitem [{\citenamefont {Ohmori}\ and\ \citenamefont
  {Shimamura}(2023)}]{Ohmori:2022rzz}%
  \BibitemOpen
  \bibfield  {author} {\bibinfo {author} {\bibfnamefont {K.}~\bibnamefont
  {Ohmori}}\ and\ \bibinfo {author} {\bibfnamefont {S.}~\bibnamefont
  {Shimamura}},\ }\bibfield  {title} {\bibinfo {title} {{Foliated-exotic
  duality in fractonic BF theories}},\ }\href
  {https://doi.org/10.21468/SciPostPhys.14.6.164} {\bibfield  {journal}
  {\bibinfo  {journal} {SciPost Phys.}\ }\textbf {\bibinfo {volume} {14}},\
  \bibinfo {pages} {164} (\bibinfo {year} {2023})},\ \Eprint
  {https://arxiv.org/abs/2210.11001} {arXiv:2210.11001 [hep-th]} \BibitemShut
  {NoStop}%
\bibitem [{\citenamefont {Callan}\ and\ \citenamefont
  {Harvey}(1985)}]{Callan:1984sa}%
  \BibitemOpen
  \bibfield  {author} {\bibinfo {author} {\bibfnamefont {C.~G.}\ \bibnamefont
  {Callan}, \bibfnamefont {Jr.}}\ and\ \bibinfo {author} {\bibfnamefont
  {J.~A.}\ \bibnamefont {Harvey}},\ }\bibfield  {title} {\bibinfo {title}
  {{Anomalies and Fermion Zero Modes on Strings and Domain Walls}},\ }\href
  {https://doi.org/10.1016/0550-3213(85)90489-4} {\bibfield  {journal}
  {\bibinfo  {journal} {Nucl. Phys. B}\ }\textbf {\bibinfo {volume} {250}},\
  \bibinfo {pages} {427} (\bibinfo {year} {1985})}\BibitemShut {NoStop}%
\bibitem [{\citenamefont {Shirley}\ \emph {et~al.}(2019)\citenamefont
  {Shirley}, \citenamefont {Slagle},\ and\ \citenamefont {Chen}}]{foliatedcb}%
  \BibitemOpen
  \bibfield  {author} {\bibinfo {author} {\bibfnamefont {W.}~\bibnamefont
  {Shirley}}, \bibinfo {author} {\bibfnamefont {K.}~\bibnamefont {Slagle}},\
  and\ \bibinfo {author} {\bibfnamefont {X.}~\bibnamefont {Chen}},\ }\bibfield
  {title} {\bibinfo {title} {Foliated fracton order in the checkerboard
  model},\ }\href {https://doi.org/10.1103/PhysRevB.99.115123} {\bibfield
  {journal} {\bibinfo  {journal} {Phys. Rev. B}\ }\textbf {\bibinfo {volume}
  {99}},\ \bibinfo {pages} {115123} (\bibinfo {year} {2019})}\BibitemShut
  {NoStop}%
\bibitem [{\citenamefont {Devakul}\ and\ \citenamefont
  {Williamson}(2021)}]{Devakul2020b}%
  \BibitemOpen
  \bibfield  {author} {\bibinfo {author} {\bibfnamefont {T.}~\bibnamefont
  {Devakul}}\ and\ \bibinfo {author} {\bibfnamefont {D.~J.}\ \bibnamefont
  {Williamson}},\ }\bibfield  {title} {\bibinfo {title} {{Fractalizing quantum
  codes}},\ }\bibfield  {journal} {\bibinfo  {journal} {Quantum}\ }\textbf
  {\bibinfo {volume} {5}},\ \href {https://doi.org/10.22331/q-2021-04-22-438}
  {10.22331/q-2021-04-22-438} (\bibinfo {year} {2021}),\ \Eprint
  {https://arxiv.org/abs/2009.01252} {arXiv:2009.01252} \BibitemShut {NoStop}%
\bibitem [{\citenamefont {{Shao}}(2023)}]{shao2023review}%
  \BibitemOpen
  \bibfield  {author} {\bibinfo {author} {\bibfnamefont {S.-H.}\ \bibnamefont
  {{Shao}}},\ }\bibfield  {title} {\bibinfo {title} {{What's Done Cannot Be
  Undone: TASI Lectures on Non-Invertible Symmetries}},\ }\bibfield  {journal}
  {\bibinfo  {journal} {arXiv e-prints}\ }\href
  {https://doi.org/10.48550/arXiv.2308.00747} {10.48550/arXiv.2308.00747}
  (\bibinfo {year} {2023}),\ \Eprint {https://arxiv.org/abs/2308.00747}
  {arXiv:2308.00747 [hep-th]} \BibitemShut {NoStop}%
\bibitem [{\citenamefont {{Sch{\"a}fer-Nameki}}(2024)}]{sakurareview}%
  \BibitemOpen
  \bibfield  {author} {\bibinfo {author} {\bibfnamefont {S.}~\bibnamefont
  {{Sch{\"a}fer-Nameki}}},\ }\bibfield  {title} {\bibinfo {title} {{ICTP
  lectures on (non-)invertible generalized symmetries}},\ }\href
  {https://doi.org/10.1016/j.physrep.2024.01.007} {\bibfield  {journal}
  {\bibinfo  {journal} {Physics Reports}\ }\textbf {\bibinfo {volume} {1063}},\
  \bibinfo {pages} {1} (\bibinfo {year} {2024})}\BibitemShut {NoStop}%
\bibitem [{\citenamefont {Raussendorf}\ \emph {et~al.}(2005)\citenamefont
  {Raussendorf}, \citenamefont {Bravyi},\ and\ \citenamefont
  {Harrington}}]{Raussendorf2005}%
  \BibitemOpen
  \bibfield  {author} {\bibinfo {author} {\bibfnamefont {R.}~\bibnamefont
  {Raussendorf}}, \bibinfo {author} {\bibfnamefont {S.}~\bibnamefont
  {Bravyi}},\ and\ \bibinfo {author} {\bibfnamefont {J.}~\bibnamefont
  {Harrington}},\ }\bibfield  {title} {\bibinfo {title} {{Long-range quantum
  entanglement in noisy cluster states}},\ }\bibfield  {journal} {\bibinfo
  {journal} {Physical Review A - Atomic, Molecular, and Optical Physics}\
  }\textbf {\bibinfo {volume} {71}},\ \href
  {https://doi.org/10.1103/PhysRevA.71.062313} {10.1103/PhysRevA.71.062313}
  (\bibinfo {year} {2005}),\ \Eprint {https://arxiv.org/abs/0407255v2}
  {arXiv:0407255v2 [quant-ph]} \BibitemShut {NoStop}%
\end{thebibliography}%


\end{document}